\documentclass[aps,prc,showpacs,showkeys,superscriptaddress,footinbib,eqsecnum]{revtex4}
\usepackage{graphicx}
\usepackage{amsmath}
\usepackage{amsfonts}
\usepackage{amssymb}
\usepackage{bm}
\usepackage{ulem}
\newcommand{\lea}{\raisebox{-.3ex}{\small $ \
\stackrel{\textstyle <}{\sim} $ }}

\newcommand{\beq}{\begin{equation}}
\newcommand{\eeq}{\end{equation}}
\newcommand{\beqa}{\begin{eqnarray}}
\newcommand{\eeqa}{\end{eqnarray}}
\newcommand{\be}{\begin{eqnarray}}
\newcommand{\ee}{\end{eqnarray}}

\date{\today}

\begin{document}

\title{Local position-space two-nucleon potentials from leading to fourth order of chiral effective field theory}
\author{S. K. Saha}
\email{sanjoys@uidaho.edu}
\affiliation{Department of Physics, University of Idaho, Moscow, Idaho 83844, USA}
\author{D. R. Entem}
\email{entem@usal.es}
\affiliation{Grupo de F\'isica Nuclear, IUFFyM, Universidad de Salamanca, E-37008 Salamanca,
Spain}
\author{R. Machleidt}
\email{machleid@uidaho.edu}
\affiliation{Department of Physics, University of Idaho, Moscow, Idaho 83844, USA}
\author{Y. Nosyk}
\email{yevgenn@uidaho.edu}
\affiliation{Department of Physics, University of Idaho, Moscow, Idaho 83844, USA}

\begin{abstract}
We present local, position-space chiral $NN$ potentials through four orders of chiral 
effective field theory ranging from leading order (LO) to 
next-to-next-to-next-to-leading order (N$^3$LO, fourth order) of the $\Delta$-less version
of the theory.
The long-range parts of these potentials are fixed by the very accurate $\pi N$ LECs 
as determined in the Roy-Steiner equations analysis.
At the highest order (N$^3$LO),
the $NN$ data below 190 MeV laboratory energy are reproduced with the respectable
$\chi^2$/datum of 1.45. 
A comparison of the N$^3$LO potential with the phenomenological Argonne $v_{18}$ (AV18) potential
reveals substantial agreement between the two potentials in the intermediate range
ruled by chiral symmetry, thus, providing a chiral underpinning for the
phenomenological AV18 potential. 
Our chiral $NN$ potentials may serve as a
solid basis for systematic {\it ab initio} calculations of nuclear structure and reactions
that allow for a comprehensive error analysis. In particular, the order by order development of the potentials
will make possible a reliable determination of the truncation error at each 
order.
Our new family of local position-space potentials 
differs from existing potentials of this kind
by a weaker tensor force as reflected in relatively low $D$-state probabilities of the deuteron
($P_D \lea 4.0$ \% for our N$^3$LO potentials) and 
predictions for the triton binding energy above 8.00 MeV (from two-body forces alone).
As a consequence, our potentials may lead to different predictions when applied to light and
intermediate-mass nuclei in {\it ab initio} calculations and, potentially,
 help solve some of the outstanding
problems in microscopic nuclear structure.
\end{abstract}

\maketitle

\section{Introduction}
\label{sec_intro}

A primary goal of theoretical nuclear physics is to explain nuclear structure and reactions
in terms of the forces between nucleons---in present-day popular jargon dubbed the {\it ab initio} approach.
The current prevailing belief in the community is that chiral effective field theory (EFT) is best suited to 
provide those forces, because it can be related to low-energy QCD in a straight-forward way and
produces abundant three-nucleon forces (3NFs) needed for any quantitative nuclear structure
prediction~\cite{ME11,EHM09,HKK19,Heb21}.

Since chiral EFT is a low-momentum expansion, most chiral $NN$ potentials of the past
have been developed in momentum space--and are non-local. However, this feature makes
them unsuitable for a large group of {\it ab initio} few- and many-body algorithms, particularly,
the ones known as quantum Monte Carlo (QMC) methods~\cite{Car15,Lyn19}.
Variational Monte Carlo (VMC) and Green's Function Monte Carlo (GFMC) techniques
provide reliable solutions of the many-body Schr\H{o}dinger equation for, presently, up to 12
nucleons. Spectra, form factors, transitions, low-energy scattering, and response functions for light
nuclei have been successfully calculated using QMC methods~\cite{PT20}.
A further extension, the Auxiliary Field Diffusion Monte Carlo (AFDMC)~\cite{Car15,Lyn19}, additionally samples
the spin-isospin degrees of freedom, thus, making possible the study of neutron matter. 
In summary, QMC techniques have substantially contributed to the progress
in {\it ab initio} nuclear structure of the past 20+ years, and will continue to do so.
Thus, it is important that high-quality nuclear interactions are available for application
by these promising many-body methods.

An important advantage of chiral EFT is that it allows for a systematic quantification 
of the uncertainties of the predictions.
For this it is necessary to conduct calculations at different orders of the chiral expansion.
 However, so far, {\it local} chiral $NN$ potentials
have been developed only at next-to-next-to-leading order (NNLO)~\cite{Gez14} or in the hybrid format, NNLO/N$^3$LO~\cite{Pia15,Pia16},
where two-pion exchange (2PE) contributions are included up to NNLO and contact terms 
up to next-to-next-to-next-to-leading order (N$^3$LO).
To make proper uncertainty quantifications possible, local chiral $NN$ potentials at all orders
from leading order (LO) to N$^3$LO (and, if necessary, even beyond) are needed. 
It is the purpose of this work to construct such local $NN$ potentials of high quality
and make them available for QMC calculations as well as any other purposes where they can
be of use.

 We will develop these potentials within the $\Delta$-less theory, which has 
 two degrees of freedom, namely,
pions (Goldstone bosons) and nucleons, and does not include a $\Delta(1232)$-isobar degree of freedom. 
 If an explicit
$\Delta$-isobar is included in chiral EFT ($\Delta$-full theory~\cite{ORK94,ORK96,BKM97,KGW98,KEM07,KGE18}), then the
two-nucleon force (2NF) and 3NF contributions are enhanced at next-to-leading order (NLO), resulting in a smoother convergence when advancing
from leading order (LO) to NNLO. However, summing up all contributions at NNLO 
brings about
very similar results for both versions of the theory~\cite{KEM07}. The predictions of both theories
beyond NNLO are expected to be very similar~\cite{KGE18}.
In contrast
to recent claims~\cite{Jia20}, it has been shown in Ref.~\cite{NEM21} that there is no advantage 
to the $\Delta$-full theory.

This paper is organized as follows:
In Sec.~II, 
we present the expansion of the $NN$ potential through all orders from
LO to N$^3$LO.
The reproduction 
of the $NN$ scattering data and the deuteron properties are given in Sec.~III. 
Uncertainty quantification is
considered in Sec.~IV. Sec.~V concludes the paper.

\section{The chiral $NN$ potential}
\label{sec_pot}

\subsection{Effective Lagrangians}
\label{sec_lagr}

In the $\Delta$-less version of chiral EFT, which is the one we are applying, the relevant degrees o f freedom are
pions and nucleons.
Consequently, the effective Lagrangian is subdivided into the following pieces,
\begin{equation}
{\cal L}_{\rm eff}
=
{\cal L}_{\pi\pi} 
+
{\cal L}_{\pi N} 
+
{\cal L}_{NN} 
 + \, \ldots \,,
\end{equation}
where ${\cal L}_{\pi\pi}$
deals with the dynamics among pions, 
${\cal L}_{\pi N}$ 
describes the interaction
between pions and a nucleon,
and ${\cal L}_{NN}$  contains two-nucleon contact interactions
which consist of four nucleon-fields (four nucleon legs) and no
meson fields.
The ellipsis stands for terms that involve two nucleons plus
pions and three or more
nucleons with or without pions, relevant for nuclear
many-body forces.
Since the interactions of Goldstone bosons must
vanish at zero momentum transfer and in the chiral
limit ($m_\pi \rightarrow 0$), the low-energy expansion
of the effective Lagrangian is arranged in powers of derivatives
and pion masses, implying to following organization:
\begin{eqnarray}
{\cal L}_{\pi\pi} 
 & = &
{\cal L}_{\pi\pi}^{(2)} 
+
{\cal L}_{\pi\pi}^{(4)}
 + \ldots \,, \\
{\cal L}_{\pi N} 
 & = &
{\cal L}_{\pi N}^{(1)} 
+
{\cal L}_{\pi N}^{(2)} 
+
{\cal L}_{\pi N}^{(3)} 
+
{\cal L}_{\pi N}^{(4)} 
+ \ldots , \\
\label{eq_LNN}
{\cal L}_{NN} &  = &
{\cal L}^{(0)}_{NN} +
{\cal L}^{(2)}_{NN} +
{\cal L}^{(4)}_{NN} + 
\ldots \,,
\end{eqnarray}
where the superscript refers to the number of derivatives or 
pion mass insertions (chiral dimension)
and the ellipses stand for terms of higher dimensions.
We use the heavy-baryon formulation of the Lagrangians, the
explicit expressions of which can be found in Ref.~\cite{ME11}.

\subsection{Power counting}
\label{sec_pow}

Based upon the above Lagrangians, an infinite number of diagrams contributing to the interactions among 
nucleons can be drawn. 
Nuclear potentials are defined by the irreducible types of these
graphs.
By definition, an irreducible graph is a diagram that
cannot be separated into two
by cutting only nucleon lines.
These graphs are then analyzed in terms of powers of 
$Q$ with $Q=p/\Lambda_b$, 
where $p$ is generic for a momentum (nucleon three-momentum
or pion four-momentum) or a pion mass and $\Lambda_b \sim m_\rho \sim$ 0.7 GeV is the 
breakdown scale~\cite{Fur15}. Determining the power $\nu$ has become know
as power counting.

Following the Feynman rules of covariant perturbation theory,
a nucleon propagator is $p^{-1}$,
a pion propagator $p^{-2}$,
each derivative in any interaction is $p$,
and each four-momentum integration $p^4$.
This is also known as naive dimensional analysis or Weinberg counting.

Since we use the heavy-baryon formalism, we encounter terms which include factors of
$p/M_N$, where $M_N$ denotes the nucleon mass.
We count the order of such terms by the rule
\begin{equation}
p/M_N \sim (p/\Lambda_b)^2,
\label{eq_pM}
\end{equation}
for reasons explained in Ref.~\cite{Wei90}.

Applying some topological identities, one obtains
for the power of a connected irreducible diagram
involving $A$ nucleons~\cite{ME11,Wei90}
\begin{equation} \nu = -2 +2A  - 2C + 2L 
+ \sum_i \Delta_i \, ,
\label{eq_nu} 
\end{equation}
with
\begin{equation}
\Delta_i  \equiv   d_i + \frac{n_i}{2} - 2  \, ,
\label{eq_Deltai}
\end{equation}
where
$L$ denotes the number of loops in the diagram;
$d_i$ is the number of derivatives or pion-mass insertions 
and $n_i$ the number of nucleon fields (nucleon legs)
involved in vertex $i$;
the sum runs over all vertexes $i$ contained in the connected diagram 
under consideration.
Note that $\Delta_i \geq 0$
for all interactions allowed by chiral symmetry.

An important observation from power counting is that
the powers are bounded from below and, 
specifically, $\nu \geq 0$. 
This fact is crucial for the convergence of 
the low-momentum expansion.

For an irreducible 
$NN$ diagram ($A=2$, $C=1$), the
power formula collapses to the very simple expression
\begin{equation}
\nu =  2L + \sum_i \Delta_i \,,
\label{eq_nunn}
\end{equation}
which is most relevant for our current work.

In summary, the chief point of the chiral perturbation theory (ChPT) expansion of the potential is
that,
at a given order $\nu$, there exists only a finite number
of graphs. This is what makes the theory calculable.
The expression $(p/\Lambda_b)^{\nu+1}$ provides an estimate
of the relative size of the contributions left out and, thus,
of the relative uncertainty at order $\nu$.
The ability to calculate observables (in 
principle) to any degree of accuracy gives the theory 
its predictive power.

\begin{figure}[t]\centering
\vspace*{0.3cm}
\scalebox{0.80}{\includegraphics{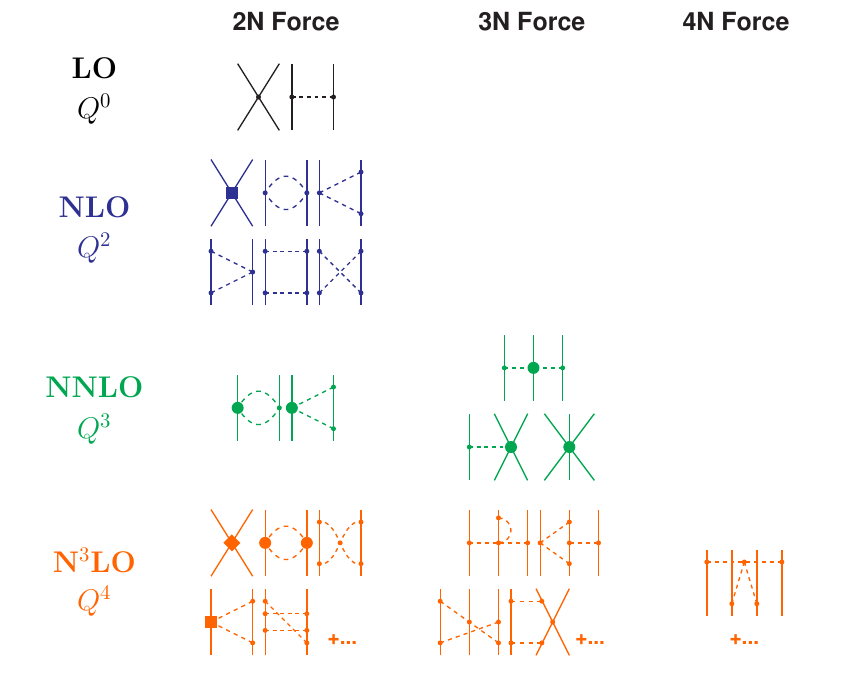}}
\caption{Hierarchy of nuclear forces in ChPT. Solid lines
represent nucleons and dashed lines pions. 
Small dots, large solid dots, solid squares, and solid diamonds
denote vertexes of index $\Delta_i= \, $ 0, 1, 2, and 4, respectively. 
$Q=p/\Lambda_b$
with $p$ a momentum or pion mass and $\Lambda_b$ the 
breakdown scale.
Further explanations are
given in the text.}
\label{fig_hi}
\end{figure}

ChPT and power counting
imply that nuclear forces evolve as a hierarchy
controlled by the power $\nu$, see Fig.~\ref{fig_hi} for an overview.
In what follows, we will focus on the 2NF.

\subsection{The long-range $NN$ potential}
\label{sec_long}

The long-range part of the $NN$ potential is built up from pion exchanges,
which are ruled by chiral symmetry.
The various pion-exchange contributions are best analyzed
by the number of pions being exchanged between the two
nucleons:
\begin{equation}
V_\pi = V_{1\pi} + V_{2\pi} + V_{3\pi} + \ldots \,,
\end{equation}
where the meaning of the subscripts is obvious
and the ellipsis represents $4\pi$ and higher pion exchanges. For each of the above terms, 
we have a low-momentum expansion:
\begin{eqnarray}
V_{1\pi} & = & V_{1\pi}^{(0)} + V_{1\pi}^{(2)} 
+ V_{1\pi}^{(3)} + V_{1\pi}^{(4)} +  \ldots \,,
\label{eq_1pe_orders}
\\
V_{2\pi} & = & V_{2\pi}^{(2)} + V_{2\pi}^{(3)} + V_{2\pi}^{(4)} + \ldots \,, \\
V_{3\pi} & = & V_{3\pi}^{(4)} +  \ldots \,,
\end{eqnarray}
where the superscript denotes the order $\nu$ of the expansion.
Higher order corrections to the one-pion exchange (1PE) are taken care of by  mass
and coupling constant renormalizations. Note also that, on 
shell, there are no relativistic corrections. Thus, 
$V_{1\pi}  =  V_{1\pi}^{(0)}$ 
 through all orders.
 The leading $3\pi$-exchange contribution that occurs at N$^3$LO,  $V_{3\pi}^{(4)}$,
has been calculated in Refs.~\cite{Kai00a,Kai00b} and found to be negligible. We, 
therefore, omit it.

Order by order, the long-range $NN$ potential then builds up as follows:
\beqa
V_\pi^{\rm LO} & = &
V_{1\pi}^{(0)} \,,
\label{eq_VLO}
\\
V_\pi^{\rm NLO} & = & V_\pi^{\rm LO} +
V_{2\pi}^{(2)} \,,
\label{eq_VNLO}
\\
V_\pi^{\rm NNLO} & = & V_\pi^{\rm NLO} +
V_{2\pi}^{(3)} \,,
\label{eq_VNNLO}
\\
V_\pi^{\rm N3LO} & = & V_\pi^{\rm NNLO} +
V_{2\pi}^{(4)} \,.
\label{eq_VN3LO}
\eeqa
We note that we add to $V_\pi^{\rm N3LO}$ the $1/M_N$ corrections of the NNLO 2PE proportional to $c_i$ (cf.\ Table~\ref{tab_lecs}).
This correction is proportional to $c_i/M_N$ 
(cf.\ Fig.~\ref{fig_dia3} and Appendix~\ref{sec_cM}, below) and appears nominally at fifth order, but we include it at fourth order.
As demonstrated in Ref.~\cite{EM02}, the 2PE football diagram proportional to $c_i^2$
that appears at N$^3$LO (Fig.~\ref{fig_dia2}(a) and Appendix~\ref{sec_football}) is unrealistically attractive, while the $c_i/M_N$ correction is large
and repulsive. Therefore, it makes sense to group these diagrams together to arrive at a more
realistic intermediate-range attraction at N$^3$LO. This is common practice and has been
done so in Refs.~\cite{EM03,EKM15,EMN17}.

The explicit mathematical expressions for the pion-exchanges up to N$^3$LO are very involved.
We have, therefore, moved them into the Appendix~\ref{sec_applong}.

\begin{table}
\caption{The $\pi N$ LECs as determined in
the Roy-Steiner-equation analysis of $\pi N$ scattering conducted in  Ref.~\cite{Hof15}.
The given orders of the chiral expansion refer to the $NN$ system. 
The $c_i$ and $\bar{d}_i$
are the LECs of the second and third order $\pi N$ Lagrangian~\cite{ME11} and are 
 in units of GeV$^{-1}$ and GeV$^{-2}$, respectively.
The uncertainties in the last digits are given in parentheses after the values.
We use the central values.}
\label{tab_lecs}
\smallskip
\begin{tabular}{crr}
\hline 
\hline 
\noalign{\smallskip}
              & \hspace{2cm} NNLO & \hspace{2cm} N$^3$LO  \\
\hline
\noalign{\smallskip}
$c_1$ & --0.74(2) & --1.07(2) \\
$c_2$ &   & 3.20(3) \\
$c_3$ & --3.61(5) & --5.32(5)  \\
$c_4$ & 2.44(3) & 3.56(3) \\
$\bar{d}_1 + \bar{d}_2$ &  & 1.04(6)  \\
$\bar{d}_3$ & & --0.48(2)  \\
$\bar{d}_5$ & & 0.14(5) \\
$\bar{d}_{14} - \bar{d}_{15}$ & & --1.90(6)  \\
\hline
\hline
\noalign{\smallskip}
\end{tabular}
\end{table}

Chiral symmetry establishes a link between the dynamics in the $\pi N$-system 
and the $N\!N$-system through common low-energy constants (LECs).
Therefore, consistency requires that we use the LECs for subleading $\pi N$-couplings as 
determined in the analysis of low-energy $\pi N$-scattering.
Currently, the most reliable $\pi N$ analysis is the one 
by Hoferichter and Ruiz de Elvira
{\it et al.}~\cite{Hof15}, in
which the Roy-Steiner equations are applied.
These LECs 
carry very small uncertainties 
(cf.\ Table~\ref{tab_lecs});
in fact, the uncertainties are so small that they are negligible for our purposes.
This makes the variation of the $\pi N$ LECs in $NN$ potential construction obsolete
and reduces the error budget in applications of these potentials. 
For the potentials constructed in this paper, the central values of Table~\ref{tab_lecs} are applied.
Other constants involved in our potential construction are shown in Table~\ref{tab_basic}.

\begin{table}[t]
\caption{Basic constants used throughout this work~\cite{PDG}.}
\label{tab_basic}
\smallskip
\begin{tabular}{lcl}
\hline 
\hline 
\noalign{\smallskip}
  quantity            &  \hspace{2cm} & Value \\
\hline
\noalign{\smallskip}
Axial-vector coupling constant $g_A$ && 1.29 \\
Pion-decay constant $f_\pi$ && 92.4 MeV \\
Charged-pion mass $m_{\pi^\pm}$ && 139.5702 MeV \\
Neutral-pion mass $m_{\pi^0}$ && 134.9766 MeV \\
Average pion-mass $\bar{m}_\pi$ && 138.0390 MeV \\
Proton mass $M_p$ && 938.2720 MeV \\
Neutron mass $M_n$ && 939.5654 MeV \\
Average nucleon-mass $\bar{M}_N$ && 938.9183 MeV \\
Conversion constant $\hbar c$                          &&  197.32698 MeV fm  \\
\hline
\hline
\noalign{\smallskip}
\end{tabular}
\end{table}

\subsection{The short-range $NN$ potential}
\label{sec_short}

\begin{figure}[t]
\scalebox{0.65}{\includegraphics{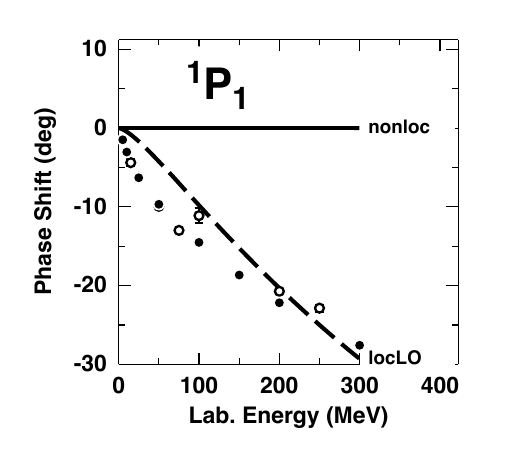}}
\scalebox{0.65}{\includegraphics{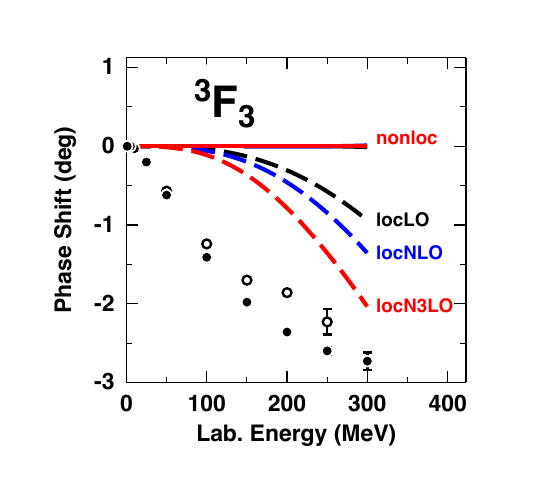}}
\vspace*{-0.3cm}
\caption{Left panel: $^1P_1$ phase shifts for the order zero (i.e., LO)  contact terms with nonlocal regulator
(solid black line, ``nonloc'') versus the same terms multiplied with a local regulator (dashed black line, ``locLO'').
Right panel: $^3F_3$ phase shifts for the central force contact terms at orders LO, NLO, and N$^3$LO with nonlocal regulator (solid red line, ``nonloc'') 
 versus the same terms multiplied with a local regulator (dashed lines at orders as denoted).
 The central force contact LECs of the N$^3$LO potential with cutoff combination
  $(R_\pi,R_{\rm ct})$=$(1.0,0.70)$ fm are applied (Table~\ref{tab_ctlecs}).
The filled and open circles represent the results from the Nijmegen multi-energy $np$ phase-shift analysis~\cite{Sto93} and the GWU single-energy $np$ analysis SP07~\cite{SP07}, respectively.
\label{fig_fierz}}
\end{figure}

The short-range $NN$ potential is described by contributions of the contact type,
which are constrained by parity, time-reversal, and the usual invariances, but not by chiral symmetry. 
Because of  parity and time-reversal only even powers of momentum
are allowed.
Thus, the expansion of the contact potential is
formally written as
\begin{equation}
V_{\rm ct} =
V_{\rm ct}^{(0)} + 
V_{\rm ct}^{(2)} + 
V_{\rm ct}^{(4)} + 
 \ldots \; ,
\label{eq_ct}
\end{equation}
where the superscript denotes the power or order.

In principle, the most general set of contact terms at each order is provided by all combinations of 
spin, isospin, and momentum operators that are allowed by the usual symmetries~\cite{OM58}
at the given order.
Two momenta are available, namely, the final and initial nucleon momenta 
in the center-of-mass system, ${\vec p}\,'$ and $\vec p$.
This can be reformulated in terms of two alternative momenta, viz.,
the momentum transfer $\vec q = {\vec p}\,' - \vec p$ 
and the average momentum $\vec k = ({\vec p}\,' + \vec p)/2$.
Functions of $\vec q $\, lead to local interactions, that is, 
to functions of the relative distance
${\vec r}$ 
between the two nucleons
after Fourier transform.
On the other hand, functions of $\vec k$ lead to nonlocal interactions.

Since ChPT is a low-momentum expansion, it requires cutting off high momenta
to avoid divergences. This is achieved by multiplying the potential with
a regulator function that suppresses the large momenta (or, equivalently, the short distances).
Depending on the type of momenta used, the regulator can be local or nonlocal.

When chiral $NN$ potentials are constructed in momentum-space and regulated by nonlocal
cutoff functions~\cite{ME11}, then it is possible
to reduce the number of contact operators (by a factor of two) due to 
Fierz ambiguity~\cite{Fie37,Hut17},
which is a consequence of the fact that nucleons are Fermions and obey the Pauli exclusion principle.
However, for the reasons stated in the Introduction,
we wish to construct $NN$ potentials which are strictly local, implying 
that we have to use local regulators.

When a local (regulator) function is applied to the contact terms, then
the Fierz rearrangemnt freedom is violated~\cite{Hut17}.
To provide a simple example of this, consider a contact operator of order zero ($\sim Q^0$, LO).
After a partial-wave decomposition and when multiplied by either no regulator or a nonlocal regulator, such operator produces no contributions for states
with orbital angular momentum $L>0$, i.e., $P$ and higher partial waves.
However, this property is violated when the operator is multiplied with a
local regulator function~\cite{Hut17}. We demonstrate this fact in Fig.~\ref{fig_fierz}, where, in the left panel,
we show phase shifts in the $^1P_1$ state: The solid line (``nonloc'')
shows the phase shifts when the LO contact terms are multiplied with a nonlocal cutoff function,
which does not violate Fierz ambiguity and, therefore, the phase shifts are zero. However,
when the LO contact terms are multiplied by a local regulator, the dashed curve (``locLO'') is obtained---obviously a severe violation.
This violation by local regulators continues through higher orders.
As an example, we show in the right panel of Fig.~\ref{fig_fierz} the phase shifts 
in an $F$-wave, where polynomial terms up to fourth order should not contribute which,
as demonstrated in the figure, is, indeed, true when a nonlocal cutoff is multiplied to 
contact terms up to fourth order (solid red curve, ``nonloc'').
However, when local functions are applied, then at orders $Q^0$, $Q^2$, and $Q^4$,
the contributions are not zero anymore as demonstrated by the dashed curves denoted by
``locLO'', ``locNLO'', ``locN3LO'', respectively; which, again, may be perceived 
as a severe violation of the Fierz rearrangement freedom.

Attempts can be undertaken to restore Fierz reordering as tried in Ref.~\cite{Hut17}
by way of contributions of higher order. However, the Fierz violations demonstrated in Fig.~\ref{fig_fierz} for $^3F_3$ cannot be compensated within the scope of this work,
since they would require contributions of sixth order. 

To make a long story short, 
our bottom-line argument is simply that it does not make sense to apply a symmetry that is invalid for the problem under consideration.
Therefore, we will not apply Fierz reordering to the contact terms and, hence,
use for the contacts all combinations of 
spin, isospin, angular momentum, and momentum $\vec q$  that are allowed by the usual symmetries, for each of the given orders.
On a historical note,
this is also the approach that was taken for the very first chiral $NN$ potentials ever 
constructed~\cite{ORK94,ORK96}.

According to standard power counting rules, only two contacts are needed at LO
while, in our approach and the one of Refs.~\cite{ORK94,ORK96},
there are four at LO. Consequently, the approach uses an overcomplete basis
implying that some parameters are redundant.
Note, however, that 
there is nothing fundamentally wrong with using
redundant parameters.
It merely means that the approach may be perceived as being inefficient (which is not the same as being wrong). Ironically, here, the inefficient approach is more efficient, since it allows to relate the contact parameters
in a one-to-one correspondence to states of well-defined total spin $S$
and total isospin $T$ [Eqs.~(\ref{eq_CST}) and (\ref{eq_CCT})] and, thus,
makes possible fitting phase shifts state-by-state.
However, as discussed, due to the local character of the regulator function,
Eq.~(\ref{eq_regct}), that is multiplied to the zero-order contacts,
$P$ and higher partial waves will be affected by LO contact terms 
(cf.\  Fig.~\ref{fig_fierz}), hence,
promoting higher order terms to LO and increasing the fit freedom when four 
independent LO parameters are available. 

At higher orders, the discussed redundancy applies to the $C_8$ term at NLO
and the $D_{10}$, $D_{12}$, and $D_{14}$ terms at N$^3$LO (see below for the detailed expressions).
As a result, at N$^3$LO we have only 11 non-redundant contact parameters,
even though according to power counting rules there should be 15.
The four ``missing'' fourth order contact terms are nonlocal 
(cf.\ Ref.~\cite{Pia15}) and, therefore,
we have to leave them out, as practiced already in Ref.~\cite{Pia16}.

Our approach overlaps with the philosophy of the Argonne $v_{18}$ potential (AV18)~\cite{WSS95},
which includes 14 charge-independent operators. Not accidentally, we will also have
14 contact operators at N$^3$LO (see below) which are all equivalent to the 14 operators of the AV18 potential. 
This fact provides another advantage to our approach, namely,
there is now a one-to-one correspondence between the terms of the
AV18 potential and the chiral potentials of this paper. This allows for a detailed comparison between the two potentials as conducted in 
Appendix~\ref{app_plots}, which turns out to be most revealing.

Next, we present the explicit expressions for the contact operators, order by order.

\subsubsection{Leading order}

In momentum-space, the LO or zeroth order charge-independent contact terms are given by
\begin{eqnarray} 
{V}_{\rm ct}^{(0)}(q) \, &  = &
 \, \left(  C_c \, + C_\tau \, \bm{\tau}_1 \cdot \bm{\tau}_2 \, 
 + \, C_\sigma \, \vec\sigma_1 \cdot \vec \sigma_2 \,
 + \, C_{\sigma \tau} \, \vec\sigma_1 \cdot \vec \sigma_2 \; \bm{\tau}_1 \cdot \bm{\tau}_2 \right) \,
 f_{\rm ct}(q) 
\label{eq_ct0q}
\end{eqnarray}
with regulator function
\begin{equation}
f_{\rm ct}(q) = e^{-(q/\Lambda)^2} 
\label{eq_regct}
\end {equation}
and $\Lambda$ a momentum cutoff.
The operators $\vec \sigma_{1,2}$ and $\bm{\tau}_{1,2}$ denote the spin 
and isospin operators for nucleon 1 and 2, respectively, with
$\bm{\tau}_{i}=(\tau_{ix},\tau_{iy},\tau_{iz})$, $i=1,2$. In the convention we apply, the proton carries an eigenvalue of $(+1)$ and the neutron an eigenvalue of $(-1)$ with regard to
$\tau_z$.

At LO, we also include charge-dependent contact terms that are defined as follows:
\begin{eqnarray} 
^{\rm CD}V_{\rm ct}^{(0)}(q) \, &  = &
 \, \left[  C_{T_{12}}^{\rm CD} \, T_{12} \, 
 + \, C_{\sigma T_{12}}^{\rm CD} \, \vec\sigma_1 \cdot \vec \sigma_2 \, T_{12}
 + \, C_{\tau_z}^{\rm CA} \, (\tau_{1z} + \tau_{2z}) \, 
 + \, C_{\sigma \tau_z}^{\rm CA} \, \vec\sigma_1 \cdot \vec \sigma_2 \,  (\tau_{1z} + \tau_{2z}) \right]
\,  f_{\rm ct}(q) \,,
\label{eq_ct0qcd}
\end{eqnarray}
with 
\begin{equation}
T_{12}= 3 \, \tau_{1z} \tau_{2z} - \bm{\tau}_1 \cdot \bm{\tau}_2
\label{eq_T12}
\end{equation}
 an isotensor operator.
Terms proportional to $T_{12}$ are charge dependent, while terms proportional to
$(\tau_{1z} + \tau_{2z})$ are charge asymmetric.

In position space, this translates into
\begin{eqnarray} 
\widetilde{V}_{\rm ct}^{(0)}(r) \, &  = &
 \, \left(  C_c \, + C_\tau \, \bm{\tau}_1 \cdot \bm{\tau}_2 \, 
 + \, C_\sigma \, \vec\sigma_1 \cdot \vec \sigma_2 \,
 + \, C_{\sigma \tau} \, \vec\sigma_1 \cdot \vec \sigma_2 \; \bm{\tau}_1 \cdot \bm{\tau}_2 \right) \,
 ^{\rm ct}\widetilde{V}_C^{(0)}(r) 
\label{eq_ct0r}
\end{eqnarray}
and
\begin{eqnarray} 
^{\rm CD}\widetilde{V}_{\rm ct}^{(0)}(r) \, &  = &
 \, \left[  C_{T_{12}}^{\rm CD} \, T_{12} \, 
 + \, C_{\sigma T_{12}}^{\rm CD} \, \vec\sigma_1 \cdot \vec \sigma_2 \, T_{12}
 + \, C_{\tau_z}^{\rm CA} \, (\tau_{1z} + \tau_{2z}) \, 
 + \, C_{\sigma \tau_z}^{\rm CA} \, \vec\sigma_1 \cdot \vec \sigma_2 \,  (\tau_{1z} + \tau_{2z}) \right]
 ^{\rm ct}\widetilde{V}_C^{(0)}(r) 
\label{eq_ct0rcd}
\end{eqnarray}
with
\begin{equation}
^{\rm ct}\widetilde{V}_C^{(0)}(r) = \widetilde f_{\rm ct} (r) = \frac{1}{\pi^{3/2} \, R_{\rm ct}^3} \, e^{-(r/R_{\rm ct})^2} \,,
\label{eq_cut0}
\end{equation}
the Fourier transform of $f_{\rm ct}(q)$, and $R_{\rm ct}=2/\Lambda$.
Note that we use units such that $\hbar=c=1$.

\subsubsection{Next-to-leading order}

In momentum-space, the NLO or second order contact contribution is
\begin{eqnarray} 
{V}_{\rm ct}^{(2)}({\vec p}~', \vec p) \, &  = &  \Big\{
 \left(  C_1 \, + C_2 \, \bm{\tau}_1 \cdot \bm{\tau}_2 \, 
 + \, C_3 \, \vec\sigma_1 \cdot \vec \sigma_2 \,
 + \, C_4 \, \vec\sigma_1 \cdot \vec \sigma_2 \; \bm{\tau}_1 \cdot \bm{\tau}_2 \right) \, 
 q^2 \nonumber \\
     &&  +
 \left( C_5 \,  + C_6 \, \bm{\tau}_1 \cdot \bm{\tau}_2 \, \right) \,
\widehat S_{12}(\vec q) \;\;   \nonumber \\
     &&   +
      \left( C_7 \,  + C_8 \, \bm{\tau}_1 \cdot \bm{\tau}_2 \, \right) \,
\left[-i \vec S \cdot (\vec q \times \vec k) \,\right]  \Big\}  \,  f_{\rm ct}(q)  \,,
\label{eq_ct2q}
\end{eqnarray}
where 
$\vec S =(\vec\sigma_1+\vec\sigma_2)/2 $ denotes the total spin
and
\begin{equation}
\widehat S_{12}(\vec q) = 3 \, \vec \sigma_1 \cdot \vec q \,\,\: \vec \sigma_2 \cdot \vec q - q^2 \, \vec \sigma_1 \cdot  \vec \sigma_2 
\end{equation}
is the spin-tensor operator in momentum-space.

Fourier transform of the above creates the second order contact contribution in position space
\begin{eqnarray} 
\widetilde{V}_{\rm ct}^{(2)}(\vec r) \, &  = &
 \, \left(  C_1 \, + C_2 \, \bm{\tau}_1 \cdot \bm{\tau}_2 \, 
 + \, C_3 \, \vec\sigma_1 \cdot \vec \sigma_2 \,
 + \, C_4 \, \vec\sigma_1 \cdot \vec \sigma_2 \; \bm{\tau}_1 \cdot \bm{\tau}_2 \right) \,\,
 ^{\rm ct}\widetilde{V}_C^{(2)}(r) \nonumber \\
     &&  +
 \left( C_5 \,  + C_6 \, \bm{\tau}_1 \cdot \bm{\tau}_2 \, \right) \,
S_{12}(\hat r) \;\;  ^{\rm ct}\widetilde{V}_T^{(2)}(r) \nonumber \\
     &&   +
      \left( C_7 \,  + C_8 \, \bm{\tau}_1 \cdot \bm{\tau}_2 \, \right) \,
(\vec L \cdot \vec S) \;\;  ^{\rm ct}\widetilde{V}_{LS}^{(2)}(r) \,,
\label{eq_ct2r}
\end{eqnarray}
where
\begin{equation}
S_{12}(\hat r) = 3 \vec \sigma_1 \cdot \hat r \,\,\: \vec \sigma_2 \cdot \hat r - \vec \sigma_1 \cdot  \vec \sigma_2 
\end{equation}
denotes the standard position-space spin-tensor operator 
with $\hat r = \vec r/r $,
and $\vec L$ is the operator of total angular momentum.
Furthermore,
\begin{eqnarray}
 ^{\rm ct}\widetilde{V}_C^{(2)}(r) &=& - \widetilde f_{\rm ct}^{(2)} (r)  - \frac{2}{r} \, \widetilde f_{\rm ct}^{(1)} (r) \,,
 \label{eq_cut2}
 \\
 ^{\rm ct}\widetilde{V}_T^{(2)}(r) &=&  - \widetilde f_{\rm ct}^{(2)} (r)  + \frac{1}{r} \, \widetilde f_{\rm ct}^{(1)} (r) \,,
 \\
^{\rm ct}\widetilde{V}_{LS}^{(2)}(r) &=&  - \frac{1}{r} \, \widetilde f_{\rm ct}^{(1)} (r) \,,
\end{eqnarray}
with
\begin{equation}
\widetilde f_{\rm ct}^{(n)} (r) = \frac{d^n  \widetilde f_{\rm ct}(r)}{dr^n} \,.
\end{equation}

\subsubsection{Next-to-next-to-next-to-leading order}
\label{sec_short4}

In momentum-space, the N$^3$LO or fourth order contact contribution is assumed to be
\begin{eqnarray} 
{V}_{\rm ct}^{(4)}({\vec p}~', \vec p)  \, &  = &   \Big\{
  \left(  D_1 \, + D_2 \, \bm{\tau}_1 \cdot \bm{\tau}_2 \, 
 + \, D_3 \, \vec\sigma_1 \cdot \vec \sigma_2 \,
 + \, D_4 \, \vec\sigma_1 \cdot \vec \sigma_2 \; \bm{\tau}_1 \cdot \bm{\tau}_2 \right) \,
q^4 \nonumber \\
     &&  +
 \left( D_5 \,  + D_6 \, \bm{\tau}_1 \cdot \bm{\tau}_2 \, \right) \, q^2 \,
\widehat S_{12}(\vec q) \;\;   \nonumber \\
     &&   +
      \left( D_7 \,  + D_8 \, \bm{\tau}_1 \cdot \bm{\tau}_2 \, \right) \, q^2 \,
 \left[-i \vec S \cdot (\vec q \times \vec k) \,\right]  \;\;   \nonumber \\
   &&   +
      \left( D_9 \,  + D_{10} \, \bm{\tau}_1 \cdot \bm{\tau}_2 \, \right) \,
 \left[-i \vec S \cdot (\vec q \times \vec k) \,\right]^2 
 \label{eq_LS2} \\
   &&  +
    \, \left(  D_{11} \, + D_{12} \, \bm{\tau}_1 \cdot \bm{\tau}_2 \, 
 + \, D_{13} \, \vec\sigma_1 \cdot \vec \sigma_2 \,
 + \, D_{14} \, \vec\sigma_1 \cdot \vec \sigma_2 \; \bm{\tau}_1 \cdot \bm{\tau}_2 \right) \,\,
\left[-i (\vec q \times \vec k) \,\right]^2 \Big\} f_{\rm ct}(q)
\label{eq_L2}
\end{eqnarray}

In position-space, the N$^3$LO or fourth order contact contribution then is
\begin{eqnarray} 
\widetilde{V}_{\rm ct}^{(4)}(\vec r) \, &  = &
 \, \left(  D_1 \, + D_2 \, \bm{\tau}_1 \cdot \bm{\tau}_2 \, 
 + \, D_3 \, \vec\sigma_1 \cdot \vec \sigma_2 \,
 + \, D_4 \, \vec\sigma_1 \cdot \vec \sigma_2 \; \bm{\tau}_1 \cdot \bm{\tau}_2 \right) \,\,
 ^{\rm ct}\widetilde{V}_C^{(4)}(r) \nonumber \\
     &&  +
 \left( D_5 \,  + D_6 \, \bm{\tau}_1 \cdot \bm{\tau}_2 \, \right) \,
S_{12}(\hat r) \;\;  ^{\rm ct}\widetilde{V}_T^{(4)}(r) \nonumber \\
     &&   +
      \left( D_7 \,  + D_8 \, \bm{\tau}_1 \cdot \bm{\tau}_2 \, \right) \,
(\vec L \cdot \vec S) \;\;  ^{\rm ct}\widetilde{V}_{LS}^{(4)}(r)  \nonumber \\
   &&   +
      \left( D_9 \,  + D_{10} \, \bm{\tau}_1 \cdot \bm{\tau}_2 \, \right) \,
(\vec L \cdot \vec S)^2 \;\;  ^{\rm ct}\widetilde{V}_{LS2}^{(4)}(r)  \nonumber \\
   &&  +
    \, \left(  D_{11} \, + D_{12} \, \bm{\tau}_1 \cdot \bm{\tau}_2 \, 
 + \, D_{13} \, \vec\sigma_1 \cdot \vec \sigma_2 \,
 + \, D_{14} \, \vec\sigma_1 \cdot \vec \sigma_2 \; \bm{\tau}_1 \cdot \bm{\tau}_2 \right) \,\,
{\vec L}^2 \, ^{\rm ct}\widetilde{V}_{LL}^{(4)}(r) \,,
\label{eq_ct4r}
\end{eqnarray}
with
\begin{eqnarray}
 ^{\rm ct}\widetilde{V}_C^{(4)}(r) &=&  \widetilde f_{\rm ct}^{(4)} (r)  + \frac{4}{r} \, \widetilde f_{\rm ct}^{(3)} (r) \,,
 \label{eq_cut4}
 \\
 ^{\rm ct}\widetilde{V}_T^{(4)}(r) &=&  \widetilde f_{\rm ct}^{(4)} (r)  + \frac{1}{r} \, \widetilde f_{\rm ct}^{(3)} (r) 
 - \frac{6}{r^2} \, \widetilde f_{\rm ct}^{(2)} (r) + \frac{6}{r^3} \, \widetilde f_{\rm ct}^{(1)} (r)  \,,
 \\
^{\rm ct}\widetilde{V}_{LS}^{(4)}(r) &=&  \frac{1}{r} \, \widetilde f_{\rm ct}^{(3)} (r) 
 + \frac{2}{r^2} \, \widetilde f_{\rm ct}^{(2)} (r) - \frac{2}{r^3} \, \widetilde f_{\rm ct}^{(1)} (r)   \,,
\\
^{\rm ct}\widetilde{V}_{LS2}^{(4)}(r) &=&   \frac{1}{r^2} \, \widetilde f_{\rm ct}^{(2)} (r) - \frac{1}{r^3} \, \widetilde f_{\rm ct}^{(1)} (r) \,,
\\
^{\rm ct}\widetilde{V}_{LL}^{(4)}(r) &=&   \frac{1}{r^2} \, \widetilde f_{\rm ct}^{(2)} (r) - \frac{1}{r^3} \, \widetilde f_{\rm ct}^{(1)} (r) \,,
\end{eqnarray}
where from the Fourier transforms of
Eqs.~(\ref{eq_LS2}) and (\ref{eq_L2})
we retained only the local terms~\cite{Pia15}.

\subsection{Charge dependence}
\label{sec_CD}

This is to summarize what charge-dependence we include.
Through all orders, we take the charge-dependence of the 1PE due to pion-mass splitting
into account, Eqs.~(\ref{eq_1pepp}) - (\ref{eq_1peCD}).
Charge-dependence is seen most prominently in the $^1S_0$ state at low energies, particularly, in the $^1S_0$ scattering lengths. Charge-dependent 1PE cannot explain it all. 
The remainder is accounted for by the LO charge-dependent contact potential
Eq.~(\ref{eq_ct0rcd}), see also Appendix~\ref{app_lectab}.
In all 2PE contributions, we apply the average pion mass, $\bar m_\pi$.
Thus, 2PE does not generate charge-dependence.
For $pp$ scattering at any order, we include the relativistic Coulomb potential~\cite{AS83,Ber88}.
We omit irreducible $\pi$-$\gamma$ 
exchange~\cite{Kol98},
which would affect the N$^3$LO  $np$ potential.
We take nucleon-mass splitting into account
in the kinetic energy
by using $M_p$ in $pp$ scattering, $M_n$ in $nn$ scattering, and $\bar M_N$
in $np$ scattering (see Table~\ref{tab_basic} for their precise values).

For a comprehensive discussion of all possible sources of charge-dependence of the $NN$
interaction, see Ref.~\cite{ME11}.

\subsection{The full potential}
\label{sec_full}

The potential $V$ is, in principal, an invariant amplitude (with relativity taken into account perturbatively) and, thus, satisfies a relativistic scattering equation, like, e.\ g., the
Blankenbeclar-Sugar (BbS) equation~\cite{BS66},
which reads explicitly,
\begin{equation}
{T}({\vec p}~',{\vec p})= {V}({\vec p}~',{\vec p})+
\int d^3p'' \:
{V}({\vec p}~',{\vec p}~'') \:
\frac{M_N^2}{E_{p''}} \:  
\frac{1}
{{ p}^{2}-{p''}^{2}+i\epsilon} \:
{T}({\vec p}~'',{\vec p}) 
\label{eq_bbs2}
\end{equation}
with $E_{p''}\equiv \sqrt{M_N^2 + {p''}^2}$ and $M_N$ the nucleon mass.
The advantage of using a relativistic scattering equation is that it automatically
includes relativistic kinematical corrections to all orders. Thus, in the scattering equation,
no propagator modifications are necessary when moving up to higher orders.

Defining
\begin{equation}
\widehat{V}({\vec p}~',{\vec p})
\equiv 
\sqrt{\frac{M_N}{E_{p'}}}\:  
{V}({\vec p}~',{\vec p})\:
 \sqrt{\frac{M_N}{E_{p}}}
\label{eq_minrel1}
\end{equation}
and
\begin{equation}
\widehat{T}({\vec p}~',{\vec p})
\equiv 
\sqrt{\frac{M_N}{E_{p'}}}\:  
{T}({\vec p}~',{\vec p})\:
 \sqrt{\frac{M_N}{E_{p}}}
\,,
\label{eq_minrel2}
\end{equation}
the BbS equation collapses into the usual, nonrelativistic
Lippmann-Schwinger (LS) equation,
\begin{equation}
 \widehat{T}({\vec p}~',{\vec p})= \widehat{V}({\vec p}~',{\vec p})+
\int d^3p''\:
\widehat{V}({\vec p}~',{\vec p}~'')\:
\frac{M_N}
{{ p}^{2}-{p''}^{2}+i\epsilon}\:
\widehat{T}({\vec p}~'',{\vec p}) \, .
\label{eq_LS}
\end{equation}
Since 
$\widehat V$ 
satisfies Eq.~(\ref{eq_LS}), 
it may be regarded as a nonrelativistic potential. By the same token, 
$\widehat{T}$ 
may be considered as the nonrelativistic 
T-matrix. 
The above momentum-space equation is equivalent to the nonrelativistic
Schr\"{o}dinger equation for the calculation of phase shifts and bound states,
the position-space techniques of which can be found in Refs.~\cite{Nag75,Shi05}.

Expanding the square-root factors in Eq.~(\ref{eq_minrel1}) up to second order
in $p/M_N$, results in
\begin{equation}
\widehat{V}({\vec p}~',{\vec p})
\approx
{V}({\vec p}~',{\vec p})
\left( 1 - \frac{p^2 + {p'}^{2}}{4 \, M_N^2} \right) \,,
\label{eq_relcorr}
\end{equation}
and similarly for $\widehat{T}({\vec p}~',{\vec p})$.
Since we count $p/M_N$ corrections the way indicated in Eq.~(\ref{eq_pM}), the 
correction displayed in Eq.~(\ref{eq_relcorr}) is four orders up from a given
potential contribution, $V$---which is beyond the order of all $\nu \leq 3$ potentials
constructed in this paper and, therefore, can be ignored in those cases.
Yet, the correction is relevant for the LO 
contributions to the N$^3$LO potentials. 
While the corrections to the LO contacts can be absorbed by the 4th order contacts,
this correction also applies to the LO (i.~e., static) 1PE.
However, because this correction term is nonlocal and---for reasons explained in the Introduction---because we wish to construct
strictly local potentials, we neglect this fourth order correction to the 1PE.
The bottom line then is that, throughout
our local potential constructions,
we employ the approximations
\begin{eqnarray}
\widehat{V}({\vec p}~',{\vec p}) & \approx & {V}({\vec p}~',{\vec p}) \,, \\
\widehat{T}({\vec p}~',{\vec p}) & \approx & {T}({\vec p}~',{\vec p}) \,.
\end{eqnarray}

The Fourier transforms of $V$ are denoted by ${\widetilde V}$ (cf.\ Appendix A).

The full $NN$ potential is the sum of the long- and the short-range potentials.
Order by order, this results into:
\beqa
{\widetilde V}^{\rm LO} & = & {\widetilde V}^{(0)}_{1\pi} + {\widetilde V}^{(0)}_{\rm ct} 
                                        + {^{\rm CD}{\widetilde V}^{(0)}_{\rm ct}} \,, \\
{\widetilde V}^{\rm NLO} & = & {\widetilde V}^{\rm LO}  + {\widetilde V}^{(2)}_{2\pi}
 + {\widetilde V}^{(2)}_{\rm ct} \,,  \\
{\widetilde V}^{\rm NNLO} & = & {\widetilde V}^{\rm NLO}  + {\widetilde V}^{(3)}_{2\pi} \,, \\
{\widetilde V}^{\rm N3LO} & = & {\widetilde V}^{\rm NNLO}  + {\widetilde V}^{(4)}_{2\pi}  
+ {\widetilde V}^{(4)}_{\rm ct}   \,,
\eeqa
where we note again that we add to ${\widetilde V}^{(4)}_{2\pi} $
 the $1/M_N$ corrections of  $ {\widetilde V}^{(3)}_{2\pi}$.
This correction is proportional to $c_i/M_N$ 
and appears nominally at fifth order, but we include it at fourth order for the reasons discussed.
The explicit mathematical expressions 
for ${\widetilde V}^{(0)}_{1\pi} $ are given in Appendix~\ref{app_1pe},
for ${\widetilde V}^{(2)}_{2\pi}$ in Appendix~\ref{app_2pe2},
for ${\widetilde V}^{(3)}_{2\pi}$ in Appendix~\ref{app_2pe3}, and
for ${\widetilde V}^{(4)}_{2\pi}$ in Appendices~\ref{app_2pe4} and \ref{sec_cM}.

\subsection{Regularization}
\label{sec_reg}

\begin{figure}[t]
\scalebox{0.42}{\includegraphics{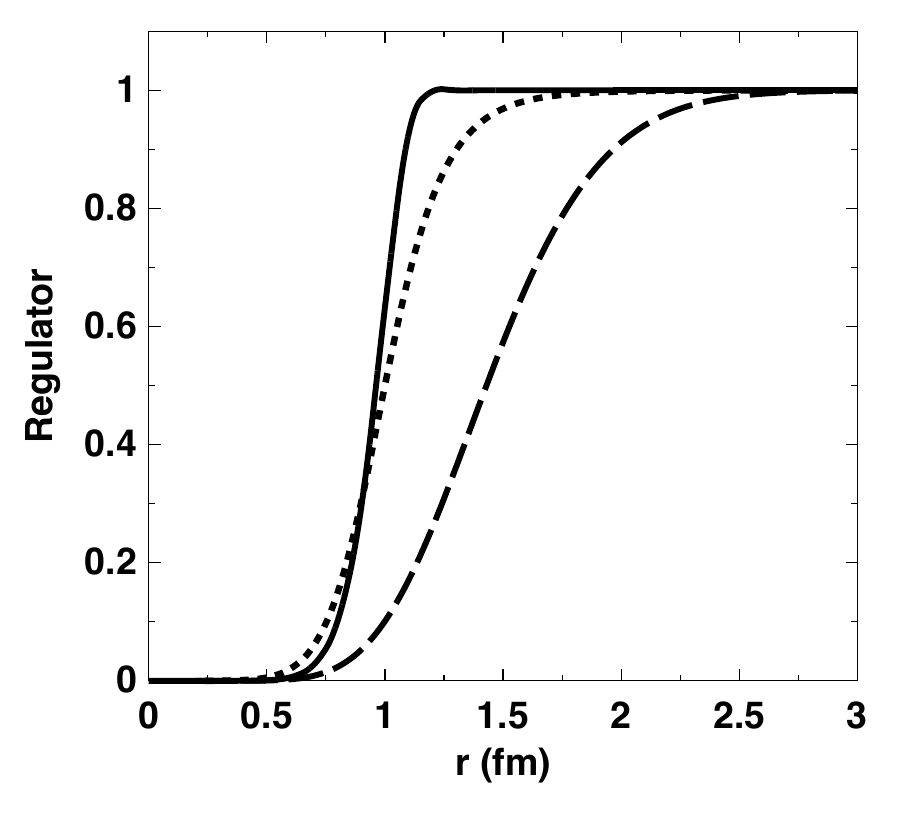}}
\vspace*{-0.3cm}
\caption{Various regulator functions 
used in the construction
of chiral position-space potentials.
 The solid, dashed, and dotted curves represent the regulators 
 $\widetilde f_{1\pi}(r)$, $\widetilde f_{2\pi}(r)$, and $\widetilde f_{\rm Pia}(r)$
 given in Eqs.~(\ref{eq_reg1pe}), (\ref{eq_reg2pe}), and (\ref{eq_regpia}), respectively.
 $R_\pi=1.0$ fm is applied in all cases.}
\label{fig_reg}
\end{figure}

All pion-exchange potentials, ${\widetilde V}_\pi (r)$, are singular at the origin and, thus, need 
regularization. For this purpose, we multiply the $ {\widetilde V}^{(0)}_{1\pi}(r)$ potential with
the regulator function
\begin{equation}
\widetilde f_{1\pi}(r)= 1 - \exp \left[ - \left( \frac{r}{R_\pi}\right)^{2n} \right]  
\label{eq_reg1pe}
\end{equation}
and all ${\widetilde V}^{(\nu)}_{2\pi}(r)$ ($\nu=2,3,4$) with~\cite{EKM15,Her15}
\begin{equation}
\widetilde f_{2\pi}(r) = \left[ 1 - \exp \left( - \frac{r^2}{R_\pi^2} \right) \right]^n 
\label{eq_reg2pe}
\end{equation}
using $n=5$ in all cases. (Notice that $n=4$ is the minimum required for ${\widetilde V}^{(4)}_{2\pi}$.) 

In the work of Piarulli {\it et al.}~\cite{Pia15,Pia16}, 
the regulator function
\begin{equation}
\widetilde f_{\rm Pia} (r) = 1 - \frac{1}{ \left( \frac{r}{R_\pi} \right)^6 \, 
 \exp \left(\frac{2(r-R_{\pi})}{R_\pi} \right) \, + \, 1 }
 \label{eq_regpia}
\end{equation}
is used for both 1PE and 2PE.

In Fig.~\ref{fig_reg} we show the shape of the different regulators for $R_\pi=1.0$ fm.
Our $\widetilde f_{1\pi}(r)$ (solid line) is similar to 
$\widetilde f_{\rm Pia} (r)$ (dotted), while
our $\widetilde f_{2\pi}(r)$ (dashed) continues to cut down in the range between 1 and 2 fm
where the other regulators have ceased to be of impact.

\begin{figure}[t]
\scalebox{0.45}{\includegraphics{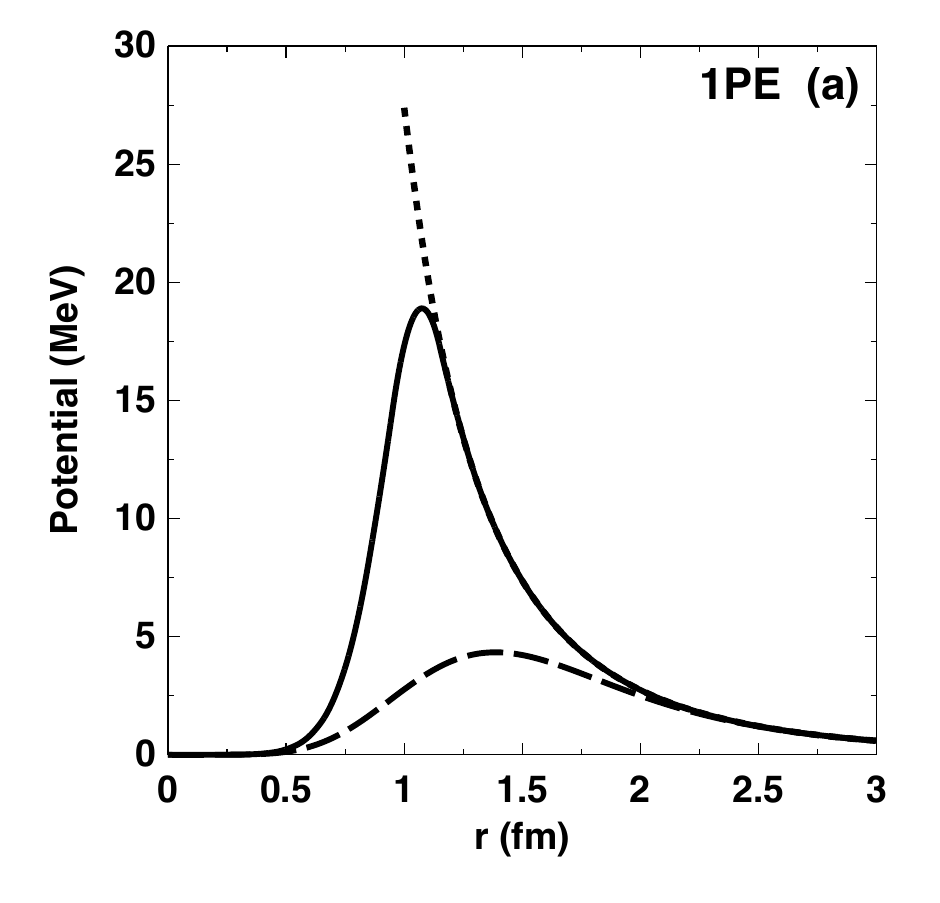}}
\scalebox{0.675}{\includegraphics{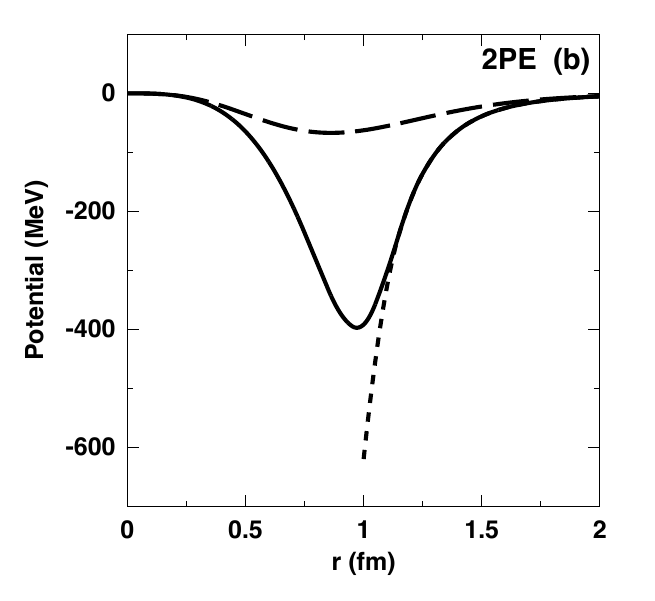}}
\vspace*{-0.3cm}
\caption{(a) The solid and dashed curves show the impact of the regulator functions
${\widetilde f}_{1\pi}(r)$ and ${\widetilde f}_{2\pi}(r)$, respectively,
on the tensor potential ${\widetilde W}_{T}(r)$ of 1PE, Eq.~(\ref{LOpot2}).
The dotted curve is obtained without regulation.
(b) Same as (a), but for the central potential ${\widetilde V}_{C}(r)$
of 2PE at N$^3$LO. $R_\pi=1.0$ fm is applied for all regulators.
}
\label{fig_reg12pe}
\end{figure}

The difference between the different regulators becomes even more evident when they
are applied to specific components of the $NN$ potential.
Therefore, we show in Fig.~\ref{fig_reg12pe}(a) the impact of ${\widetilde f}_{1\pi}(r)$ (solid line)
and ${\widetilde f}_{2\pi}(r)$ (dashed) on the 1PE tensor potential ${\widetilde W}_{T}(r)$, 
Eq.~(\ref{LOpot2}).
Both regulators suppress 1PE below 1 fm, but differ substantially above.
While the regulator ${\widetilde f}_{1\pi}(r)$ leaves the 1PE essentially unchanged above 1 fm,
${\widetilde f}_{2\pi}(r)$ suppresses 1PE drastically in the range 1 to 2 fm.
 It is well established that the 1PE at intermediate and long-range gets the physics right 
 (in particular the one of the deuteron)~\cite{Sup56,ER83} and, therefore, should not be suppressed in that range.
Consequently, the regulator ${\widetilde f}_{2\pi}(r)$ (dashed line) is inappropriate for 1PE, since it cuts out too much in the region 1 to 2 fm.

In Fig.~\ref{fig_reg12pe}(b) we show the corresponding situation for 2PE by way of the
central potential ${\widetilde V}_{C}(r)$ produced by 2PE at N$^3$LO. 
The situation with the 2PE is very different from 1PE. 

It is well known that,
in conventional meson theory,
the 2PE contribution
to the $NN$ interaction always comes out too attractive at short and intermediate range.
For a conventional field-theoretic model~\cite{Mac89,MHE87}, this is demonstrated
in Fig.~10 of Ref.~\cite{ME11}. It is also
true for the dispersion theoretic derivation of the 2PE that was pursued by the Paris 
group (see, e.~g., the predictions for $^1D_2$, $^3D_2$, and $^3D_3$
in Fig.~8 of Ref.~\cite{Vin79} which are all too attractive). 
In conventional meson theory~\cite{Mac89,MHE87}, this surplus
attraction is compensated by heavy-meson exchanges ($\rho$-, $\omega$-, and $\pi\rho$-exchanges) 
which, however, have no place in chiral EFT.
Instead, a drastic regulator has to be invoked that is also effective in the intermediate range.
This is the case with the regulator ${\widetilde f}_{2\pi}(r)$
(dashed curve in Fig.~\ref{fig_reg12pe}(b))
which,  therefore, is our choice for 2PE.

\section{$NN$ scattering and the deuteron}

Based upon the formalism presented in the previous section, we have constructed
$NN$ potentials at four different orders, namely, LO, NLO, NNLO, and N$^3$LO, 
cf.\ Sec.~\ref{sec_full}.
At each order, we apply three different cutoff combinations $(R_\pi, R_{\rm ct})$,
see  Secs.~\ref{sec_reg} and \ref{sec_short}, respectively, for their definitions.
Specifically, we use the combinations (1.0, 0.70) fm, (1.1, 0.72) fm, and (1.2, 0.75) fm.
Since we take charge dependence into account, each $NN$ potential comes in three versions:
$pp$, $np$, and $nn$.
In this section, we will present the predictions by these potentials for $NN$ scattering and the deuteron.

\subsection{$NN$ scattering}

\begin{figure}[th]
\scalebox{0.37}{\includegraphics{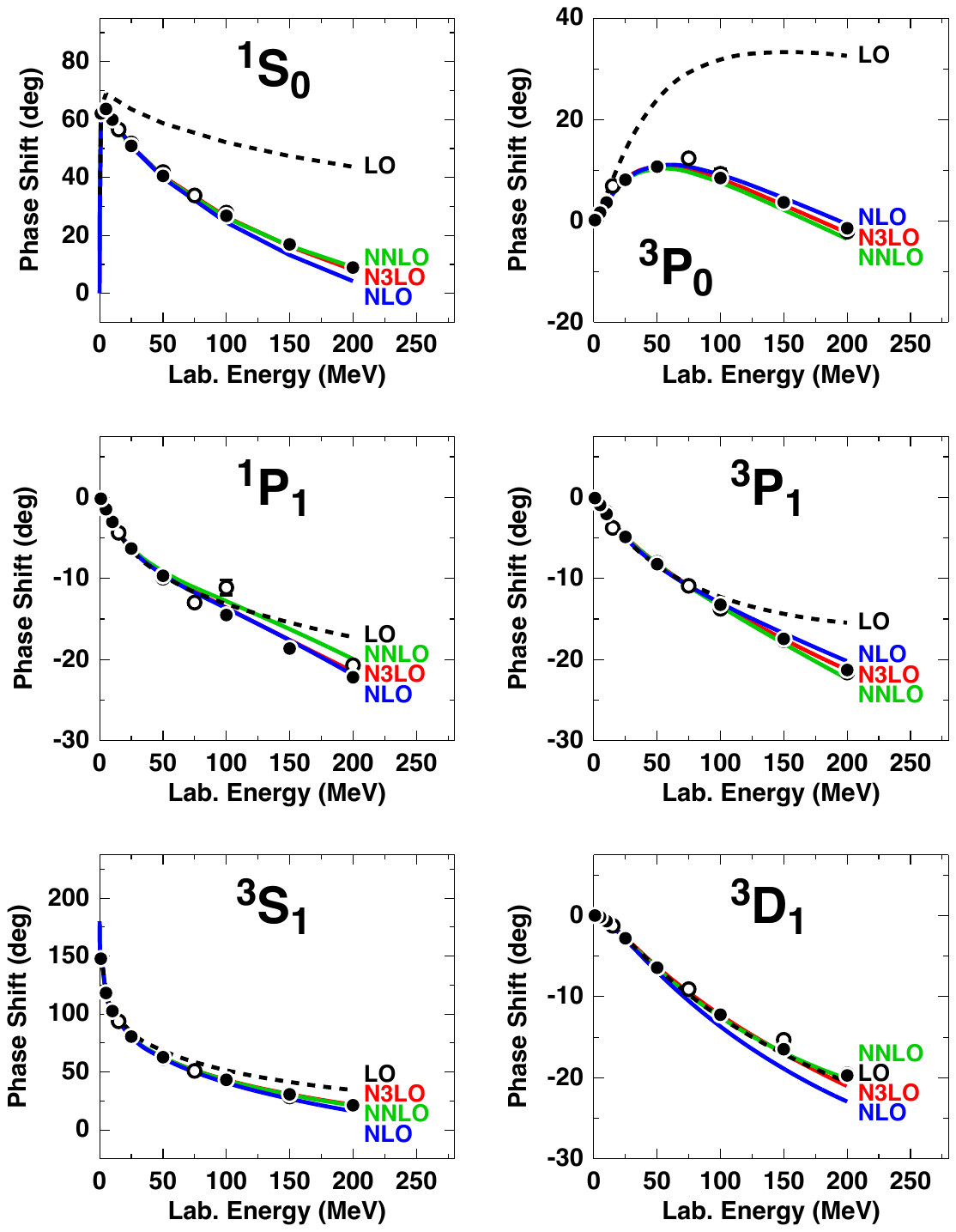}}
\hspace*{1.0cm}
\scalebox{0.37}{\includegraphics{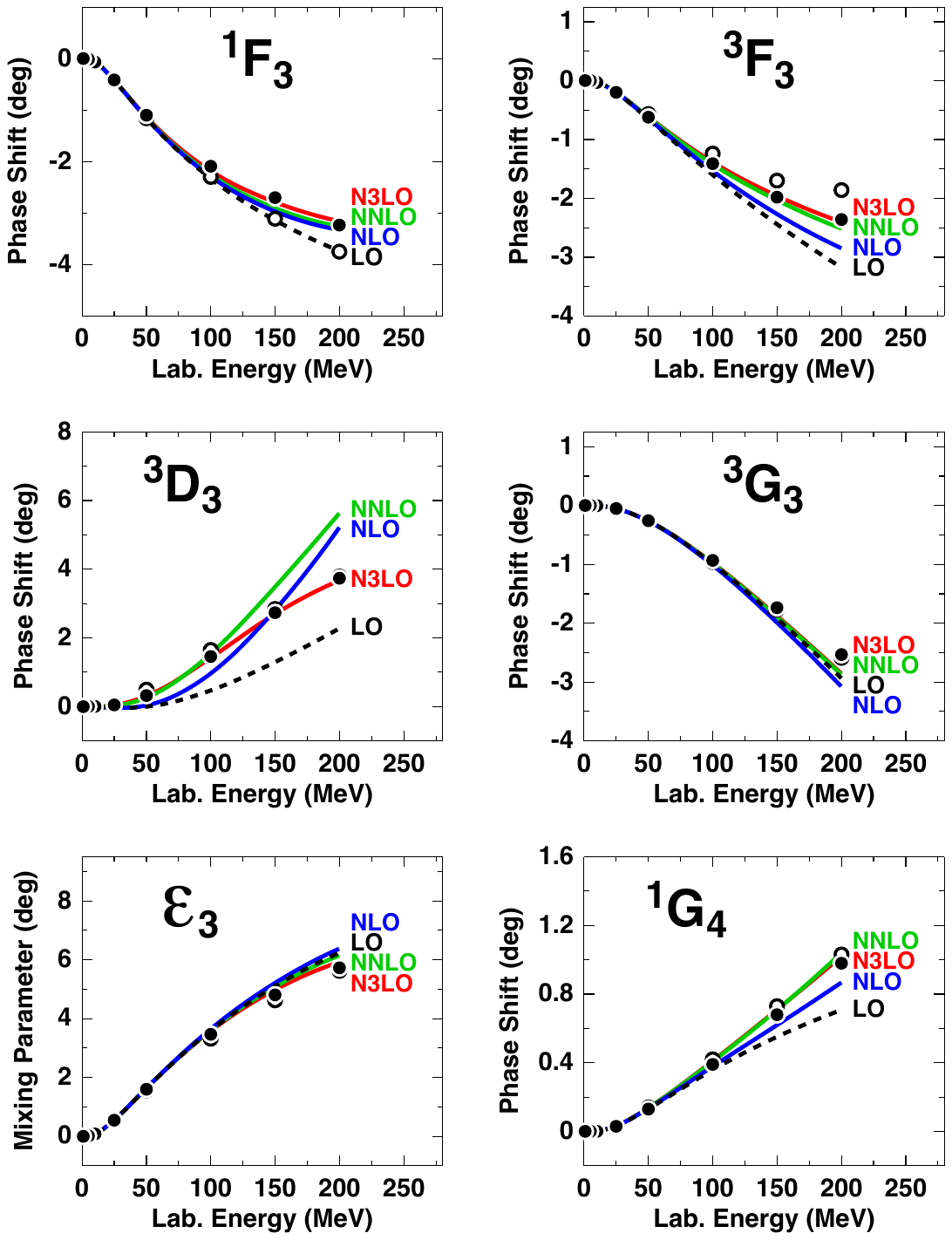}}

\scalebox{0.37}{\includegraphics{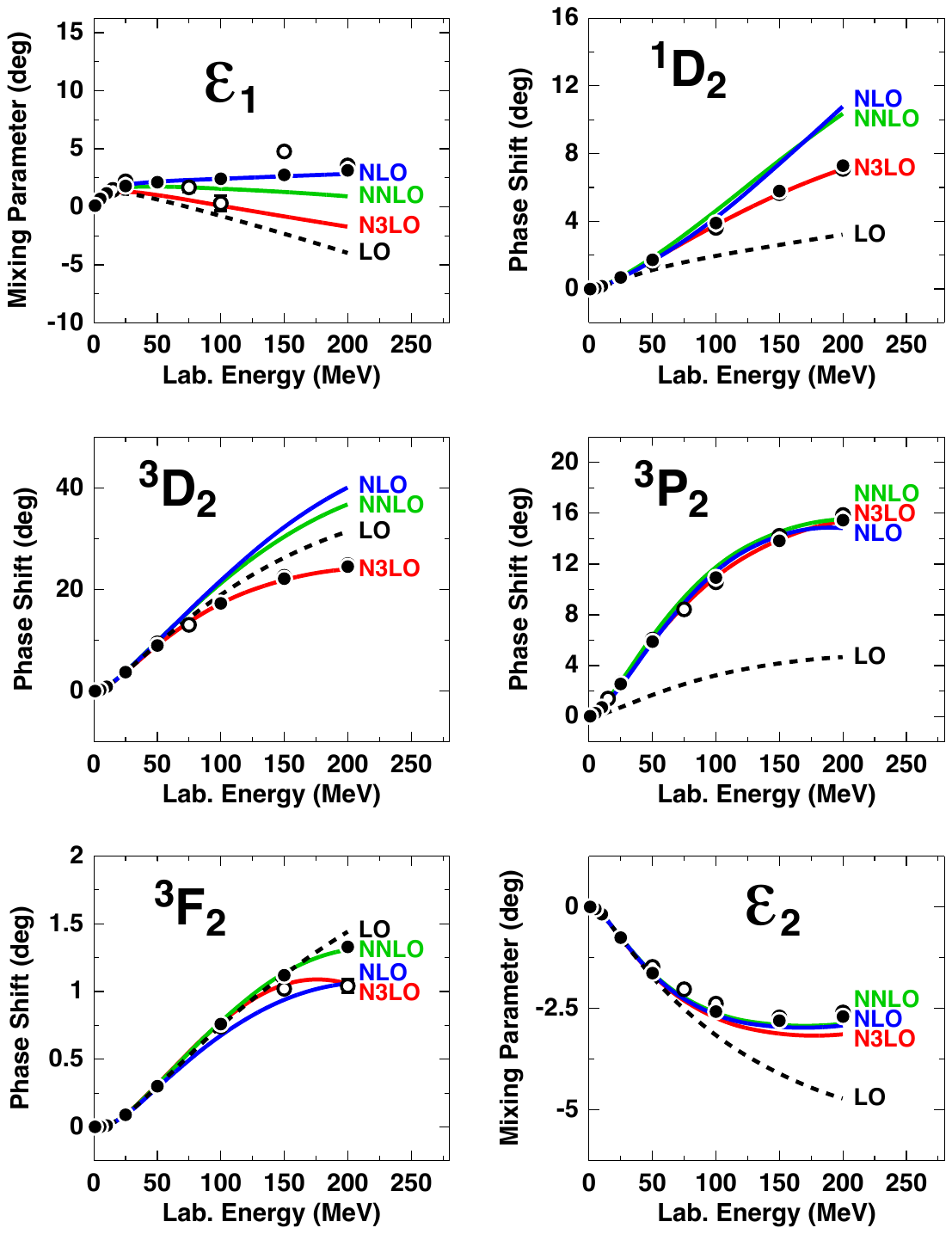}}
\hspace*{1.0cm}
\scalebox{0.37}{\includegraphics{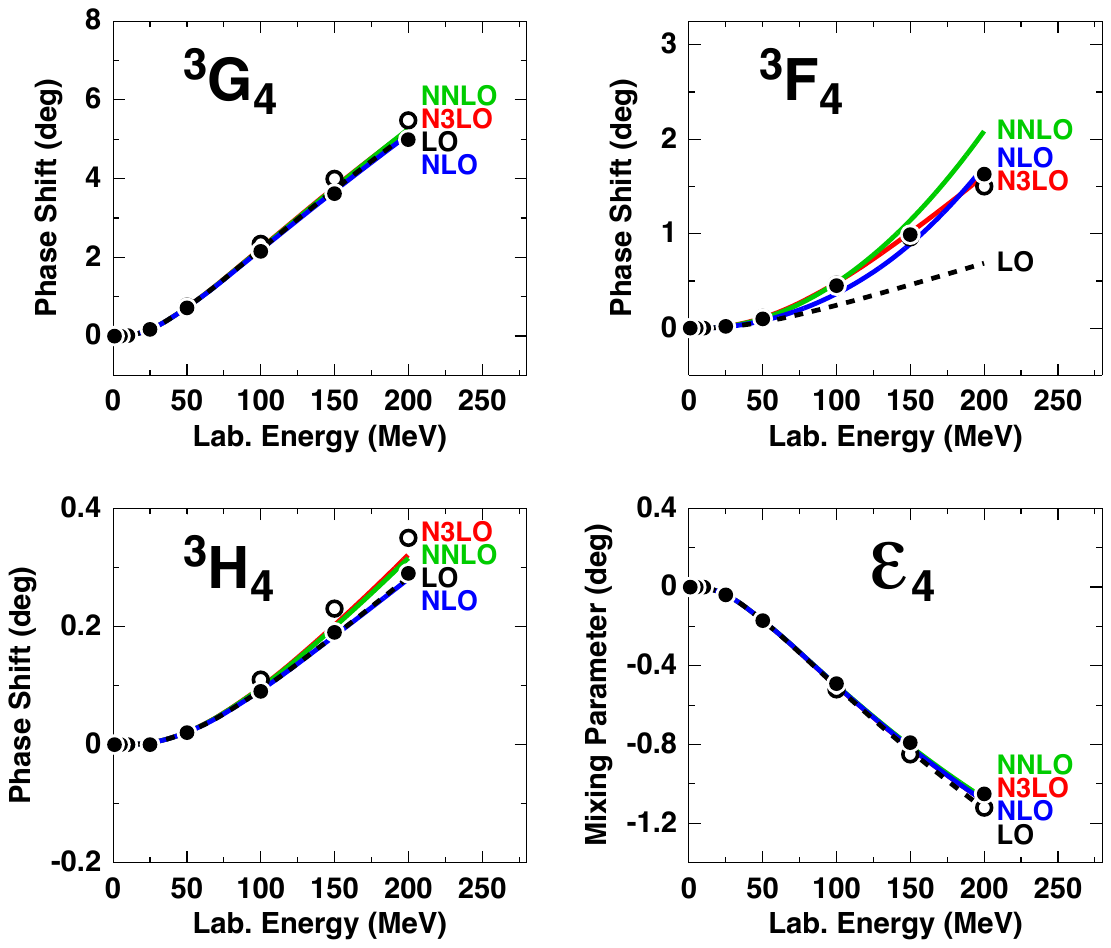}}
\vspace*{-0.2cm}
\caption{
Chiral expansion of neutron-proton scattering as represented by the phase parameters
for $J \leq 4$.
 Four orders ranging from LO to N$^3$LO are shown as denoted.
The cutoff combination $(R_\pi,R_{\rm ct})=(1.0,0.70)$ fm
is applied in all cases.
The filled and open circles represent the results from the Nijmegen multi-energy $np$ phase-shift analysis~\cite{Sto93} and the GWU single-energy $np$ analysis SP07~\cite{SP07}, respectively.
\label{fig_figph1}}
\end{figure}

\begin{table}
\caption{$\chi^2/$datum for several energy intervals as obtained from the fit of the 2016 $NN$ database~\cite{EMN17} by $NN$ potentials at various orders of chiral EFT applying the cutoff combination $(R_\pi,R_{\rm ct})=(1.0,0.70)$ fm. Note that the $\chi^2$ is always minimized
for the interval 0--190 MeV.
\label{tab_chi}}
\smallskip
\begin{tabular*}{\textwidth}{@{\extracolsep{\fill}}cccccc}
\hline 
\hline 
\noalign{\smallskip}
 $T_{\rm lab}$ bin (MeV) & No.\ of data & LO & NLO & NNLO & N$^3$LO  \\
\hline
\noalign{\smallskip}
\multicolumn{6}{c}{\bf proton-proton} \\
0--100 & 795 & 433 & 1.85  & 2.64   &  1.32\\
0--190 & 1206  & 363 & 4.60  & 7.84 & 1.33 \\
0--290 & 2132 & 341 & 16.2 & 18.1 & 1.69 \\
\hline
\noalign{\smallskip}
\multicolumn{6}{c}{\bf neutron-proton} \\
0--100 & 1180 & 211 & 1.58  &  2.34  & 1.59  \\
0--190 & 1697 &  157 & 15.0  &  10.2 &  1.53  \\
0--290 & 2721 & 109 & 35.4 & 21.4 & 1.99 \\
\hline
\noalign{\smallskip}
\multicolumn{6}{c}{\boldmath $pp$ plus $np$} \\
0--100 & 1975 & 300 &  1.68 &  2.45  & 1.48   \\
0--190 & 2903 & 243 &  10.7 &  9.23 & 1.45  \\
0--290 & 4853 & 203 & 26.9 & 20.0 & 1.86 \\
\hline
\hline
\noalign{\smallskip}
\end{tabular*}
\end{table}

The free (fit) parameters of our theory are the coefficients of the contact terms presented in
Sec.~\ref{sec_short}.
The other set of parameters involved in $NN$ potential construction are the $\pi N$ LECs.
 We apply the ones from the 
very accurate Roy-Steiner analysis of Ref.~\cite{Hof15} given in Table~\ref{tab_lecs}.
We use the central values and, thus, the $\pi N$ LECs are precisely fixed from
the outset and no fit parameters.

Fitting proceeds in two steps. First we fit phase shifts, where the adjustment is done
to the Nijmegen multi-energy analysis~\cite{Sto93}, which we perceive as the most reliable one. 
In the second step, the potential predictions are confronted with
the experimental $NN$ data---calculating the $\chi^2$ as follows.

The experimental data are broken up into
groups (sets) of data, $A$, with
$N_A$ data points and 
an experimental over-all normalization uncertainty $\Delta n_A^{exp}$. 
For datum $i$ of set $A$,
$x^{exp}_{A,i}$ is the experimental value,
$\Delta x^{exp}_{A,i}$ the experimental uncertainty, and
$x^{mod}_{A,i}$ the model prediction.
When fitting the data of group $A$ by a model (or a phase shift
solution), the over-all normalization, $n_A^{mod}$, is floated and finally
chosen such as to minimize the $\chi^2$ for this group.
The $\chi^2$ is then calculated from~\cite{Ber88}
\begin{equation}
\chi^2= \sum_A \left\{ \sum^{N_A}_{i=1} \left[
\frac{n_A^{mod} \,\, x^{mod}_{A,i} \, -x^{exp}_{A,i}}{\Delta x^{exp}_{A,i}}
\right]^2
+ \, \left[ \frac{n_A^{mod} \, -1}{\Delta n_A^{exp}} \right]^2
\right\} \; ;
\label{eq_chi2}
\end{equation}
that is, the over-all normalization of a group is treated as
an additional parameter.
For groups of data without normalization uncertainty
($\Delta n_A^{exp}=0$), $n_A^{mod}=1$ is used and the second
term on the r.h.s.\ of Eq.~(\ref{eq_chi2}) is dropped.
The total number of data is
\begin{equation}
N_{dat}=N_{obs}+N_{ne}
\end{equation}
where $N_{obs}$ denotes the total number of measured data points (observables), i.~e.,
$N_{obs}=\sum_A N_A$; and
$N_{ne}$ is the number of experimental normalization
uncertainties.
We state results in terms of $\chi^2/N_{dat} \equiv  \chi^2/$datum, where we use
for the experimental $NN$ data the
 ``2016 database'' defined in Ref.~\cite{EMN17}.

Each of the two steps described above, is done in two parts. In part one, we adjust the $pp$ potential, which fixes the $T=1$ partial waves (where $T$ denotes the total isospin
of the two-nucleon system). In part two, the charge-dependence described in Sec.~\ref{sec_CD}
is applied to obtain the $np$ $T=1$ phase shifts from the $pp$ ones. 
The $np$ $T=0$ partial-waves are then
pinned down by first fitting phase shifts and, after that,  minimizing the $\chi^2$ in regard to the $np$ data. During this last step, we allowed for minor changes of the $T=1$ parameters 
(which also modifies the $pp$ potential) to obtain an even lower overall $\chi^2$. 
We always minimize the $\chi^2$ for the energy range 0-190 MeV
laboratory energy ($T_{\rm lab}$). 
For more details on the $NN$ database and the fitting procedure, see Ref.~\cite{EMN17}.

The $nn$ potential is obtained by starting from the $pp$ version, replacing the proton mass
by the neutron mass in the kinetic energy, leaving out Coulomb, and adjusting the zeroth-order contacts such as
to reproduce the empirical $nn$ $^1S_0$ scattering length of --18.95 fm~\cite{Gon06,Che08}.

The contact LECs that result from our best fits at N$^3$LO are tabulated in 
Appendix~\ref{app_lectab}.

Plots of the various components of the chiral potentials in comparison to more traditional
potentials are shown and discussed in Appendix~\ref{app_plots}.

The $\chi^2$/datum for the reproduction of the $NN$ data at various orders of chiral EFT 
are shown in Table~\ref{tab_chi} for different energy intervals below $T_{\rm lab} = 290$ MeV.
The most relevant energy interval is the one from 0--190 MeV, for which 
the $\chi^2$/datum 
is 10.7 at NLO and 9.2 at NNLO for the $pp$ plus $np$ data.
Note that the number of $NN$ contact terms is the same for both orders,
which may naively explain why there is essentially no change.
However, for nonlocal momentum-space potentials~\cite{EMN17} the $\chi^2$
at NNLO turns out to be substantially lower than at NLO, because of a large 2PE contribution at NNLO providing the proper intermediate-range attraction for the nuclear force.
The fact that this is not happening for the present local potentials may have the following explanation:
First note that our $\chi^2$ at NLO is already unusually low as compared to what nonlocal momentum-space potentials (cf., e.g., Ref.~\cite{EMN17}) generate at that order leaving not much room for improvement at NNLO. The unusually good results at NLO may be due to the fact that the iteration of a locally regularized 1PE creates a larger 2PE contribution than the iteration of a nonlocal one. After all, the reason why NLO is in general not doing well is a lack of a sizable 2PE contribution.

Finally, moving on to N$^3$LO, 14 more contacts are added [Eq.~(\ref{eq_ct4r})]
that affect, in particular, the the $^1D_2$ and $^3D_2$ waves, which typically come out far too attractive at NLO and NNLO (Fig.~\ref{fig_figph1}).  This improves the $\chi^2$/datum to 1.45 at N$^3$LO, a respectable value.

All $np$ phase shifts up to $J=4$ and $T_{\rm lab}=200$ MeV are displayed in Fig.~\ref{fig_figph1}, which reflects what just has been
said in the context of the the $\chi^2$.
At this point, it is instructive to talk about the uncertainties of the phase shift predictions. As discussed in Sec.~\ref{sec_uncert} below, the truncation error creates the largest uncertainty, for which the simplest formula is given by 
Eq.~(\ref{eq_err}). Following this prescription, the error at a certain order is the difference between the given order and the next higher one. For example, the uncertainties of our NNLO phase shifts are given by the differences between the (green) NNLO curves and the (red) N3LO curves in Fig.~\ref{fig_figph1}. For the
uncertainty at N3LO, Eq.~(\ref{eq_err2}) has to be invoked.
The factor $Q$ in this formula is, of course, energy dependent but, as a simple rule of thumb, one may assume  $Q \approx 1 / 3$.

\begin{table}
\caption{Scattering lengths ($a$) and effective ranges ($r$) in units of fm as predicted by $NN$ potentials at various orders of chiral EFT  applying the cutoff combination $(R_\pi,R_{\rm ct})=(1.0,0.70)$ fm.
($a_{pp}^C$ and $r_{pp}^C$ refer to the $pp$ parameters in the presence of
the Coulomb force. $a^N$ and $r^N$ denote parameters determined from the
nuclear force only and with all electromagnetic effects omitted.)
$a_{nn}^N$, and $a_{np}$ are fitted, all other quantities are predictions.
\label{tab_lep}}
\smallskip
\begin{tabular*}{\textwidth}{@{\extracolsep{\fill}}cccccc}
\hline 
\hline 
\noalign{\smallskip}
 & LO & NLO & NNLO & N$^3$LO & Empirical \\
\hline
\noalign{\smallskip}
\multicolumn{6}{c}{\boldmath $^1S_0$} \\
$a_{pp}^C$  &--7.8161 & --7.8134& --7.8147 & --7.8136  
  &--7.8196(26)~\cite{Ber88} \\
  &&&&& --7.8149(29)~\cite{SES83} \\
$r_{pp}^C$  & 2.009   & 2.715 & 2.764 & 2.748
  & 2.790(14)~\cite{Ber88}  \\
  &&&&& 2.769(14)~\cite{SES83} \\
$a_{pp}^N$  &--- & --17.364 & --17.466 & --17.391  & --- \\
$r_{pp}^N$  &  ---  & 2.788& 2.834 & 2.818  & --- \\
$a_{nn}^N$  &--18.950 & --18.950& --18.950 & --18.950 
&--18.95(40)~\cite{Gon06,Che08} \\
$r_{nn}^N$  &  1.985  & 2.761 & 2.807 & 2.790  &  2.86(10)~\cite{Mal22} \\
$a_{np}  $  &--23.738 & --23.738 & --23.738 & --23.738 
 &--23.740(20)~\cite{Mac01} \\
$r_{np}  $  &  1.888  & 2.653 & 2.695 & 2.679 & [2.77(5)]~\cite{Mac01}   \\
\hline
\noalign{\smallskip}
\multicolumn{6}{c}{\boldmath $^3S_1$} \\
$a_t$     &  5.299   & 5.414 & 5.413 & 5.420 & 5.419(7)~\cite{Mac01}  \\
$r_t$     &  1.586   & 1.750 & 1.747 & 1.756  & 1.753(8)~\cite{Mac01}  \\
\hline
\hline
\noalign{\smallskip}
\end{tabular*}
\end{table}

The low-energy scattering parameters, order by order for the cutoff combination $(R_\pi,R_{\rm ct})=(1.0,0.70)$ fm,  are shown in Table~\ref{tab_lep}.
For $nn$ and $np$, the effective range expansion without any electromagnetic interaction is used.
In the case of $pp$ scattering, the quantities $a_{pp}^C$ and $r_{pp}^C$
are obtained by using the effective range expansion appropriate in the presence of the
Coulomb force (cf.\ Appendix A4 of Ref.~\cite{Mac01}). Note that the empirical values for
$a_{pp}^C$ and $r_{pp}^C$
in Table~\ref{tab_lep} were obtained by subtracting from the corresponding electromagnetic values the effects due to two-photon exchange and vacuum polarization. Thus, the comparison between theory and experiment for these two quantities is conducted correctly.
$a_{nn}^N$, and $a_{np}$ are fitted, all other quantities are predictions.
Note that the $^3S_1$ effective range parameters $a_t$ and $r_t$ are not fitted.
But the deuteron binding energy is fitted  and that essentially fixes
$a_t$ and $r_t$.

\subsection{Electromagnetic effects}
\label{sec_em}

The full scattering amplitude for $NN$ scattering consists of two parts: the strong-interactions (nuclear) amplitude plus the electromagnetic (em) amplitude.
Following the way the Nijmegen partial-wave analysis was conducted~\cite{Ber88,Sto93,SS90},
the em amplitude includes relativistic Coulomb, two-photon exchange, vacuum polarization, and
magnetic moment (MM) interactions. The nuclear amplitude is parametrized in terms of the strong
nuclear phase shifts which are to be calculated in the presence of the em interaction, i.~e.,
with respect to em wave functions. In the case of $pp$ scattering, it is in general a good
approximation to just use
the phase shifts of the nuclear plus relativistic Coulomb interaction with respect
to Coulomb wave functions. The exception are the $^1S_0$ $pp$
phase shifts below 30 MeV, where
electromagnetic phase shifts are to be used, which are obtained by correcting the Coulomb phase shifts for the distorting effects from two-photon exchange, vacuum polarization, and MM interactions as calculated by the Nijmegen group~\cite{Ber88,Sto95}.
In the case of $np$ and $nn$ scattering,
 the phase shifts  from the nuclear interaction
with respect to Riccati-Bessel functions are applied. More technical details of our phase shift calculations
can be found in Appendix A3 of Ref.~\cite{Mac01}.

The $NN$ potentials constructed in this paper represent the strong nuclear interaction
between two nucleons.
Electromagnetic interactions are not provided, because they are well known and readily available
elsewhere~\cite{WSS95}. In applications of the potentials in the nuclear many-body problem, one would add at least the Coulomb interaction between protons. Other more subtle em interactions between protons, like, two-photon-exchange, vacuum polarization, and MM interactions, 
can also be added to our nuclear $pp$ potentials. However their effects are, in general,
very small and, in fact, much smaller than the effects from off-shell differences between different
strong nuclear potentials. Thus, in most applications, there is no significance to their inclusion.

A special word is called-for concerning our $np$ potentials. 
Following tradition~\cite{Sto93,Sto94,Mac01,EMN17,RKE18,NAA13},
we fit the experimental $^1S_0$ $np$ scattering length, $a_{np}=-23.74$ fm (cf.\
Table~\ref{tab_lep}), and the experimental deuteron binding energy, $B_d=2.22458$ MeV.
This implies that we tacitly include the $np$ MM interaction in our 
strong interaction $np$ potentials. This is not unreasonable, because, e.~g. in $^1S_0$,
only a MM contact term with the range of the $\rho$ meson contributes, which 
is naturally absorbed by the contacts of the EFT potentials.
Therefore, no em interactions must be added to our $np$ potentials.

The bottom line is that, in typical nuclear many-body calculations, all that needs to be
added to our strong $NN$ potentials is the Coulomb force between protons (and 
nuclear three-nucleon forces).

\subsection{The deuteron and triton}
\label{sec_deu}

The evolution of the deuteron properties from LO to N$^3$LO  of chiral EFT are shown in Table~\ref{tab_deu}.
In all cases, we fit the deuteron binding energy ($B_d$) to its empirical value of 2.22458 MeV
using the LO contact parameters. All other deuteron properties are predictions.
Note, however, that the asymptotic $S$ state, $A_S$, and the $^3S_1$ effective range parameter, $r_t$, are related~\cite{KMS84,MV84,STS95} and, furthermore, the $r_t$ is strongly correlated with $B_d$.
Thus, the fact that, at NLO and up, $A_S$ falls essentially within the empirical range is no real freelance prediction.
In contrast, the asymptotic $D/S$ state, $\eta$, is more versatile. While at LO, NLO, and NNLO, the predictions agree with experiment, the value at N$^3$LO is low and outside the N$^3$LO truncation error. This phenomenon  is most likely related to the local character of the present potentials, since such 
underprediction is not happening with nonlocal potentials at N$^3$LO (and N$^4$LO)~\cite{EMN17}. It represents an interesting topic for future investigations
(see also the $\epsilon_1$ discussion, below).

At the bottom of Table~\ref{tab_deu}, we also show the predictions for the triton binding
as obtained in 34-channel charge-dependent Faddeev calculations using only 2NFs. The result is around 8.1 MeV at N$^3$LO. This contribution from the 2NF will require only a moderate 3NF. 
The relatively low deuteron $D$-state probabilities ($\approx 4$\% at N$^3$LO) and the concomitant generous triton binding energy predictions are
a reflection of the fact that our $NN$ potentials have a weaker tensor force than
 commonly used local position-space potentials.
This can also be seen in the predictions for the $\epsilon_1$ mixing parameter that
is a measure for the strength of the mixing of the $^3S_1$ and $^3D_1$ states due to the tensor force.
Our predictions for $\epsilon_1$ at NNLO and N$^3$LO are on the lower side for lab.\ energies above 100 MeV (Fig.~\ref{fig_figph1}). 
However, there is agreement with the GWU analysis~\cite{SP07} at 100 MeV.
Note that the average relative momentum in nuclear matter at normal density
is equivalent to $T_{\rm lab} \approx 50$ MeV. Thus, the properties of $NN$ potentials 
for $T_{\rm lab} \lea 100$ MeV are the most important ones for nuclear structure applications.
Moreover, the discrepancies between the Nijmegen~\cite{Sto93} and the GWU~\cite{SP07} analyses for $\epsilon_1$
may be seen as an indication that this parameter is not as well determined
as the uncertainties quoted in the analyses suggest. The $\chi^2$/datum of our
N$^3$LO potential is 1.45, which is a typical value achieved in the GWU phase shift
analyses.
Furthermore, the $\chi^2$/datum (for the energy range  0--$\approx$200 MeV)
for the well-established  and highly appreciated N$^3$LO  potentials
 of  Refs.~\cite{Pia16,EMN17,RKE18} are
1.40, 1.35, and 1.50, respectively.
The fact that our $\chi^2$/datum is the same as for the referenced potentials,
while our $\epsilon_1$ differs, implies that our $\epsilon_1$ prediction is as consistent with the data as the alternatives
 and may simply be viewed as
another valid phase shift analysis.

We finally note that the observation that a weak tensor force (low $P_D$) causes a low
$\epsilon_!$ at intermediate energies is a typical feature of {\it local} $NN$ potentials.
For {\it nonlocal} potentials there is not necessarily such a trend as the
weak-tensor force potentials of Ref.~\cite{EMN17} demonstrate.

\begin{table}
\small
\caption{Two- and three-nucleon bound-state properties as predicted by
  $NN$ potentials at various orders of chiral EFT  applying the cutoff combination $(R_\pi,R_{\rm ct})=(1.0,0.70)$ fm.
(Deuteron: Binding energy $B_d$, asymptotic $S$ state $A_S$,
asymptotic $D/S$ state $\eta$,
quadrupole moment $Q$, $D$-state probability $P_D$; the prediction for
 $Q$ is without meson-exchange current contributions
and relativistic corrections. Triton: Binding energy $B_t$.)
$B_d$ is fitted, all other quantities are predictions.
\label{tab_deu}}
\smallskip
\begin{tabular*}{\textwidth}{@{\extracolsep{\fill}}llllll}
\hline 
\hline 
\noalign{\smallskip}
 & LO & NLO & NNLO & N$^3$LO &  Empirical$^a$ \\
\hline
\noalign{\smallskip}
{\bf Deuteron} \\
$B_d$ (MeV) &
 2.22458& 2.22458 &
 2.22458 & 2.22458 & 2.224575(9) \\
$A_S$ (fm$^{-1/2}$) &
 0.8613& 0.8833 &
0.8836 & 0.8852 &  0.8846(9)  \\
$\eta$         & 
 0.0254& 0.0259 &
0.0252& 0.0242 &  0.0256(4) \\
$Q$ (fm$^2$) &
 0.264& 0.284&
 0.274 & 
 0.260 & 
 0.2859(3)  \\
$P_D$ (\%)    & 
 5.08& 5.67&
5.02 & 4.03  & --- \\
\hline
\noalign{\smallskip}
{\bf Triton} \\
$B_t$ (MeV) & 11.88  & 7.87  & 7.98 & 8.09   & 8.48 \\
\hline
\hline
\noalign{\smallskip}
\end{tabular*}
\footnotesize
$^a$See Table XVIII of Ref.~\cite{Mac01} for references.
\\
\end{table}

\subsection{Cutoff variations}

\begin{figure}[th]
\hspace*{-0.7cm}
\scalebox{0.43}{\includegraphics{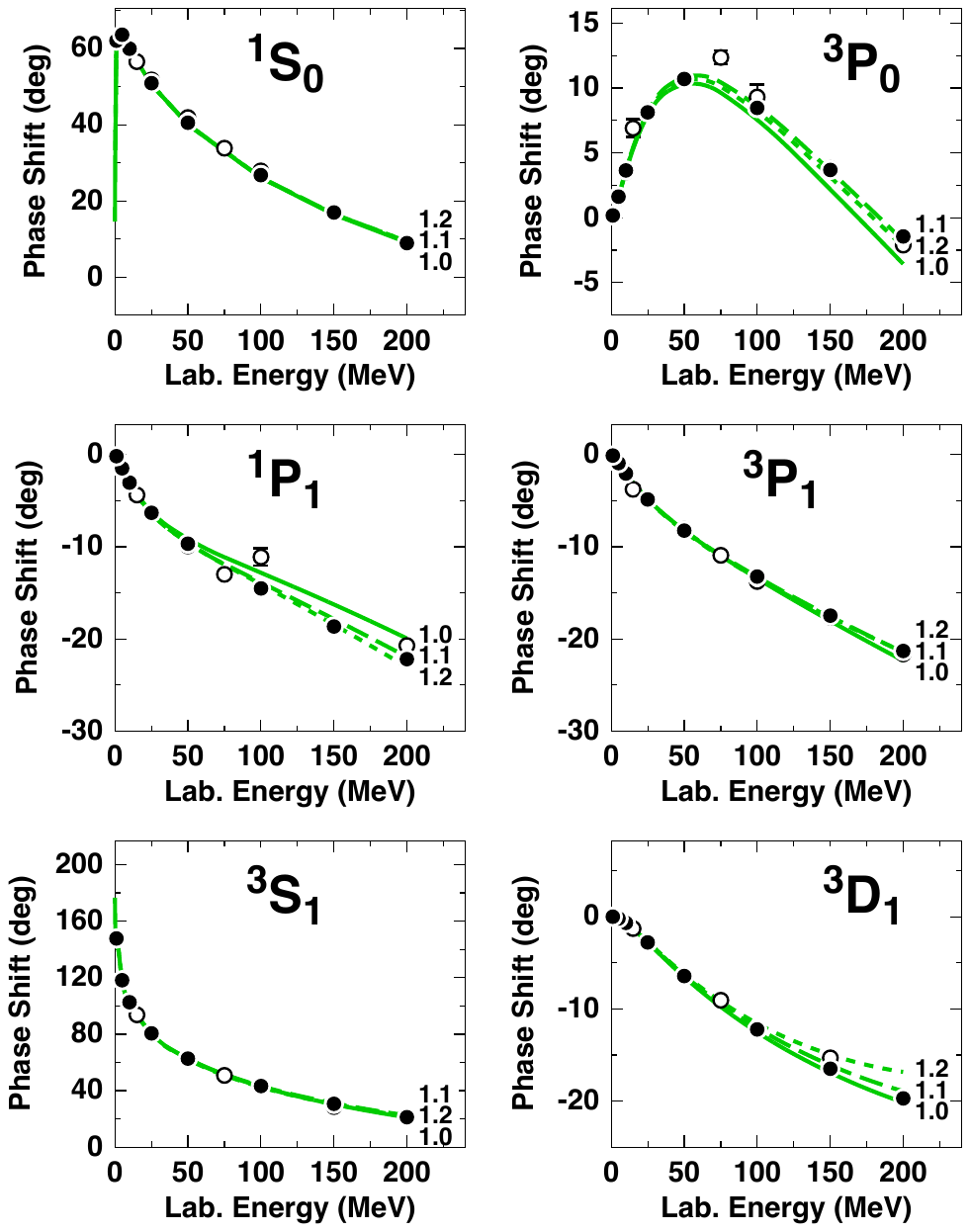}}
\hspace*{1.0cm}
\scalebox{0.43}{\includegraphics{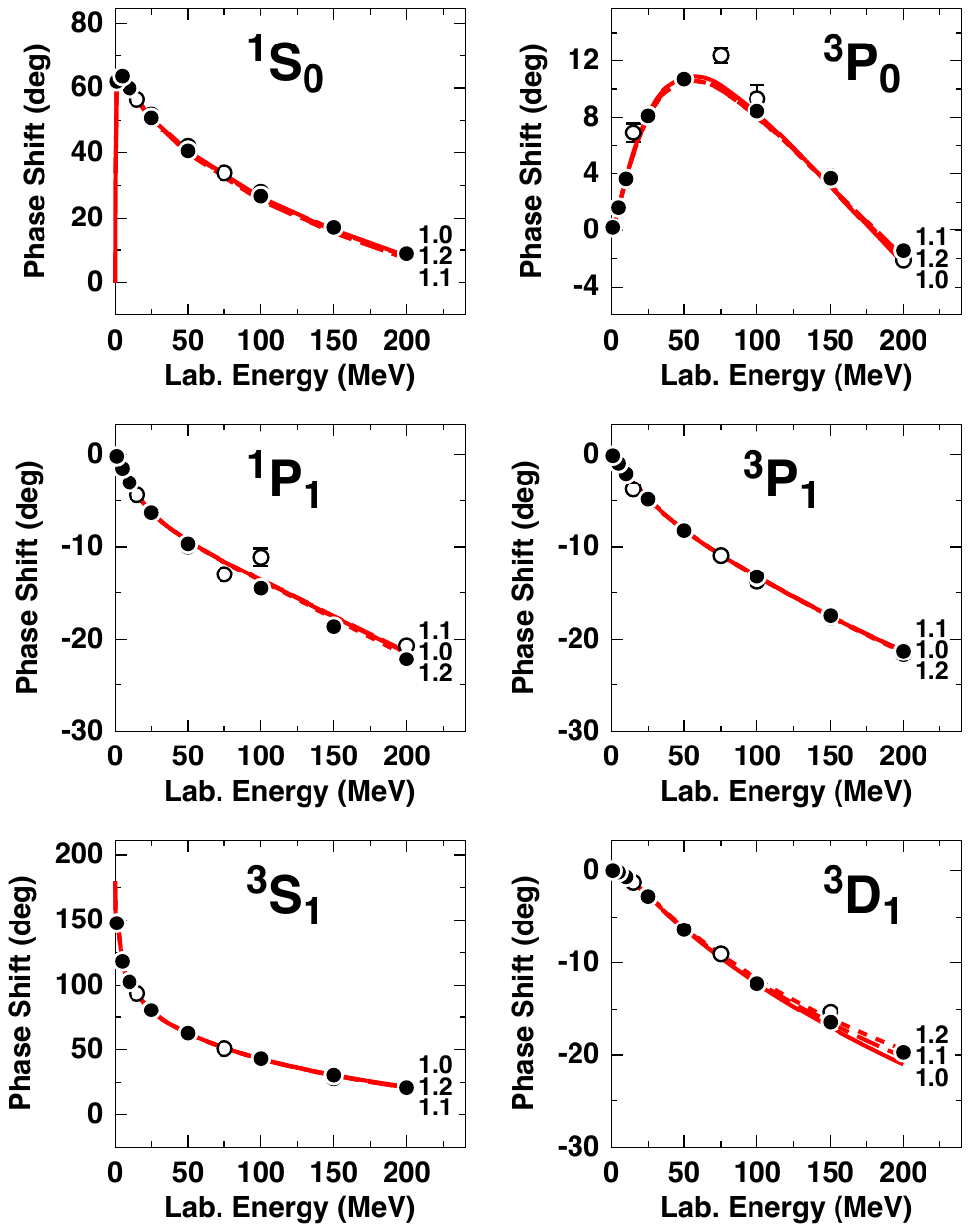}}

\vspace*{-0.4cm}
\hspace*{-0.7cm}
\scalebox{0.43}{\includegraphics{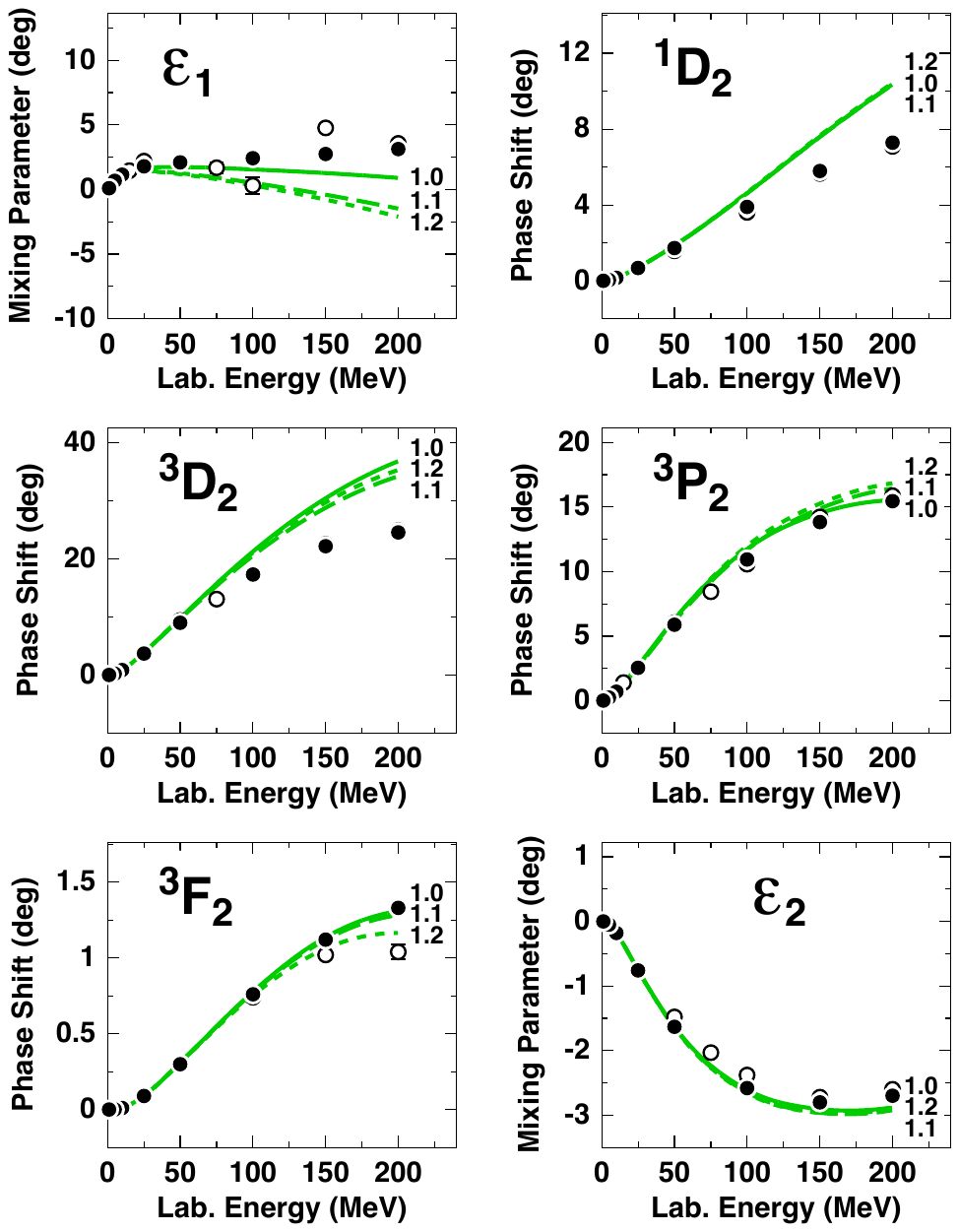}}
\hspace*{1.0cm}
\scalebox{0.43}{\includegraphics{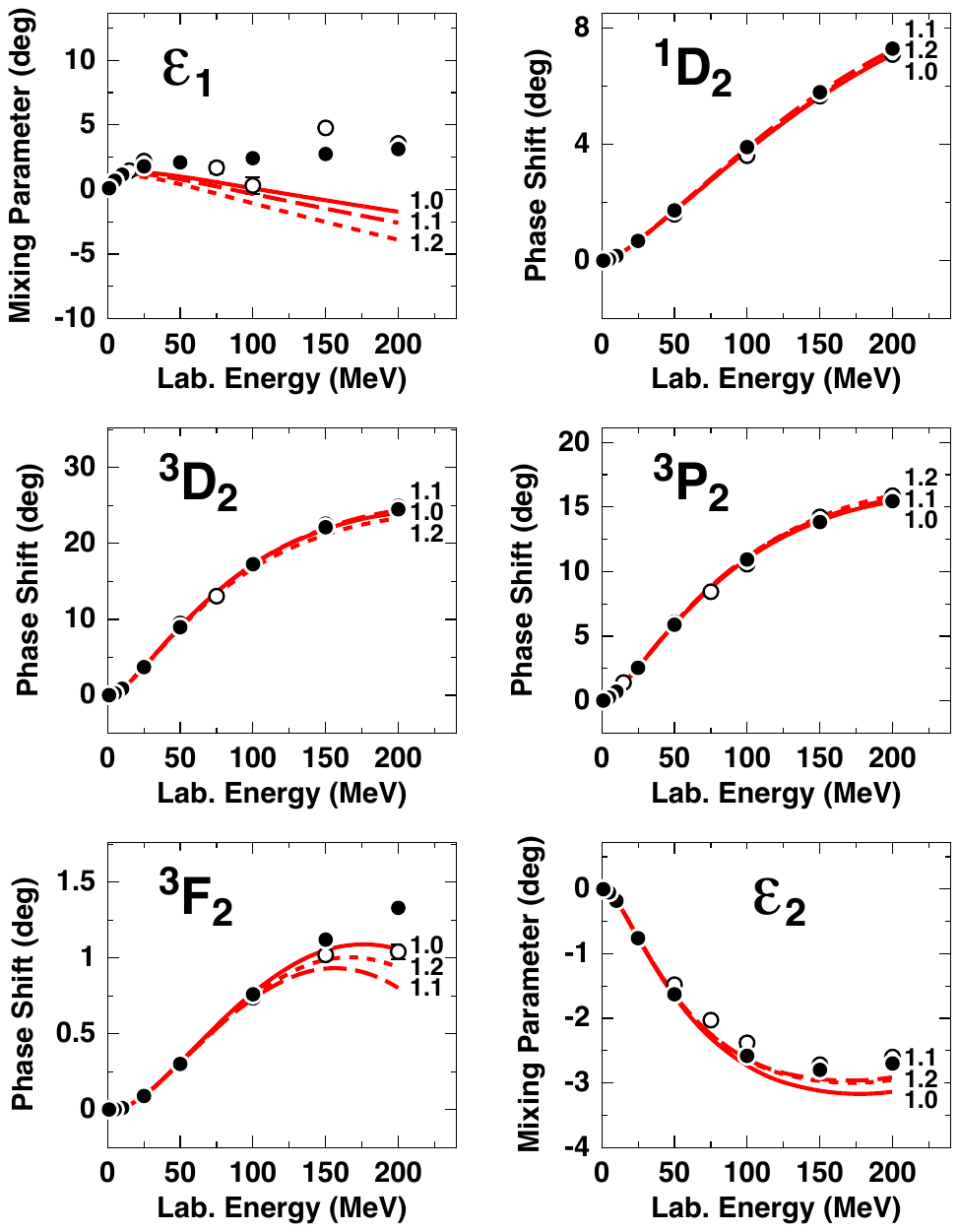}}
\vspace*{-0.5cm}
\caption{
Cutoff variations of the $np$ phase shifts at NNLO (left side, green lines) and N$^3$LO (right side, red lines). Solid, dashed, and dotted lines represent the results obtained with
 the cutoff combinations $(R_\pi, R_{\rm ct})$ =
 (1.0, 0.70) fm, (1.1, 0.72) fm, and (1.2, 0.75) fm, respectively,
as also indicated by the curve labels which state the $R_\pi$ value.
Filled and open circles as in Fig.~\ref{fig_figph1}.
\label{fig_figph2}}
\end{figure}

\begin{table}
\small
\caption{$\chi^2$/datum for the fit of the $pp$ plus $np$ data up to 100 MeV and two- and three-nucleon bound-state properties as produced by
  $NN$ potentials at NNLO and N$^3$LO with the cutoff combinations
 $(R_\pi, R_{\rm ct})= (1.2,0.75)$ fm, $(1.1,0.72)$ fm, and $(1.0,0.70)$ fm. 
 In the column headings, we use the $R_\pi$ value to identify the different cases.
  For some of the notation, see Table~\ref{tab_deu}, where also empirical information on the deuteron and triton can be found.
\label{tab_deu2}}
\smallskip
\begin{tabular*}{\textwidth}{@{\extracolsep{\fill}}lllllll}
\hline 
\hline 
\noalign{\smallskip}
  & \multicolumn{3}{c}{NNLO} & \multicolumn{3}{c}{N$^3$LO} \\
               \cline{2-4}                            \cline{5-7}
   & $R_\pi=1.2$ fm  & $R_\pi=1.1$ fm & $R_\pi=1.0$ fm & 
   $R_\pi=1.2$ fm  & $R_\pi=1.1$ fm & $R_\pi=1.0$ fm \\
  \hline
\noalign{\smallskip}
{\bf\boldmath $\chi^2$/datum $pp$ \& $np$} \\
0--100 MeV (1975 data) &
 2.75 & 2.39 & 2.45 &
 1.75 & 1.56 & 1.48 \\
\hline
\noalign{\smallskip}
{\bf Deuteron} \\
$B_d$ (MeV) &
 2.22458& 2.22458&
 2.22458 & 2.22458 & 2.22458 & 2.22458 \\
$A_S$ (fm$^{-1/2}$) &
0.8862 & 0.8835 & 0.8836 & 
 0.8842 & 0.8851 & 0.8852  \\
$\eta$         & 
0.0244 & 0.0246 & 0.0252 &
 0.0234 & 0.0239 & 0.0242 \\
$Q$ (fm$^2$) &
0.263  & 0.265 & 0.274 & 
 0.248 & 0.255 & 0.260  \\
$P_D$ (\%)    & 
 3.98 & 4.27 & 5.02 &
3.22 & 3.65 & 4.03 \\
\hline
\noalign{\smallskip}
{\bf Triton} \\
$B_t$ (MeV) & 
 8.31 & 8.25 & 7.98 & 
8.40  & 8.18 & 8.09 \\
\hline
\hline
\noalign{\smallskip}
\end{tabular*}
\end{table}

As noted before, besides
 the cutoff combination $(R_\pi,R_{\rm ct})=(1.0, 0.70)$ fm, 
 we have also constructed potentials with the combinations
 (1.1, 0.72) fm, and (1.2, 0.75) fm, to allow for systematic
studies of the cutoff dependence.
In Fig.~\ref{fig_figph2}, we display the variations of the $np$ phase shifts for different cutoffs at NNLO (left half of figure, green curves) and at N$^3$LO (right half of figure, red curves).
Fig.~\ref{fig_figph2} demonstrates nicely how cutoff dependence
diminishes with increasing order---a reasonable trend.
Another point that is evident from this figure is that 
(1.2, 0.75) fm
should be considered
as an upper limit for cutoffs, because obviously cutoff artifacts start showing up.

In Table~\ref{tab_deu2}, we show the cutoff dependence for three selected aspects that are of great interest: the $\chi^2$ for the fit of the $NN$ data below 100 MeV, the deuteron properties, and the triton binding energy. The $\chi^2$ does not change substantially as a function of cutoff.  Thus, we can make the interesting observation that the reproduction of $NN$ observables
is not much affected by the cutoff variations.
However, the $D$-state probability of the deuteron, $P_D$, which is not an observable, changes
substantially as a function of cutoff.
As discussed, $P_D$ is intimately related to the strength of
the tensor force of a potential
and so are the binding energies of few-body systems. 
In particular, the cutoff combination $(R_\pi,R_{\rm ct})=(1.1, 0.72)$ fm and (1.2, 0.75) fm
at NNLO as well as N$^3$LO generate the substantial triton binding energies 
between 8.20 and 8.40 MeV and, therefore, differ significantly from other
local position-space potentials that are commonly in use. On these grounds one
can expect that results for light and intermediate-mass nuclei may differ 
considerably when applying our potentials in {\it ab initio} calculations.
It will be interesting to see if this may solve some of the problems that some
 {\it ab initio} calculations with local potentials are currently beset with.

\section{Uncertainty quantifications}
\label{sec_uncert}

 In {\it ab initio} calculations applying chiral two- and many-body forces,
major sources of uncertainties are~\cite{FPW15}:
\begin{enumerate}
\item
Experimental errors of the input $NN$ data that the 2NFs are based upon and the input
few-nucleon data to which the 3NFs are adjusted.
\item
Uncertainties in the Hamiltonian due to 
       \begin{enumerate}
       \item
       uncertainties in the determination of the $NN$ and $3N$ contact LECs,
       \item
       uncertainties in the $\pi N$ LECs, 
       \item
       regulator dependence, 
       \item
       EFT truncation error.
       \end{enumerate}
\item
Uncertainties associated with the few- and many-body methods applied.
\end{enumerate}

The experimental errors in the $NN$ scattering and deuteron data propagate into the                     
$NN$ potentials that are adjusted to reproduce those data.
 To systematically investigate this error propagation,
the Granada group has constructed smooth local
potentials~\cite{PAA14}, the parameters of which carry the uncertainties implied by the errors in the $NN$ data. 
Applying 205 Monte Carlo samples of these potentials, they find an uncertainty of 15 keV 
for the triton binding energy~\cite{Per14}.
In a more recent study~\cite{Per15}, in which only 33 Monte Carlo samples were used, 
the Granada group reproduced the uncertainty of 15 keV 
for the triton binding energy and, in addition, determined the uncertainty
for the $^4$He binding energy to be 55 keV.
The conclusion is that the statistical error propagation from the $NN$ input data to
the binding energies of light nuclei is negligible as compared to uncertainties from other sources (discussed below). Thus, this source of error can be safely neglected at this time.
Furthermore, we need to consider the propagation of experimental errors from the experimental
few-nucleon data that the 3NF contact terms are fitted to. Also this will be negligible as long as 
the 3NFs are adjusted to data with very small experimental errors; for example the empirical
binding energy of the triton is $8.481795 \pm 0.000002$ MeV, which will definitely lead to
negligible propagation. 

Now turning to the Hamiltonian, we have to, first, account for uncertainties in the $NN$ and
$3N$ LECs due to the way they are fixed.
Based upon our experiences from Ref.~\cite{Mar13} and the fact that chiral EFT is
a low-energy expansion, we have fitted the $NN$ contact LECs to the $NN$ data below
100 MeV at LO and NLO and below 190 MeV at NNLO and N$^3$LO.
One could think of choosing these fit-intervals slightly different and
a systematic investigation of the impact of such variation on the $NN$ LECs 
is still outstanding. However, we do not anticipate 
that large uncertainties would emerge from this source of error.

The story is different for the 3NF contact LECs, since several, 
very different procedures are in use for how to fix them.
The 3NF at NNLO has 
two free parameters (known as the $c_D$ and $c_E$ parameters). 
To fix them, two data are needed.
In most procedures, one of them is the triton binding energy.
For the second datum, the following choices have been made:
 the $nd$ doublet scattering length $^2a_{nd}$~\cite{Epe02},
the binding energy of $^4$He~\cite{Nog06},
the point charge radius radius of $^4$He~\cite{Heb11},
the Gamow-Teller matrix element of tritium $\beta$-decay~\cite{GP06,GQN09,Mar12}.
Alternatively,  the $c_D$ and $c_E$ parameters have also been pinned down by just
an optimal over-all fit of the properties of light nuclei~\cite{Nav07a}.
3NF contact LECs determined by different procedures will lead to different predictions
for the observables that were not involved in the fitting procedure. The differences in those results
establish the uncertainty.  
Specifically, it would be of interest to investigate the differences that occur
for the properties of intermediate-mass nuclei and nuclear matter when 3NF LECs fixed by different protocols
are applied.

The uncertainty in the $\pi N$ LECs used to be a large source of 
uncertainty, in particular, for predictions for many-body systems~\cite{Kru13,DHS16,Dri16}.
With the new, high-precision determination of the $\pi N$ LECs 
in the Roy-Steiner equations analysis~\cite{Hof15} (cf.\ Table~\ref{tab_lecs})
this large uncertainty is essentially eliminated, which is great progress, since it substantially reduces the error budget. We have varied the $\pi N$ LECs within the errors given in 
Table~\ref{tab_lecs} and find that the changes caused by these variations can easily be compensated by small readjustments of the $NN$ LECs resulting in essentially identical
phase shifts and $\chi^2$ for the fit of the data. Thus, this source of error is essentially negligible. The $\pi N$ LECs also appear in the 3NFs,
which also include contacts that can be used for readjustment. Future calculations of finite nuclei and nuclear matter should investigate what residual changes remain after such readjustment
(that would represent the uncertainty).
We expect this to be small.

The choice of the regulator function and its cutoff parameter create uncertainty.  
Originally, cutoff variations were perceived as a demonstration of the uncertainty at a given order
(equivalent to the truncation error).
However, in various investigations~\cite{Sam15,EKM15} it has been demonstrated that this is not correct and that cutoff variations,
in general, underestimate this uncertainty. 
Therefore, the truncation error is better determined by sticking literally to what
 `truncation error' means, namely, the error due to
 omitting the contributions from orders beyond the given order $\nu$. 
 The largest such contribution is the one of order $(\nu + 1)$,
 which one may, therefore, consider as representative for the magnitude of what is left out.
This suggests that the truncation error at order $\nu$ can reasonably be defined as
\begin{equation}
\Delta X_\nu (p)= |X_\nu (p) - X_{\nu+1} (p) | \,, 
\label{eq_err}
\end{equation}
where $X_\nu (p)$ denotes the prediction for observable $X$ at order $\nu$ and momentum $p$. If $X_{\nu+1}$ is not available, then one may use, 
\begin{equation}
\Delta X_\nu (p) = |X_{\nu-1} (p) - X_\nu (p) | Q \,,
\label{eq_err2}
\end{equation}
 with the expansion parameter $Q$ chosen as
  \begin{equation}
  Q = \max \left\{ \frac{m_\pi}{\Lambda_b}, \; \frac{p}{\Lambda_b} \right\} \,,
  \end{equation}
  where $p$ is the characteristic center-of-mass (cms) momentum scale and $\Lambda_b$
  the breakdown scale.
 
Alternatively, one may also apply the more elaborate scheme suggested in Ref.~\cite{EKM15}
where  the truncation error at, e.g.,  N$^3$LO is  calculated in the following way:
   \begin{eqnarray}
  \Delta X_{\rm N^3LO}(p) &=& \max \left\{ Q^5 \times \left| X_{\rm LO}(p) \right|, \;\;
  Q^3 \times \left| X_{\rm LO}(p) - X_{\rm NLO}(p) \right|, \;\;
  Q^2 \times \left| X_{\rm NLO}(p) - X_{\rm NNLO}(p) \right|, \;\;
        \right.      \\  
        && \left.
   Q \times \left| X_{\rm NNLO}(p) - X_{\rm N^3LO}(p) \right|
   \right\} \,,
   \label{eq_error}
  \end{eqnarray}
  with $X_{\rm N^3LO}(p)$ denoting the N$^3$LO prediction for observable $X(p)$, etc..

Note that one should not add up (in quadrature) the uncertainties due to regulator dependence and the truncation error, because they are not independent. In fact, it is appropriate to 
leave out the uncertainty due to regulator dependence entirely and just focus on the truncation error~\cite{EKM15}. The latter should be estimated using the same cutoff 
in all orders considered.

Finally, the last uncertainty to be taken into account is the uncertainty in the few- and many-body methods applied in the {\it ab initio} calculation.
This source of error has nothing to do with EFT.
Few-body problems are nowadays exactly solvable such that the error is negligible in those cases.
For heavier nuclei and nuclear matter, there are definitely uncertainties no matter what method is used. These uncertainties need to be estimated by the practitioners of those methods. But with the improvements of algorithms and the increase of computing power these errors are decreasing.

The conclusion is that the most substantial uncertainty is represented by the truncation error. This is the dominant source of (systematic) error that should be carefully estimated for any calculation applying chiral 2NFs and 3NFs up to a given order.

\section{Summary and Conclusions}
\label{sec_concl}

We have constructed local, position-space chiral $NN$ potentials through four orders of chiral EFT ranging from
LO to N$^3$LO. The construction may be perceived as consistent, because the same power 
counting scheme as well as the same cutoff procedures are applied in all orders.
Moreover, the long-range parts of these potentials are fixed by the very accurate $\pi N$ LECs 
as determined in the Roy-Steiner equations analysis of Ref.~\cite{Hof15}. In fact,
the uncertainties of these LECs are so small that a variation within the errors leads to effects
that are essentially negligible at the current level of precision.
Another aspect that has to do with precision is that, at least at the highest order (N$^3$LO),
the $NN$ data below 190 MeV laboratory energy are reproduced with the respectable
$\chi^2$/datum of 1.45. 

The $NN$ potentials presented in this paper may serve as a
solid basis for systematic {\it ab initio} calculations of nuclear structure and reactions
that allow for a comprehensive error analysis. In particular, the order by order development of the potentials
will make possible a reliable determination of the truncation error at each 
order.

Our new family of local position-space potentials 
differs from the already available potentials of this kind~\cite{Gez14,Pia15,Pia16} 
by a weaker tensor force as reflected in relatively low $D$-state probabilities of the deuteron
($P_D \lea 4.0$ \% for our N$^3$LO potentials) and 
predictions for the triton binding energy above 8.00 MeV (from two-body forces alone).
As a consequence, our potentials will also lead to different predictions when applied to light and
intermediate-mass nuclei in {\it ab initio} calculations~\cite{note2}.
It will be interesting to see if this will help solving some of the outstanding
problems in microscopic nuclear structure.

\acknowledgments
One of the authors (R.M.) would like to thank L. E. Marcucci and R. B. Wiringa for useful communications.
The work by S.K.S., R.M., and Y.N.\ was supported in part by the U.S. Department of Energy
under Grant No.~DE-FG02-03ER41270.
 The contributions by D.R.E.\ have been partially funded through the Ministerio de Ciencia e Innovaci\'on under Contract No. PID2019-105439GB-C22/AEI/10.13039/501100011033 and by the EU Horizon 2020 research and innovation program, STRONG-2020 project under grant agreement No 824093.

\appendix

\section{The long-range $NN$ potential}
\label{sec_applong}

For each order,
we will state, first, the momentum-space functions and then the 
corresponding position-space potentials as obtained by Fourier transform.
Note that all long-range potentials are local.

In momentum space, we use the following decomposition of the long-range potential, 
\begin{eqnarray} 
V_\pi({\vec p}~', \vec p) &  = &
 \,\:\:\:\:\, V_C(q) \:\, + \bm{\tau}_1 \cdot \bm{\tau}_2 \, W_C(q) 
\nonumber \\ && + 
\left[ \, V_S(q) \:\, + \bm{\tau}_1 \cdot \bm{\tau}_2 \, W_S(q) 
\,\:\, \right] \,
\vec\sigma_1 \cdot \vec \sigma_2
\nonumber \\ &&+
\left[ \, V_T(q) \:\,     + \bm{\tau}_1 \cdot \bm{\tau}_2 \, W_T(q) 
\,\:\, \right] \,
\vec \sigma_1 \cdot \vec q \,\, \vec \sigma_2 \cdot \vec q  
\nonumber \\ &&+
\left[ \, V_{LS}(q) + \bm{\tau}_1 \cdot \bm{\tau}_2 \, W_{LS}(q)    
\right] 
\left(-i \vec S \cdot (\vec q \times \vec k) \,\right)
\, .
\label{eq_nnq}
\end{eqnarray}
For notation, see Sec.~\ref{sec_short}.
The position-space potential is represented as follows:
\begin{eqnarray} 
\widetilde{V}_\pi( \vec r) &  = &
 \,\,\:\:\:\:\, \widetilde{V}_C(r) \:\, + \bm{\tau}_1 \cdot \bm{\tau}_2 \, \widetilde{W}_C(r) 
\nonumber \\ && + 
\left[ \, \widetilde{V}_S(r) \:\, + \bm{\tau}_1 \cdot \bm{\tau}_2 \, \widetilde{W}_S(r) 
\,\:\, \right] \,
\vec\sigma_1 \cdot \vec \sigma_2
\nonumber \\ &&+
\left[ \, \widetilde{V}_T(r) \:\,     + \bm{\tau}_1 \cdot \bm{\tau}_2 \, \widetilde{W}_T(r) 
\,\:\, \right] \,
S_{12}(\hat r)
\nonumber \\ &&+
\left[ \, \widetilde{V}_{LS}(r) + \bm{\tau}_1 \cdot \bm{\tau}_2 \, \widetilde{W}_{LS}(r)    
\right] 
\vec L \cdot \vec S 
\, ,
\label{eq_nnr}
\end{eqnarray}
where the operator for total orbital angular momentum is denoted by $\vec L$.

The 2PE potentials in spectral representation are given in momentum space by
\begin{eqnarray}
	V_{C,S}(q) &=& - \frac{2q^6}{\pi} \int_{2m_\pi}^\infty d\mu \frac{{\rm Im} V_{C,S}(i\mu)}{\mu^5(\mu^2+q^2)} \,,  \nonumber
	\\
	V_{T,LS}(q) &=&   \frac{2q^4}{\pi} \int_{2m_\pi}^\infty d\mu \frac{{\rm Im} V_{T,LS}(i\mu)}{\mu^3(\mu^2+q^2)} \,,
\end{eqnarray}
and similarly for $W_{C,S,T,LS}$.
Their Fourier transforms are
\begin{eqnarray}
	\widetilde V_C(r) &=& \frac{1}{2\pi^2 r} \int_{2m_\pi}^\infty d\mu \mu e^{-\mu r} {\rm Im} V_C(i\mu) \,, \nonumber
\label{eq_fc} \\
	\widetilde V_S(r) &=& -\frac{1}{6\pi^2 r} \int_{2m_\pi}^\infty d\mu \mu e^{-\mu r} 
	\left[ \mu^2 {\rm Im} V_T(i\mu)-3 {\rm Im} V_S(i\mu) \right] \,, \nonumber
\label{eq_fs} \\
	\widetilde V_T(r) &=& -\frac{1}{6\pi^2 r^3} \int_{2m_\pi}^\infty d\mu \mu e^{-\mu r} 
	(3+3\mu r+\mu^2r^2)  {\rm Im} V_T(i\mu) \,,  \nonumber
\label{eq_ft} \\
	\widetilde V_{LS}(r) &=& \frac{1}{2\pi^2 r^3} \int_{2m_\pi}^\infty d\mu \mu e^{-\mu r} 
	(1+\mu r) {\rm Im} V_{LS}(i\mu) \,,
	\label{eq_fls}
\end{eqnarray}
and similarly for $\widetilde W_{C,S,T,LS}$.

\subsection{Leading order}
\label{app_1pe}

\begin{figure}
\scalebox{1.0}{\includegraphics{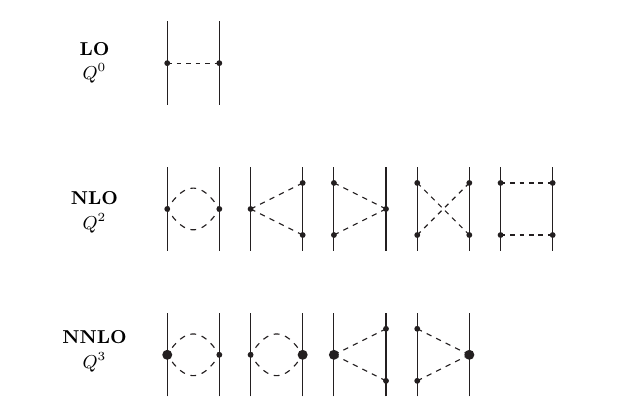}}
\caption{
LO, NLO, and NNLO pion-exchange contributions to the $NN$ interaction.
Notation as in Fig.~\ref{fig_hi}.
\label{fig_dia1}}
\end{figure}

At leading order, only 1PE contributes to the long range, cf.\ Fig.~\ref{fig_dia1}.
The charge-independent 1PE is given in momentum space by
\begin{equation}
W_{T} (q) = - 
\frac{g_A^2}{4f_\pi^2}
\: 
\frac{1}{q^2 + m_\pi^2} 
\,,
\label{eq_1PEci}
\end{equation}
where
$g_A$, $f_\pi$, and $m_\pi$ denote the axial-vector coupling constant,
pion-decay constant, and the pion mass, respectively. See Table~\ref{tab_basic}
for their values.  
Fourier transform yields:
\begin{eqnarray}
\widetilde W_S(r)&=&\frac{g_A^2 m_\pi^2}{48 \pi f_\pi^2} \; \frac{e^{-x}}{r} \,, \label{LOpot1}    \\
\widetilde W_T(r)&=&\frac{g_A^2}{48 \pi f_\pi^2} \; \frac{e^{-x}}{r^3} (3+3x+x^2) \,, \label{LOpot2}
\end{eqnarray}
with $x = m_\pi r$.

For the $NN$ potentials constructed in this paper, we take the charge-dependence of the 1PE due to pion-mass splitting into account. 
For this, we define:
\begin{eqnarray}
\widetilde V_S(m_\pi)&=&\frac{g_A^2 m_\pi^2}{48 \pi f_\pi^2} \; \frac{e^{-x}}{r} \,,     \\
\widetilde V_T(m_\pi)&=&\frac{g_A^2}{48 \pi f_\pi^2} \; \frac{e^{-x}}{r^3} (3+3x+x^2) \,.
\end{eqnarray}
The proton-proton ($pp$) and neutron-neutron ($nn$) potentials are then given by:
\begin{eqnarray}
\widetilde V_S^{(pp)} (r) &=& 
\widetilde V_S^{(nn)} (r) =
\widetilde V_S (m_{\pi^0}) \,, 
\label{eq_1pepp}
\\
\widetilde V_T^{(pp)} (r) &=& 
\widetilde V_T^{(nn)} (r) =
\widetilde V_T (m_{\pi^0}) 
\,,
\end{eqnarray}
and the neutron-proton ($np$) potentials are:
\begin{eqnarray}
\widetilde V_S^{(np)} (r) 
&=& - \widetilde V_S(m_{\pi^0}) + (-1)^{T+1}\, 2\, \widetilde V_S (m_{\pi^\pm})
\,, \\
\widetilde V_T^{(np)} (r) 
&=& - \widetilde V_T(m_{\pi^0}) + (-1)^{T+1}\, 2\, \widetilde V_T (m_{\pi^\pm})
\,,
\label{eq_1penp}
\end{eqnarray}
where $T=0,1$ denotes the total isospin of the two-nucleon system.
See Table~\ref{tab_basic} for the precise values of the pion masses.
Formally speaking, the charge-dependence of the 1PE exchange is of order 
NLO~\cite{ME11}, but we include it also at leading order to make the comparison with
the (charge-dependent) phase-shift analyses meaningful.

Alternatively, the charge-dependent 1PE can also be stated in terms of a
``charge-independent'' 1PE,
\begin{eqnarray}
\widetilde W_S^{\rm CI} (r) 
&=& \frac13 \left[ \widetilde V_S(m_{\pi^0}) + 2\, \widetilde V_S (m_{\pi^\pm}) \right]
\,, \\
\widetilde W_T^{\rm CI} (r) 
&=& \frac13 \left[ \widetilde V_T(m_{\pi^0}) +  2\, \widetilde V_T (m_{\pi^\pm}) \right]
\,,
\label{eq_1peci2}
\end{eqnarray}
plus charge-dependent contributions given by,
\begin{eqnarray}
\widetilde V^{\rm CD} (r) 
&=& \frac13 \left[ \widetilde V_S(m_{\pi^0}) - \widetilde V_S (m_{\pi^\pm}) \right]
\vec\sigma_1 \cdot \vec \sigma_2 \, T_{12}
\,, \\
& + & \frac13  \left[ \widetilde V_T(m_{\pi^0}) - \widetilde V_T (m_{\pi^\pm})  \right]
S_{12} \, T_{12}
\,,
\label{eq_1peCD}
\end{eqnarray}
with the isotensor operator $T_{12}$ defined in Eq.~(\ref{eq_T12}).

\subsection{Next-to-leading order}
\label{app_2pe2}

The 2PE $NN$ diagrams that occur at NLO (cf.\ Fig.~\ref{fig_dia1})
contribute---in momentum space--- in the following way~\cite{KBW97}:
\begin{eqnarray} 
W_C(q) &=&{L(q)\over384\pi^2 f_\pi^4} \left[4m_\pi^2(1+4g_A^2-5g_A^4)
+q^2(1+10g_A^2-23g_A^4) - {48g_A^4 m_\pi^4 \over w^2} \right] \,,  
\label{eq_2C}
\\   
V_T(q) &=& -{1\over q^2} V_{S}(q) \; = \; -{3g_A^4 \over 64\pi^2 f_\pi^4} L(q)\,,
\label{eq_2T}
\end{eqnarray}  
with the logarithmic loop function
\begin{equation}
 L(q) =  {w\over q} 
\ln {\frac{w+q}{2m_\pi}} \,
\end {equation}
and $ w = \sqrt{4m_\pi^2+q^2}$.
Note that
we apply dimensional renormalization for all loop diagrams.
 Moreover,
   in all 2PE contributions, we use the average pion-mass, i.~e.,
$m_\pi = \bar m_\pi$ (cf.\ Table~\ref{tab_basic}).

These expressions imply the spectral functions
\begin{eqnarray} 
{\rm Im}W_C(i\mu) &=&-{1\over768\pi f_\pi^4} \frac{\sqrt{\mu^2-4m_\pi^2}}{\mu} \left[4m_\pi^2(1+4g_A^2-5g_A^4)
-\mu^2(1+10g_A^2-23g_A^4) - {48g_A^4 m_\pi^4 \over 4m_\pi^2-\mu^2} \right] \,,  
\label{eq_2Cim}
\\   
{\rm Im}V_T(i\mu) &=& {1\over \mu^2} {\rm Im}V_{S}(i\mu) \; =
 \; {3g_A^4 \over 128 \pi f_\pi^4} \frac{\sqrt{\mu^2-4m_\pi^2}}{\mu} \,.
\label{eq_2Tim}
\end{eqnarray}  
Via Fourier transform, Eq.~(\ref{eq_fls}), the equivalent position-space potentials are:
\begin{eqnarray}
\widetilde{W}_C(r)&=&\frac{m_\pi}{128\pi^3 f_\pi^4} \frac{1}{r^4} \Bigl\{ \left[ 1 + 2g_A^2(5+2x^2) - g_A^4(23+12x^2)\right] K_1(2x)  
\nonumber \\
&&+ x\left[ 1 + 10g_A^2 - g_A^4(23+4x^2)\right] K_0(2x) \Bigl\} \,, \\
\widetilde{V}_S(r)&=&\frac{g_A^4 m_\pi}{ 32\pi^3 f_\pi^4} \frac{1}{r^4} \left[  3x K_0(2x) + (3+2x^2)K_1(2x) \right] \,, \\
\widetilde{V}_T(r)&=&-\frac{g_A^4 m_\pi}{128\pi^3 f_\pi^4} \frac{1}{r^4} \left[12xK_0(2x) + (15+4x^2)K_1(2x) \right]  \,,
\end{eqnarray}
where $K_0$ and $K_1$ denote the modified Bessel functions.

\subsection{Next-to-next-to-leading order}
\label{app_2pe3}

The 2PE NNLO contribution (cf.\ Fig.~\ref{fig_dia1})  is given by~\cite{KBW97}:
\begin{eqnarray} 
V_C &=&  {3g_A^2 \over 16\pi f_\pi^4} \left[2m_\pi^2(c_3- 2c_1)+c_3 q^2 \right](2m_\pi^2+q^2) 
A(q) \,, \label{eq_3C}
\\
W_T &=&-{1\over q^2}W_{S} =-{g_A^2 \over 32\pi f_\pi^4} c_4 w^2  A(q)\,,
\label{eq_3T}
\end{eqnarray}   
with the loop function 
\begin{equation}
A(q) =  {1\over 2q} \arctan{q \over 2m_\pi} \,.
\end {equation}
The associated spectral functions are
\begin{eqnarray} 
{\rm Im} V_C(i\mu) &=& {3g_A^2 \over 64\mu f_\pi^4} \left[2m_\pi^2(c_3- 2c_1)-c_3 \mu^2 \right](2m_\pi^2-\mu^2) \,, 
\\
{\rm Im} W_T(i\mu) &=&{1\over \mu^2} {\rm Im} W_{S}(i\mu) =-{g_A^2 \over 128\mu f_\pi^4} c_4 (4m_\pi^2-\mu^2) \,;
\end{eqnarray}   
which, by way of Eq.~(\ref{eq_fls}), yield the position-space expressions
\begin{eqnarray}
\widetilde V_C(r)&=&\frac{3g_A^2}{32 \pi^2 f_\pi^4 } \frac{e^{-2x}}{r^6} \left[2c_1 x^2(1+x)^2 + c_3(6+12x+10x^2+4x^3+x^4) \right] \,, \\
\widetilde W_S(r)&=&\frac{g_A^2}{48 \pi^2 f_\pi^4 } \frac{e^{-2x}}{r^6} c_4(1+x)(3+3x+2x^2) \,, \\
\widetilde W_T(r)&=& - \frac{g_A^2}{48 \pi^2 f_\pi^4 } \frac{e^{-2x}}{r^6} c_4(1+x)(3+3x+x^2) \,.
\end{eqnarray}

\subsection{Next-to-next-to-next-to-leading order}
\label{app_2pe4}

\begin{figure}
\scalebox{0.80}{\includegraphics{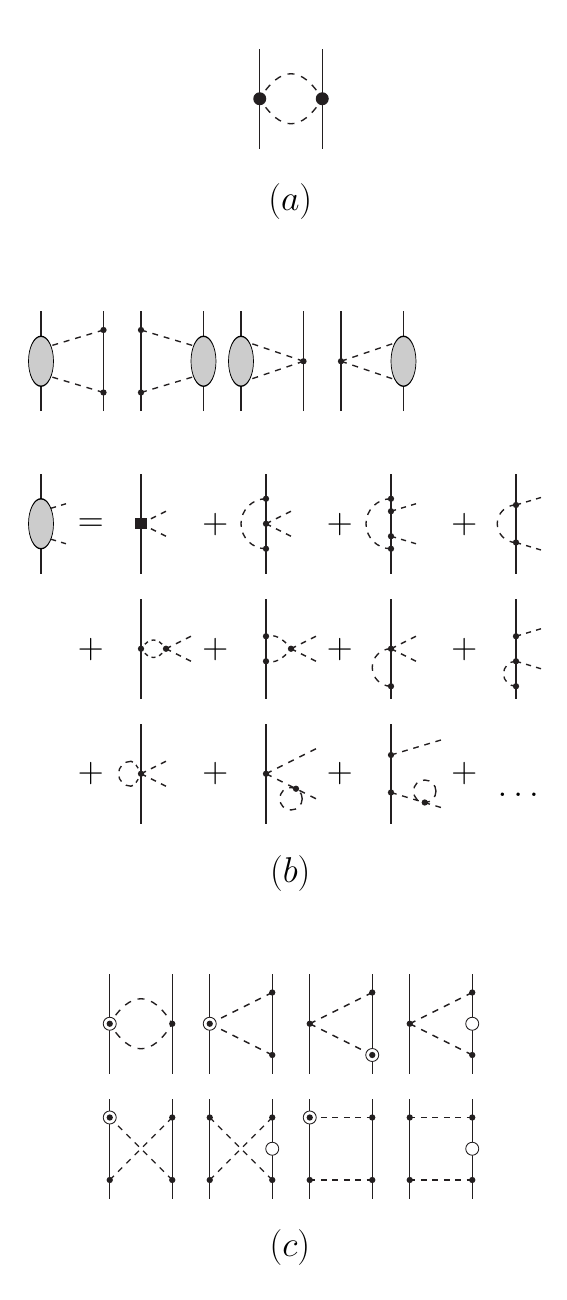}}
\caption{Two-pion exchange contributions at N$^3$LO
with (a) the N$^3$LO football diagram, (b) the leading 2PE two-loop contributions,
and (c) the leading relativistic corrections.
Basic notation as in Fig.~\ref{fig_hi}.
The shaded disc stands for all one-loop $\pi N$ graphs as illustrated.
Open circles  are relativistic $1/M_N$ corrections.}
\label{fig_dia2}
\end{figure}

\subsubsection{Football diagram at N$^3$LO}
\label{sec_football}

The N$^3$LO football diagram, Fig.~\ref{fig_dia2}(a), generates~\cite{Kai01a}:
\begin{eqnarray}
\mbox{Momentum-space potentials:} && \nonumber \\
	V_C(q) &=& \frac{3 \, L(q)}{16\pi^2 f_\pi^4} \bigg[
		\bigg(\frac{c_2}{6} w^2+c_3 (2m_\pi^2+q^2)-4c_1m_\pi^2\bigg)^2
	+\frac{c_2^2}{45} w^4 \bigg] \,, \\
	W_T(q) &=& -\frac{1}{q^2} W_S(q) = \frac{c_4^2 \, w^2 \, L(q)}{96 \, \pi^2 f_\pi^4} \,. \\
\mbox{Spectral functions:} && \nonumber \\
	{\rm Im} V_C(i\mu) &=& -\frac{3}{32\pi f_\pi^4} \frac{\sqrt{\mu^2-4m_\pi^2}}{\mu} \bigg[
		\bigg(\frac{c_2}{6} (4m_\pi^2-\mu^2)+c_3 (2m_\pi^2-\mu^2)-4c_1m_\pi^2\bigg)^2
		\nonumber \\ &&
	+\frac{c_2^2}{45} (4m_\pi^2-\mu^2)^2 \bigg] \,, \\
	{\rm Im} W_T(i\mu) &=& \frac{1}{\mu^2} {\rm Im} W_S(i\mu)  = 
	 \frac{c_4^2}{192\pi f_\pi^4} \frac{(\mu^2-4m_\pi^2)^{3/2}}{\mu}  \,.  \\
\mbox{Position-space potentials:} && \nonumber \\
	\widetilde V_C(r) &=& -\frac{3m_\pi^7}{32\pi^3f_\pi^4} \frac{1}{x^5}
	\bigg[ \left(3c_2^2+20 c_2c_3+60c_3^2 + 4(2c_1+c_3)^2 x^2 \right) x K_1(2x)
	\nonumber \\ &&
+ 2 \left(3c_2^2+20 c_2c_3+60c_3^2 + 2(2c_1+c_3)(c_2+6c_3) x^2 \right) K_2(2x) \bigg] \,, \\
\widetilde W_S(r) &=& \frac{c_4^2 m_\pi^7}{24 \pi^3 f_\pi^4} \frac{1}{x^4}
	\bigg[ 2xK_2(2x) + 5 K_3(2x) \bigg] \,,
	\\
	\widetilde W_T(r) &=& -\frac{c_4^2 m_\pi^7}{96 \pi^3 f_\pi^4} \frac{1}{x^5}
	\bigg[ (3+4x^2) K_2(2x) + 16 x K_3(2x) \bigg]  \,,
\end{eqnarray}
where $K_2(z)=K_0(z)+\frac 2 z K_1(z)$
and $K_3(z)=K_1(z)+\frac 4 z K_2(z)=\frac 4 z K_0(z)+(\frac{8}{z^2}+1) K_1(z)$.

\subsubsection{Leading 2PE two-loop diagrams}

The leading-order $2\pi$-exchange two-loop diagrams are shown in 
Fig.~\ref{fig_dia2}(b). The various contributions are~\cite{Kai01a}:

{\bf Isoscalar central potential:}
\begin{eqnarray}
\mbox{Spectral functions:} && \nonumber \\
	{\rm Im} V_C^{(a)}(i\mu) &=& -\frac{3g_A^4(\mu^2-2m_\pi^2)}{\pi\mu (4f_\pi)^6}
	\bigg\{ (m_\pi^2-2\mu^2) 2m_\pi  
	+ 4 g_A^2 m_\pi (2m_\pi^2-\mu^2) \bigg\}  \,,
\\
	{\rm Im} V_C^{(b)}(i\mu) &=& -\frac{3g_A^4(\mu^2-2m_\pi^2)}{\pi\mu (4f_\pi)^6}
	(m_\pi^2-2\mu^2) 
	\frac{2m_\pi^2-\mu^2}{2\mu} \ln \frac{\mu+2m_\pi}{\mu-2m_\pi}  \,.
\\
\mbox{Position-space potentials:} && \nonumber \\
	\widetilde V_C^{(a)}(r) &=& \frac{3m_\pi^7 g_A^4}{2048\pi^3 f_\pi^6} \frac{e^{-2x}}{x^6} \bigg\{
	24+48 x+43x^2 +22x^3+7x^4
	\nonumber \\ &&
	+4g_A^2(6+12x+10x^2+4x^3+x^4)
	\bigg\} \,,
	\\
	\widetilde V_C^{(b)}(r) &=& -\frac{3m_\pi^7 g_A^4}{8192\pi^3 f_\pi^6} \frac{e^{-2x}}{x^7} \bigg\{
	(120+240x+213x^2+106x^3+32x^4+8x^5)(\ln(4x)+\gamma_E)
	\nonumber \\ &&
	-(120-240x+213x^2-106x^3+32x^4-8x^5)e^{4x} {\rm Ei}(-4x)
	\nonumber \\ &&
	-4x(96+72x+38x^2+7x^3)
	\bigg\}
	\nonumber \\ &&
	+\frac{3m_\pi^7 g_A^4}{4096\pi^3 f_\pi^6} \frac{\bar I_{-1}(2x)}{x}  \,,
\end{eqnarray}
where ${\rm Ei}(-z)$ denotes the exponential integral function defined by
\begin{equation}
{\rm Ei}(-z) = - \int^\infty_z dt \, \frac{e^{-t}}{t} \,,
\end{equation}
and
\begin{equation}
	\bar I_{-1}(z) =
	 \int_{1}^\infty dt \frac{e^{- z t}}{t} 
	\ln\bigg(\frac{t+1}{t-1}\bigg) \,.
\end{equation}
The double precision value for Euler's constant is
$\gamma_E=0.5772156649015329$.
\\
\\
\\
{\bf Isovector central potential:} 
\begin{eqnarray}
\mbox{Spectral functions:} && \nonumber \\
	{\rm Im} W_C^{(a)}(i\mu) &=& -\frac{2\kappa}{3\mu(8\pi f_\pi^2)^3}
	\int_0^1 dz \left[ g_A^2(2m_\pi^2-\mu^2)+2(g_A^2-1) \kappa^2 z^2 \right]
	\nonumber \\ && \times \bigg\{
	\left[ 4m_\pi^2(1+2g_A^2)-\mu^2(1+5g_A^2)\right] \frac{\kappa}{\mu} \ln \frac{\mu+2\kappa}{2m_\pi}
	+\frac{\mu^2}{12}(5+13g_A^2) 
	\nonumber \\ &&
	-2m_\pi^2(1+2g_A^2) + 96\pi^2f_\pi^2\left[(2m_\pi^2-\mu^2)(\bar d_1+ \bar d_2)-2\kappa^2z^2 \bar d_3+4m_\pi^2 \bar d_5\right]
        \bigg\}
	\nonumber \\ &=&
	-\frac{2\kappa}{3\mu(8\pi f_\pi^2)^3}
	\left[ g_A^2(2m_\pi^2-\mu^2)+\frac23(g_A^2-1) \kappa^2 \right]
	\nonumber \\ && \times \bigg\{
	\left[ 4m_\pi^2(1+2g_A^2)-\mu^2(1+5g_A^2)\right] \frac{\kappa}{\mu} \ln \frac{\mu+2\kappa}{2m_\pi}
	+\frac{\mu^2}{12}(5+13g_A^2) 
	\nonumber \\ &&
	-2m_\pi^2(1+2g_A^2) + 96\pi^2f_\pi^2\left[(2m_\pi^2-\mu^2)(\bar d_1+\bar d_2)+4m_\pi^2 \bar d_5\right]
        \bigg\}
	\nonumber \\ &&
	-\frac{\kappa^3}{\mu 4\pi f_\pi^4}
	\left[ \frac 1 3 g_A^2(2m_\pi^2-\mu^2)+\frac 2 5(g_A^2-1) \kappa^2 \right] \bar d_3 \,,
\end{eqnarray}
\begin{eqnarray}
	{\rm Im} W_C^{(b)}(i\mu) &=& -\frac{2\kappa}{3\mu(8\pi f_\pi^2)^3}
	\int_0^1 dz \left[ g_A^2(2m_\pi^2-\mu^2)+2(g_A^2-1) \kappa^2 z^2 \right] 
	\nonumber \\ && \times \bigg\{
	-3\kappa^2 z^2 + 6\kappa z \sqrt{m_\pi^2+\kappa^2 z^2} 
	\ln \frac{\kappa z+\sqrt{m_\pi^2+\kappa^2z^2}}{m_\pi} +
	\nonumber \\ &&
	g_A^4(\mu^2-2\kappa^2 z^2-2m_\pi^2)\bigg[
	\frac 5 6 +\frac{m_\pi^2}{\kappa^2 z^2} -\bigg(1+\frac{m_\pi^2}{\kappa^2 z^2} \bigg)^{3/2}
	\ln \frac{\kappa z+\sqrt{m_\pi^2+\kappa^2z^2}}{m_\pi} 
	\bigg] \bigg\} \,,
	\nonumber \\ &&
\end{eqnarray}
with $\kappa=\sqrt{\mu^2/4-m_\pi^2}$.

In Ref.~\cite{EM02} it was found that the contribution from $W_C^{(b)}$ is negligible.
Therefore, we include only $W_C^{(a)}$, which we divide it into three parts:
\begin{eqnarray}
	{\rm Im} W_C^{(a_1)}(i\mu) &=& 
	-\frac{2\kappa}{3\mu(8\pi f_\pi^2)^3}
	\left[ g_A^2(2m_\pi^2-\mu^2)+\frac23(g_A^2-1) \kappa^2 \right]
	\nonumber \\ && \times
	\left[ 4m_\pi^2(1+2g_A^2)-\mu^2(1+5g_A^2)\right] \frac{\kappa}{\mu} \ln \frac{\mu+2\kappa}{2m_\pi}  \,,
	\\
	{\rm Im} W_C^{(a_2)}(i\mu) &=& 
	-\frac{2\kappa}{3\mu(8\pi f_\pi^2)^3}
	\left[ g_A^2(2m_\pi^2-\mu^2)+\frac23(g_A^2-1) \kappa^2 \right]
	\bigg\{
	\frac{\mu^2}{12}(5+13g_A^2) 
	\nonumber \\ &&
	-2m_\pi^2(1+2g_A^2) + 96\pi^2f_\pi^2\left[(2m_\pi^2-\mu^2)(\bar d_1+\bar d_2)+4m_\pi^2 \bar d_5\right]
        \bigg\}  \,,
	\\
	{\rm Im} W_C^{(a_3)}(i\mu) &=& 
	\frac{\kappa^3}{\mu 4\pi f_\pi^4}
	\left[ \frac 1 3 g_A^2(2m_\pi^2-\mu^2)+\frac 2 5(g_A^2-1) \kappa^2 \right] \bar d_3  \,, \\
\mbox{Position-space potentials:} && \nonumber \\
	\widetilde W_C^{(a_1)}(r) &=& -\frac{m_\pi^7}{9216 \pi^5 f_\pi^6}
	\frac{1}{x^7} \bigg\{ 
	\bigg[ 30+89x^2-8x^4 + g_A^2(300+926x^2-32x^4)
	\nonumber \\ &&
	+ g_A^4(750+2405x^2+76x^4) \bigg] K_0(2x)
	+ \bigg[ 137+8x^2+8x^4 
	\nonumber \\ &&
	+ 2g_A^2(685+106x^2+16x^4)
	+ g_A^4(3425+860x^2+32x^4) \bigg] x K_1(2x)
	\bigg\}
	\nonumber \\ &&
	+\frac{m_\pi^7}{576\pi^5 f_\pi^6}(1+2g_A^2)^2 \frac{\widetilde I_{-1}(2x)}{x}  \,,
	\\
	\widetilde W_C^{(a_2)}(r) &=& -\frac{m_\pi^7}{8\pi^3 f_\pi^4} \bigg\{
	-\frac{2g_A^2 x K_1(2x)+(1+5g_A^2)K_2(2x)}{x^3} 2 \bar d_5
	\nonumber \\ &&
	+ \frac{(5+g_A^2(25+2x^2))x K_1(2x)+(10+x^2+g_A^2(50+11x^2))K_2(2x)}{x^5}(\bar d_1+\bar d_2)
	\bigg\}
	\nonumber \\ &&
	+\frac{m_\pi^7}{9216\pi^5f_\pi^6} \frac{1}{x^5} \bigg\{
	(25+g_A^2(190-4x^2)+g_A^4(325+4x^2)) x K_1(2x)
	\nonumber \\ &&
	+2(25-x^2+g_A^2(190+11x^2)+g_A^4(325+44x^2)) K_2(2x)
	\bigg\}  \,,
	\\
	\widetilde W_C^{(a_3)}(r) &=& -\frac{m_\pi^7}{16\pi^3f_\pi^4}
	\frac{2g_A^2 x K_2(2x)+(3+7g_A^2)K_3(2x)}{x^4} 
	\bar d_3  \,,
\end{eqnarray}
with
\begin{equation}
	\widetilde I_{-1} (z) = 
	 \int_1^\infty dt \, \frac{e^{-z t}}{t} \ln(t+\sqrt{t^2-1})
\end{equation}
\\
\\
\\
{\bf Isoscalar spin-spin and tensor potentials:} 
\begin{eqnarray}
\mbox{Spectral functions:} && \nonumber \\
	{\rm Im} V_S^{(a)}(i\mu) &=& \mu^2 \, {\rm Im} V_T^{(a)}(i\mu) =
	-\frac{g_A^2 \kappa^3\mu}{8\pi f_\pi^4}(\bar d_{14}-\bar d_{15})  \,,
	\\
	{\rm Im} V_S^{(b)}(i\mu) &=& \mu^2 \, {\rm Im} V_T^{(b)}(i\mu) 
	\nonumber \\ &=& 
	-\frac{2g_A^6 \kappa^3\mu}{(8\pi f_\pi^2)^3}
	\int_0^1 dz (1-z^2) \bigg[ -\frac 1 6 +\frac{m_\pi^2}{\kappa^2 z^2} 
		-\bigg(1+\frac{m_\pi^2}{\kappa^2 z^2} \bigg)^{3/2} 
	\ln \frac{\kappa z+\sqrt{m_\pi^2+\kappa^2 z^2}}{m_\pi} \bigg]  \,.
\end{eqnarray}
In Ref.~\cite{EM02} it was found that the contribution from $V_S^{(b)}$ and $V_T^{(b)}$
are negligible.
Therefore, we include only $V_S^{(a)}$ and $V_T^{(a)}$, which yield the position-space potentials:
\begin{eqnarray}
	\widetilde V_S^{(a)}(r) &=& -\frac{g_A^2 m_\pi^7}{8\pi^3 f_\pi^4 x^4} (\bar d_{14}-\bar d_{15})
	(2xK_2(2x)+5K_3(2x))  \,,
	\\
	\widetilde V_T^{(a)}(r) &=& \frac{g_A^2 m_\pi^7}{32\pi^3 f_\pi^4 x^5} (\bar d_{14}-\bar d_{15})
	\left[ (3+4x^2)K_2(2x)+16 x K_3(2x) \right]  \,.
\end{eqnarray}
\\
\\
\\
{\bf Isovector spin-spin and tensor potentials:} 
\begin{eqnarray}
\mbox{Spectral functions:} && \nonumber \\
	{\rm Im} W_S(i\mu) &=& -\frac{g_A^4(\mu^2-4m_\pi^2)}{\pi(4f_\pi)^6} \bigg\{
		\bigg[m_\pi^2-\frac{\mu^2}{4}\bigg] \ln\bigg(\frac{\mu+2m_\pi}{\mu-2m_\pi}\bigg)
	+(1+2g_A^2) \mu m_\pi \bigg\} \,, \\
{\rm Im} W_T^{(a)}(i\mu) &=& -\frac{1}{\mu^2} \frac{g_A^4(\mu^2-4m_\pi^2)}{\pi(4f_\pi)^6} 
	(1+2g_A^2) \mu m_\pi  \,,
	\\
	{\rm Im} W_T^{(b)}(i\mu) &=& -\frac{1}{\mu^2} \frac{g_A^4(\mu^2-4m_\pi^2)}{\pi(4f_\pi)^6} 
		\bigg[m_\pi^2-\frac{\mu^2}{4}\bigg] \ln\bigg(\frac{\mu+2m_\pi}{\mu-2m_\pi}\bigg) \,.  \\
\mbox{Position-space potentials:} && \nonumber \\
\widetilde W_S(r) &=& \frac{g_A^4 m_\pi^7}{6144\pi^3 f_\pi^6} \frac{e^{-2x}}{x^7} \bigg\{
		(15+30x+24x^2+8x^3)(\ln(4x)+\gamma_E)
		\nonumber \\ && +
		(-15+30x-24x^2+8x^3)e^{4x} {\rm Ei}(-4x)
		\nonumber \\ && -
		4x(15+15x+8x^2+2x^3)
		\nonumber \\ &&  -
		8g_A^2x(3+6x+5x^2+2x^3)
	\bigg\}  \,,  \\
\widetilde W_T^{(a)}(r) &=& \frac{g_A^4(1+2g_A^2)m_\pi^7}{1536\pi^3f_\pi^6}
	\frac{e^{-2x}}{x^6} (3+6x+4x^2+x^3)  \,,
	\\
\widetilde W_T^{(b)}(r) &=& -\frac{g_A^4 m_\pi^7}{49152 \pi^3 f_\pi^6}
	\frac{e^{-2x}}{x^7} \bigg\{
	-324x -228x^2 -48x^3 
	\nonumber \\ &&
	+5(21+42x+30x^2+4x^3)(\ln(4x)+\gamma_E)
	\nonumber \\ &&
	+5(-21+42x-30x^2+4x^3)e^{4x} {\rm Ei}(-4x) \bigg\}
	\nonumber \\ &&
	-\frac{g_A^4 m_\pi^7}{2048 \pi^3 f_\pi^6} \frac{1}{x^3} \bar I_{-1}(2x)   \,.
\end{eqnarray}

\subsubsection{Leading relativistic corrections}

The leading relativistic corrections, which are shown in Fig.~\ref{fig_dia2}(c),
count as N$^3$LO and are given by~\cite{Ent15a}:

\begin{eqnarray}
\mbox{Momentum-space potentials:} && \nonumber \\
	V_C(q) &=& \frac{3g_A^4}{128 \pi f_\pi^4 M_N} \left[
\frac{m_\pi^5}{2 w^2} + (2m_\pi^2+q^2) (q^2-m_\pi^2) A(q) \right]  \,,
\\
	W_C(q) &=& \frac{g_A^2}{64 \pi f_\pi^4 M_N} \left\{
\frac{3g_A^2m_\pi^5}{2 w^2} +  \left[ (g_A^2(3m_\pi^2+2q^2)-q^2-2m_\pi^2 \right] (2m_\pi^2+q^2) A(q) \right\} \,, 
\\
V_T(q) &=& -\frac{1}{q^2} V_S(q) = \frac{3g_A^4}{256 \pi f_\pi^4 M_N} (5m_\pi^2+2q^2) A(q)  \,,
	\\
	W_T(q) &=& -\frac{1}{q^2} W_S(q) = \frac{g_A^2}{128 \pi f_\pi^4 M_N} \left[g_A^2(3m_\pi^2+q^2)-w^2 \right] A(q)  \,, \\
V_{LS}(q) &=&  {3g_A^4  \over 32\pi f_\pi^4 M_N} \, (2m_\pi^2+q^2) A(q)
 \,,\\  
W_{LS}(q) &=& {g_A^2(1-g_A^2)\over 32\pi f_\pi^4 M_N} \, w^2 A(q) \,. 
\end{eqnarray}
\begin{eqnarray}
\mbox{Spectral functions:} && \nonumber \\
	{\rm Im} V_C(i\mu) &=& \frac{3g_A^4}{512 f_\pi^4 M_N} \left[
2m_\pi^5 \delta(\mu^2-4m_\pi^2) - \frac{(2m_\pi^2-\mu^2)(m_\pi^2+\mu^2)}{\mu} \right]  \,,
\\
	{\rm Im} W_C(i\mu) &=& \frac{g_A^2}{256 f_\pi^4 M_N} \left\{
6 g_A^2 m_\pi^5 \delta(\mu^2-4m_\pi^2) + \frac{(2m_\pi^2-\mu^2)\left[\mu^2-2m_\pi^2+g_A^2(3m_\pi^2-2\mu^2)\right]}{\mu} \right\}  \,,
\nonumber \\
\\
	{\rm Im} V_S(i\mu) &=& \mu^2 \, {\rm Im} V_T(i\mu) =
	 \frac{3g_A^4 \mu}{1024 f_\pi^4 M_N} (5m_\pi^2-2\mu^2)  \,,
	\\
	{\rm Im} W_S(i\mu) &=& \mu^2 \, {\rm Im} W_T(i\mu) =
	 \frac{g_A^2\mu}{512 f_\pi^4 M_N} (g_A^2(3m_\pi^2-\mu^2)+\mu^2-4m_\pi^2) \,,
	\\
	{\rm Im} V_{LS} (i\mu) &=&  {3g_A^4  \over 128 \mu f_\pi^4 M_N} \, (2m_\pi^2-\mu^2)
 \,,\\  
 {\rm Im} W_{LS}(i\mu) &=& {g_A^2(1-g_A^2)\over 128 \mu f_\pi^4 M_N} \, (4m_\pi^2-\mu^2) \,. 
 \end{eqnarray}
\begin{eqnarray}
	\mbox{Position-space potentials:} && \nonumber \\
		\widetilde V_C(r) &=& \frac{3g_A^4 m_\pi^6}{1024 \pi^2 f_\pi^4 M_N} \frac{e^{-2x}}{x^6}
(24+48 x+46 x^2+28x^3+10 x^4+x^5)  \,,
\\
	\widetilde W_C(r) &=& \frac{g_A^2 m_\pi^6}{512 \pi^2 f_\pi^4 M_N} \frac{e^{-2x}}{x^6}
(24(2g_A^2-1)(1+2x)+(82g_A^2-40)x^2 +
\nonumber \\ &&
(36g_A^2-16)x^3+(10g_A^2-4)x^4+3g_A^2x^5) \,, 
\\
\widetilde V_S(r) &=& -\frac{g_A^4 m_\pi^6}{512 \pi^2 f_\pi^4 M_N} \frac{e^{-2x}}{x^6}
	(24+48 x+43x^2+22x^3+6x^4)  \,,
	\\
	\widetilde V_T(r) &=& \frac{g_A^4 m_\pi^6}{1024 \pi^2 f_\pi^4 M_N} \frac{e^{-2x}}{x^6}
	(48+96 x+76x^2+31x^3+6x^4)  \,,
        \\
	\widetilde W_S(r) &=& -\frac{g_A^2 m_\pi^6}{1536 \pi^2 f_\pi^4 M_N} \frac{e^{-2x}}{x^6} 
	(24(g_A^2-1)(1+2x)
	\nonumber \\ &&
	+2(21g_A^2-20)x^2+4(5g_A^2-4)x^3+4g_A^2x^4)  \,,
	\\
	\widetilde W_T(r) &=& \frac{g_A^2 m_\pi^6}{3072 \pi^2 f_\pi^4 M_N} \frac{e^{-2x}}{x^6} 
	(48(g_A^2-1)(1+2x)
	\nonumber \\ &&
	+8(9g_A^2-8)x^2+2(13g_A^2-8)x^3+4g_A^2x^4)  \,, \\
	\widetilde V_{LS}(r) &=& -\frac{3 g_A^4 m_\pi^6}{64 \pi^2 f_\pi^4 M_N} \frac{e^{-2x}}{x^6}
	(1+x)(2+2x+x^2)  \,, \\
	\widetilde W_{LS}(r) &=& \frac{g_A^2(g_A^2-1) m_\pi^6}{32 \pi^2 f_\pi^4 M_N} \frac{e^{-2x}}{x^6} (1+x)^2  \,.
\end{eqnarray}
In all $1/M_N$ corrections, we use the average nucleon mass, i.~e. $M_N=\bar{M}_N$
(cf.\ Table~\ref{tab_basic}), to avaoid randomly generated charge-dependence.

\subsection{Relativistic $c_i/M_N$ corrections}
\label{sec_cM}

\begin{figure}
\scalebox{0.7}{\includegraphics{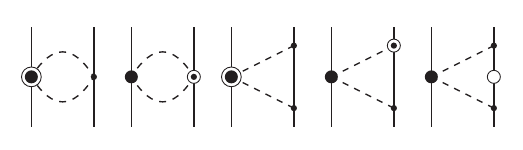}}
\caption{Relativistic corrections of NNLO diagrams.
Notation as in Fig.~\ref{fig_hi}.
Open circles  are relativistic $1/M_N$ corrections.
}
\label{fig_dia3}
\end{figure}

At N$^3$LO, we add the $1/M_N$ correction of the NNLO 2PE proportional to $c_i$.
This correction is proportional to $c_i/M_N$ (Fig.~\ref{fig_dia3}) and appears nominally at fifth order.
As discussed, the 2PE bubble diagram proportional to $c_i^2$
that appears at N$^3$LO is unrealistically attractive, while the $c_i/M_N$ correction is large
and repulsive. Therefore, it makes sense to group these diagrams together to arrive at a more
realistic intermediate attraction at N$^3$LO.
The contribution is given by~\cite{Kai01a}:
\begin{eqnarray}
\mbox{Momentum-space potentials:} && \nonumber \\
V_C(q) &=& -\frac{g_A^2 L(q)}{32 \pi^2 M_N f_\pi^4} \bigg[
		(c_2-6c_3)q^4+4(6c_1+c_2-3c_3) q^2 m_\pi^2
		\nonumber \\ &&
		+6(c_2-2c_3)m_\pi^4
	+24(2c_1+c_3)m_\pi^6 w^{-2} \bigg] \,,  \\
W_C(q) &=& -\frac{c_4 q^2 L(q)}{192 \pi^2 M_N f_\pi^4} \bigg[
	g_A^2(8m_\pi^2+5q^2)+w^2
	\bigg]  \,,  \\
W_T(q) &=& -\frac{1}{q^2} W_S(q) = - \frac{c_4 L(q)}{192\pi^2 M_N f_\pi^4}
	\bigg[ g_A^2(16m_\pi^2+7q^2)-w^2 \bigg]  \,,  \\
V_{LS}(q) &=& \frac{c_2 g_A^2}{8\pi^2 M_N f_\pi^4} w^2 L(q) \,, \\
W_{LS}(q) &=& -\frac{c_4 L(q)}{48\pi^2 M_N f_\pi^4} \left[ g_A^2(8m_\pi^2+5q^2)+w^2 \right] \,.
\end{eqnarray}
\begin{eqnarray}
\mbox{Spectral functions:} && \nonumber \\
{\rm Im} V_C(i\mu) &=& \frac{g_A^2}{64 \pi M_N f_\pi^4} 
	\frac{\sqrt{\mu^2-4m_\pi^2}}{\mu} \bigg[
		(c_2-6c_3)\mu^4-4(6c_1+c_2-3c_3) \mu^2 m_\pi^2
		\nonumber \\ &&
		+6(c_2-2c_3)m_\pi^4
	-24(2c_1+c_3) \frac{m_\pi^6}{\mu^2-4m_\pi^2} \bigg] \,, \\
{\rm Im} W_C(i\mu) &=& -\frac{c_4}{384 \pi M_N f_\pi^4} 
	\mu\sqrt{\mu^2-4m_\pi^2} \bigg[
	g_A^2(8m_\pi^2-5\mu^2)-\mu^2+4m_\pi^2
	\bigg]  \,,  \\
	{\rm Im} W_T(i\mu) &=& \frac{1}{\mu^2} {\rm Im} W_S(i\mu) = \frac{c_4}{384 \pi M_N f_\pi^4} 
	\frac{\sqrt{\mu^2-4m_\pi^2}}{\mu}
	 \bigg[ \mu^2-4m_\pi^2+g_A^2(16m_\pi^2-7\mu^2) \bigg]  \,,  \\
{\rm Im} V_{LS}(i\mu) &=& \frac{c_2 g_A^2}{16\pi M_N f_\pi^4} \frac{(\mu^2-4m_\pi^2)^{3/2}}{\mu}  \,, \\
{\rm Im} W_{LS}(i\mu) &=& \frac{c_4}{96\pi M_N f_\pi^4} \frac{\sqrt{\mu^2-4m_\pi^2}}{\mu}
	\left[ g_A^2(8m_\pi^2-5\mu^2)+ 4m_\pi^2-\mu^2 \right] \,.
\end{eqnarray}
\begin{eqnarray}
\mbox{Position-space potentials:} && \nonumber \\
\widetilde V_C(r) &=& \frac{3g_A^2 m_\pi^7}{32\pi^3 M_N f_\pi^4} \frac{1}{x^6} \bigg[
	\bigg( 20(c_2-6c_3)-4(6c_1-c_2+9c_3)x^2
	\nonumber \\ &&
	-2(2c_1+c_3)x^4 \bigg) x K_0(2x) 
	+\bigg(20(c_2-6c_3)-2(12c_1-7c_2+48c_3)x^2
	\nonumber \\ &&
	-(16c_1-c_2+10c_3)x^4\bigg) K_1(2x) 
	\bigg]  \,,  \\
\widetilde W_C(r) &=& \frac{c_4 m_\pi^7}{32\pi^3 M_N f_\pi^4} \frac{1}{x^5} \bigg[
	\bigg( 5+25 g_A^2+4g_A^2 x^2 \bigg) x K_1(2x)
	\nonumber \\ &&
	+ 2\bigg( 5+25 g_A^2+(1+8g_A^2)x^2 \bigg) K_2(2x)
	\bigg]  \,,  \\
\widetilde W_S(r) &=& \frac{c_4 m_\pi^7}{48\pi^3 M_N f_\pi^4} \frac{1}{x^5}
	\bigg[ \bigg( 5-35g_A^2-4g_A^2x^2 \bigg) x K_1(2x)
	\nonumber \\ &&
	+2\bigg( 5(1-7g_A^2)+(1-10g_A^2)x^2 \bigg) K_2(2x) \bigg]  \,,
	\\
\widetilde W_T(r) &=& \frac{c_4 m_\pi^7}{192\pi^3 M_N f_\pi^4} \frac{1}{x^5}
	\bigg[ 2\bigg( -8+59g_A^2+4g_A^2x^2 \bigg) x K_1(2x)
	\nonumber \\ &&
	-\bigg( 35(1-7g_A^2)+4(1-13g_A^2)x^2 \bigg) K_2(2x) \bigg]  \,,  \\
\widetilde V_{LS}(r) &=& \frac{3c_2 g_A^2 m_\pi^7}{8\pi^3 M_N f_\pi^4} \frac{1}{x^5}
	\bigg[ K_2(2x)+2xK_3(2x) \bigg]  \,, \\
\tilde W_{LS}(r) &=& -\frac{c_4 m_\pi^7}{16\pi^3 M_N f_\pi^4} \frac{1}{x^5}
	\bigg[ (1+6g_A^2)2xK_1(2x)+(5+25g_A^2+4g_A^2 x^2) K_2(2x) \bigg] \,.
\end{eqnarray}

\section{The LECs of the contact terms}
\label{app_lectab}

\begin{table}
\caption{Values for the contact LECs of the N$^3$LO potentials with cutoff combination $(R_\pi,R_{\rm ct})=(1.2,0.75)$ fm, $(1.1,0.72)$ fm, and $(1.0,0.70)$ fm. 
In the column headings, we use the $R_\pi$ value to identify the different cases.
The notation ($\pm n$) stands for $\times 10^{\pm n} $.
\label{tab_ctlecs}}
\smallskip
\begin{tabular*}{\textwidth}{@{\extracolsep{\fill}}lccc}
\hline 
\hline 
\noalign{\smallskip}
 LECs &  $R_{\pi}=1.2$ fm &  $R_{\pi}=1.1$ fm &  $R_{\pi}=1.0$ fm  \\
\hline
\noalign{\smallskip}
$C_{c}$ (${\rm fm}^{2}$) &0.28808881 (+1)   &0.39582494 (+1) &0.68583069 (+1) \\
$C_{\tau}$ (${\rm fm}^{2}$) &0.26865444  &0.37170364 &0.84621879\\
$C_{\sigma}$ (${\rm fm}^{2}$) & 0.37304419 (-1)   &0.13087859 &0.45593912 \\
$C_{\sigma \tau}$ (${\rm fm}^{2}$) &0.99745306 &0.86768636 &0.9008921 \\
\noalign{\smallskip}
$C_{1}$ (${\rm fm}^{4}$) &0.20339187 (-1) &-0.69958000 (-1)  &-0.19849806 \\
$C_{2}$ (${\rm fm}^{4}$) &-0.26911188 (-1)   & -0.73932500 (-2)  &0.27128125 (-2) \\
$C_{3}$ (${\rm fm}^{4}$) &-0.78260937 (-1) &-0.57466500 (-1) &-0.26448938 (-1) \\
$C_{4}$ (${\rm fm}^{4}$) &-0.35220625 (-2)  &-0.13702250 (-1) &-0.89698125 (-2) \\
$C_{5}$ (${\rm fm}^{4}$) &-0.10596750 (-1)  &-0.80355000 (-2) &-0.54697500 (-2) \\
$C_{6}$ (${\rm fm}^{4}$) & 0.31287500 (-2) &0.39985000 (-2) &0.48457500 (-2) \\
$C_{7}$ (${\rm fm}^{4}$) &-0.84559075  &-0.83002375 &-0.82673000 \\
$C_{8}$ (${\rm fm}^{4}$) &-0.11612925 &-0.10974825 &-0.10887000 \\
\noalign{\smallskip}
$D_{1}$ (${\rm fm}^{6}$) &0.27843312 (-1) &0.31251437 (-1)  &0.35406750 (-1) \\
$D_{2}$ (${\rm fm}^{6}$) &-0.11181250 (-3)   &0.30660625 (-2)   &0.64797500 (-2) \\
$D_{3}$ (${\rm fm}^{6}$) &0.17309375 (-2)   &0.39478125 (-2) &0.28025000 (-2) \\
$D_{4}$ (${\rm fm}^{6}$) &-0.25564375 (-2) &-0.11373125 (-2)  &-0.84200000 (-3) \\
$D_{5}$ (${\rm fm}^{6}$) &-0.22787500 (-2) &-0.17605000 (-2)  &0.13175000 (-3) \\
$D_{6}$ (${\rm fm}^{6}$) & -0.76425000 (-3)  &-0.58650000 (-3) &0.44250000 (-4) \\
$D_{7}$ (${\rm fm}^{6}$) &0.40027500 (-2)  &0.11374250 (-1)  &0.70485000 (-2) \\
$D_{8}$ (${\rm fm}^{6}$) &-0.26426750 (-1) &-0.22689250 (-1) &-0.29755500 (-1) \\
$D_{9}$ (${\rm fm}^{6}$) &-0.42584000 (-1) &-0.50699750 (-1) &-0.57539750 (-1) \\
$D_{10}$ (${\rm fm}^{6}$) &-0.14453000 (-1) &-0.16889250 (-1) &-0.19163250 (-1) \\
$D_{11}$ (${\rm fm}^{6}$) &-0.18565375 (-1)  &-0.27816625 (-1)  &-0.63730625 (-2) \\
$D_{12}$ (${\rm fm}^{6}$) & 0.16119625 (-1)  &0.11181125 (-1)  &0.20284813 (-1)  \\
$D_{13}$ (${\rm fm}^{6}$) & 0.54308750 (-2)   &0.25901250 (-2)  &0.77255625 (-2) \\
$D_{14}$ (${\rm fm}^{6}$) & 0.92428750 (-2) &0.76783750 (-2) &0.10042688 (-1)  \\
\noalign{\smallskip}
$C^{\rm CD}_{T_{12}}$ (fm$^2$) & 0.30527375 (-2)  & 0.3081975 (-2)  & 0.2791292 (-2)  \\
$C^{\rm CD}_{\sigma T_{12}}$ (fm$^2$) & -0.30527375 (-2) &  -0.3081975 (-2) & -0.2791292 (-2)  \\
$C^{\rm CA}_{\tau_z}$ (fm$^2$) & 0.17322500 (-2)  &  0.20032500 (-2) & 0.1817375 (-2)  \\
$C^{\rm CA}_{\sigma \tau_z}$ (fm$^2$) &  -0.17322500 (-2) &  -0.20032500 (-2) & -0.1817375 (-2)  \\
\hline
\noalign{\smallskip}
\end{tabular*}
\end{table}

In this Appendix, we show in Table~\ref{tab_ctlecs} 
the LECs of the contact terms defined in Sec.~\ref{sec_short}
for our N$^3$LO potentials. The shown LECs are the coefficients of the various contact operators
displayed in Sec.~\ref{sec_short}.

For the fitting of the phase shifts of the different states,
it is more convenient to fit to states with well-defined total spin $S$ and total isospin $T$,
the  (charge-independent)  LO coefficients of which we denote by $C_{ST}$.
From these $C_{ST}$, one obtains the LECs for the operators used in Eq.~(\ref{eq_ct0r})
via:
\begin{equation}
\left( {\begin{array}{l}
   C_{c}\\ 
  C_{\tau}\\
  C_{\sigma}\\
  C_{\sigma \tau}
  \end{array} } \right)
 = \frac{1}{16}
  \left( {\begin{array}{rrrr}
  1 &3 &3 & \;\; 9 \\  
  -1 &1 &-3 &3 \\ 
  -1 &-3 &1 &3 \\
  1 &-1 &-1 &1 
  \end{array} } \right)
  \left( {\begin{array}{c}
   C_{00}\\ 
  C_{01}\\
  C_{10}\\
  C_{11}
  \end{array} } \right)
  \label{eq_CST}
\end{equation}
Similar relations apply to the central force LECs of higher order, like the
$C_1$ to $C_4$ of Eq.~(\ref{eq_ct2r}) and the $D_1$ to $D_4$ of Eq.~(\ref{eq_ct4r});
as well to
the coefficients of the four ${\vec L}^2$ terms, $D_{11}$ to $D_{14}$
 [Eq.~(\ref{eq_ct4r})].
 
 Vice versa, the spin-isospin coefficients can be obtained from the operator LECs via:
\begin{equation}
\left( {\begin{array}{c}
   C_{00}\\ 
  C_{01}\\
  C_{10}\\
  C_{11}
  \end{array} } \right)
 =
  \left( {\begin{array}{rrrr}
  1 &-3 &-3 &9 \\ 
  1 &1 &-3 &-3 \\ 
  1 &-3 &1 &-3 \\
  1 &1 &1 &1 
  \end{array} } \right)
  \left( {\begin{array}{l}
   C_{c}\\ 
  C_{\tau}\\
  C_{\sigma}\\
  C_{\sigma \tau}
  \end{array} } \right)
  \label{eq_CCT}
\end{equation}

Tensor, spin-orbit, and quadratic spin-orbit terms exist only in $S=1$ states, 
such that one needs to distinguish only between a $T=0$ and $T=1$ channel.
For example, in the case of the NLO tensor force, the relations are:
\begin{eqnarray}
C_5 \, \equiv \, C_{S_{12}}  &=&  \frac14 \left( C_{10}^{(S_{12})} + 3 C_{11}^{(S_{12})} \right) \,, 
\nonumber \\
C_6 \, \equiv \, C_{S_{12} \tau}  &=&  \frac14 \left( -C_{10}^{(S_{12})} + C_{11}^{(S_{12})} \right) \label{eq_chan_op} \,,
\end{eqnarray}
and vice versa
\begin{eqnarray}
C_{10}^{(S_{12})} &=& C_{S_{12}} - 3C_{S_{12} \tau} = C_5 - 3 C_6 \,, \nonumber \\
C_{11}^{(S_{12})} &=& C_{S_{12}} + C_{S_{12} \tau} = C_5 + C_6 \,,
\end{eqnarray}
and similarly for the other cases that appear only at $S=1$.

To reproduce the three charge dependent $^1S_0$ scattering lenghts,
the LO contact LEC with $(S,T)=(0,1)$  is fit in a charge-dependent way. Thus, this LEC
comes in three versions: $C_{01}^{pp}$, $C_{01}^{np}$, and $C_{01}^{nn}$.
In tune with Eqs.~(\ref{eq_ct0r}) and (\ref{eq_ct0rcd}), the charge-dependent LEC
can be represented by
\begin{eqnarray}
C_{01}^{NN} &=& C_{01} + C_{01}^{\rm CD}  T_{12} + C_{01}^{\rm CA} (\tau_{1z} + \tau_{2z}) 
\end{eqnarray}
with $T_{12}$ defined in Eq.~(\ref{eq_T12}).
$C_{01}$ denotes the charge-independent value, which is fixed by
\begin{eqnarray}
C_{01} &=& \frac13 \left( C_{01}^{pp} + C_{01}^{np} + C_{01}^{nn} \right) \,,
\end{eqnarray}
while the charge-dependent ones are
\begin{eqnarray}
C_{01}^{\rm CD} &=& \frac16 \left[ \frac12 \left( C_{01}^{pp} + C_{01}^{nn} \right)
 - C_{01}^{np} \right] \,\, \mbox{\rm and} \\
 C_{01}^{\rm CA} &=&  \frac14 \left( C_{01}^{pp} - C_{01}^{nn} \right) \,.
\end{eqnarray}
By analogy to Eqs.~(\ref{eq_chan_op}), 
 the operator LECs used in Eq.~(\ref{eq_ct0rcd})
can be obtained from  the channel LECs 
through:
\begin{eqnarray}
C_{T_{12}}^{\rm CD}  &=&  \frac14 \left( C_{01}^{\rm CD} + 3 C_{11}^{\rm CD} \right) \,, 
\nonumber \\
C_{\sigma T_{12}}^{\rm CD}  &=&  \frac14 \left( -C_{01}^{\rm CD} + C_{11}^{\rm CD} \right) \,.
\end{eqnarray}
We do not assume any charge dependence for the contacts in $S=1,T=1$ states 
(triplet $P$-waves); therefore, we have $C_{11}^{\rm CD} = 0$.
Thus,
\begin{eqnarray}
C_{T_{12}}^{\rm CD}  &=&  \frac14 \left( C_{01}^{\rm CD} \right) \,, \nonumber \\
C_{\sigma T_{12}}^{\rm CD}  &=&  \frac14 \left( -C_{01}^{\rm CD} \right) \,.
\end{eqnarray}
Similar relations apply to charge asymmetry,
\begin{eqnarray}
C_{\tau_z}^{\rm CA}  &=&  \frac14 \left( C_{01}^{\rm CA} + 3 C_{11}^{\rm CA} \right) \,, 
\nonumber \\
C_{\sigma \tau_z}^{\rm CA}  &=&  \frac14 \left( -C_{01}^{\rm CA} + C_{11}^{\rm CA} \right) \,.
\end{eqnarray}
Also here, we do not assume any charge asymmetry for the contacts in $S=1,T=1$ states; 
thus, $C_{11}^{\rm CA} = 0$; hence
\begin{eqnarray}
C_{\tau_z}^{\rm CA}  &=&  \frac14 \left( C_{01}^{\rm CA}  \right) \,, \nonumber \\
C_{\sigma \tau_z}^{\rm CA}  &=&  \frac14 \left( -C_{01}^{\rm CA}  \right) \,.
\end{eqnarray}

A final aspect to discuss is the question to what extend  the LECs are natural.
LECs may be perceived as natural if they are of the following magnitudes:
\begin{eqnarray}
|C_{c,\tau,\sigma,\sigma\tau}| &\sim& \frac{1}{f_\pi^2} \approx 5 \,\, \mbox{\rm fm}^2 \,, \\
|C_{i}| &\sim& \frac{1}{f_\pi^2 \, \Lambda_b^2} \approx 0.4 \,\, \mbox{\rm fm}^4 \,, \\
|D_{i}| &\sim& \frac{1}{f_\pi^2 \, \Lambda_b^4} \approx 0.03 \,\, \mbox{\rm fm}^6 \,,
\end{eqnarray}
with $\Lambda_b \approx m_\rho \approx $ 0.7 GeV the 
breakdown scale~\cite{Fur15}.

Comparing these estimates with the values shown in Table~\ref{tab_ctlecs} reveals that our contact LECs are, in general, 
natural.
At zeroth order, $C_c$ is certainly of the right order, and the 
$C_{\tau,\sigma,\sigma\tau}$ are around one, which is close enough to the estimate.
At second order, $C_1$ and the $LS$ force parameters, $C_7$ and $C_8$  are of the right size, while the other LECs are on the smaller side.
Finally at fourth order, $D_1$, the $LS$ parameter $D_8$, the
$(LS)^2$ parameters $D_9$ and $D_{10}$,
and the $L^2$ LECs $D_{12}$ and $D_{14}$ come out natural,
whereas the other $D_i$ emerge in small format.

\section{Potential plots}
\label{app_plots}

In this appendix, we show figures for the various components of the chiral $NN$ potentials
and contrast them with some well known traditional phenomenology.

\begin{figure}[t]
\scalebox{0.4}{\includegraphics{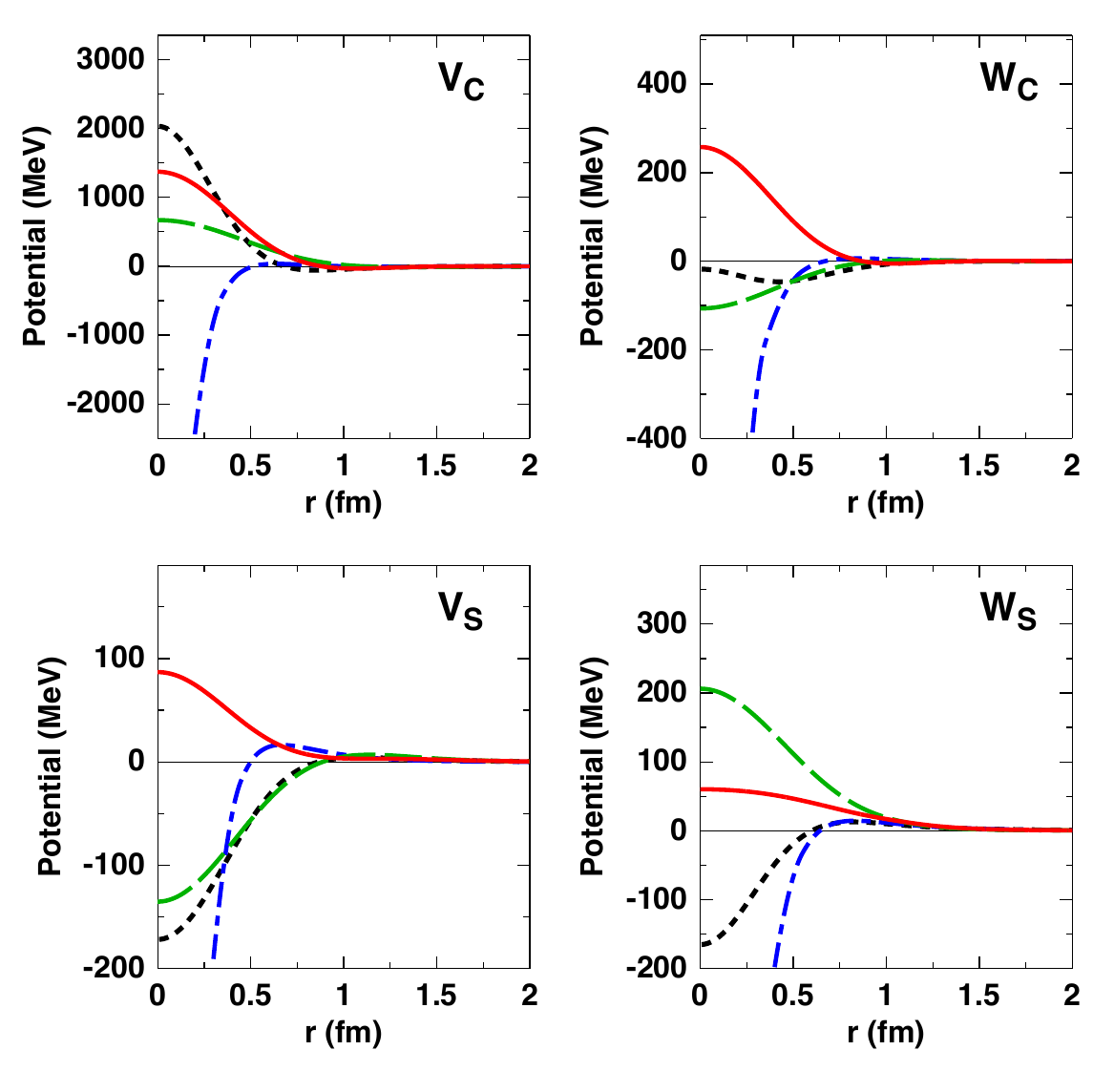}}
\hspace*{1.0cm}
\scalebox{0.4}{\includegraphics{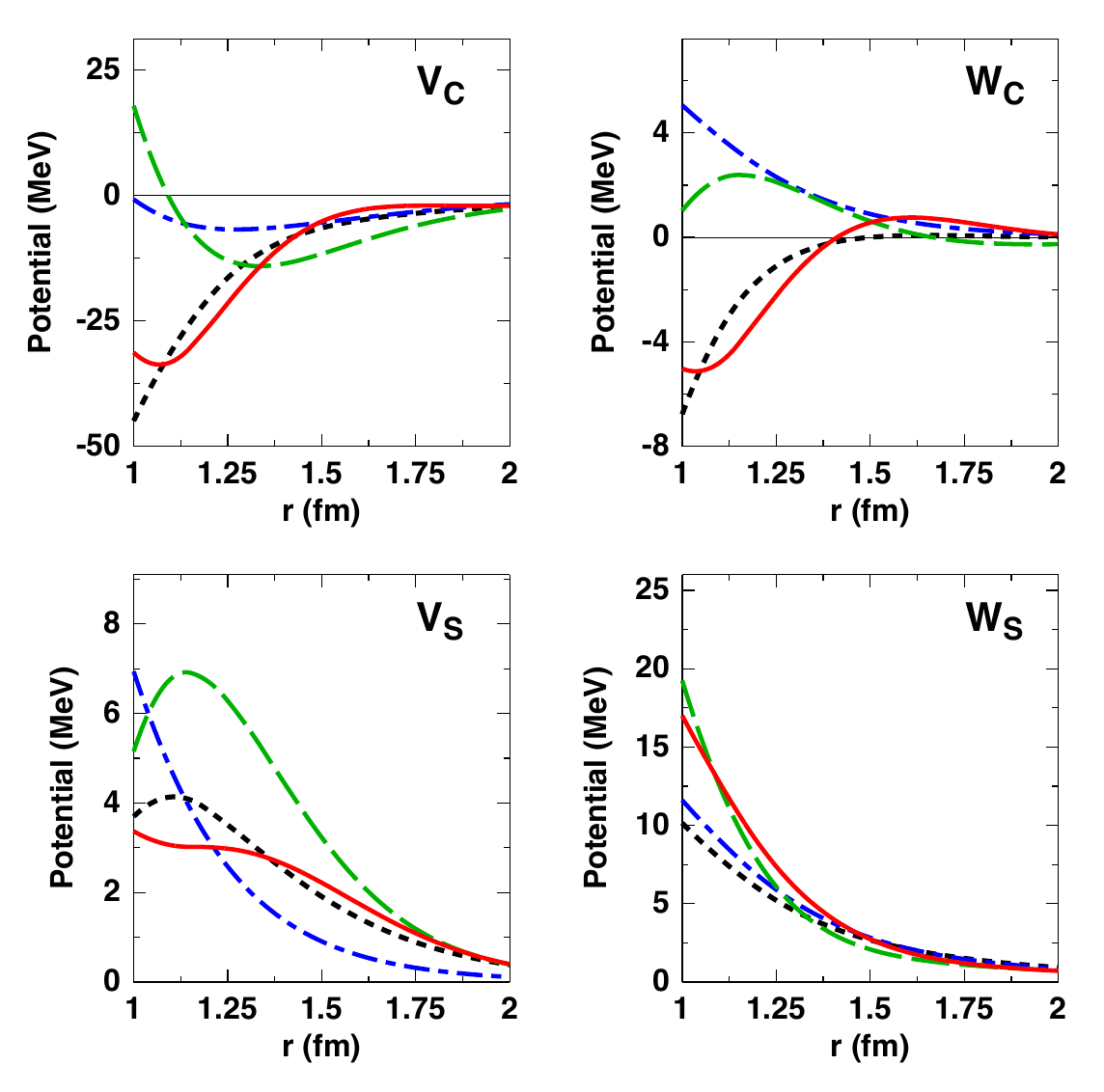}}
\caption{Left side: The four central potentials in the range 0 to 2 fm.
Right side: The same in the range 1 to 2 fm.
To mark the various potential components, the notation of Eq.~(\ref{eq_nnr}) is used
(with the tilde omitted).
Predictions are shown for the AV18 potential~\cite{WSS95} (black dotted line),
an OBEP~\cite{note1} (blue dash-dotted line),
the chiral NNLO potential of this work (green dashed line), and
the chiral N$^3$LO potential of this work (red solid line).
For the chiral potentials, the cutoff combination $(R_\pi,R_{\rm ct})=(1.0,0.70)$ fm is used.
}
\label{fig_0a}
\end{figure}

In Fig.~\ref{fig_0a}, we compare the four central-potential components
(notation as in Eq.~(\ref{eq_nnr}), but without the tilde)
as predicted by the chiral potentials at NNLO and N$^3$LO  (green dashed and red solid lines,
respectively) with two phenomenological potentials, namely,
the AV18 potential~\cite{WSS95} and a one-boson-exchange potential (OBEP)~\cite{note1}
(black dotted and blue dash-dotted lines, respectively).
The chiral potentials apply the cutoff combination $(R_\pi,R_{\rm ct})=(1.0,0.70)$ fm.
While at short range ($r< 1$ fm) there are large differences between the models,
there is qualitative agreement between most models in the (more important) intermediate range 
($1<r<2$ fm) as revealed in the right side of Fig.~\ref{fig_0a}. In particular, there
is good agreement between the N$^3$LO potential and AV18, providing
support from chiral EFT for the AV18 potential.

From the left $V_C$ panel of Fig.~\ref{fig_0a}, it may appear that OBEP (blue dash-dotted curve) does not create
a hard core (repulsive short range force). This is misleading, because a hard core
is needed for the $S$-wave states. For the $^1S_0$ state, the $V_S$ and the $W_S$
potentials are multiplied by a factor of $(-3)$, which creates strong short-range repulsion
(cf.\ Fig.~\ref{fig_0c} below). For $^3S_1$, $W_C$ and $W_S$ are multiplied by $(-3)$
producing the hard core.

\begin{figure}
\scalebox{0.5}{\includegraphics{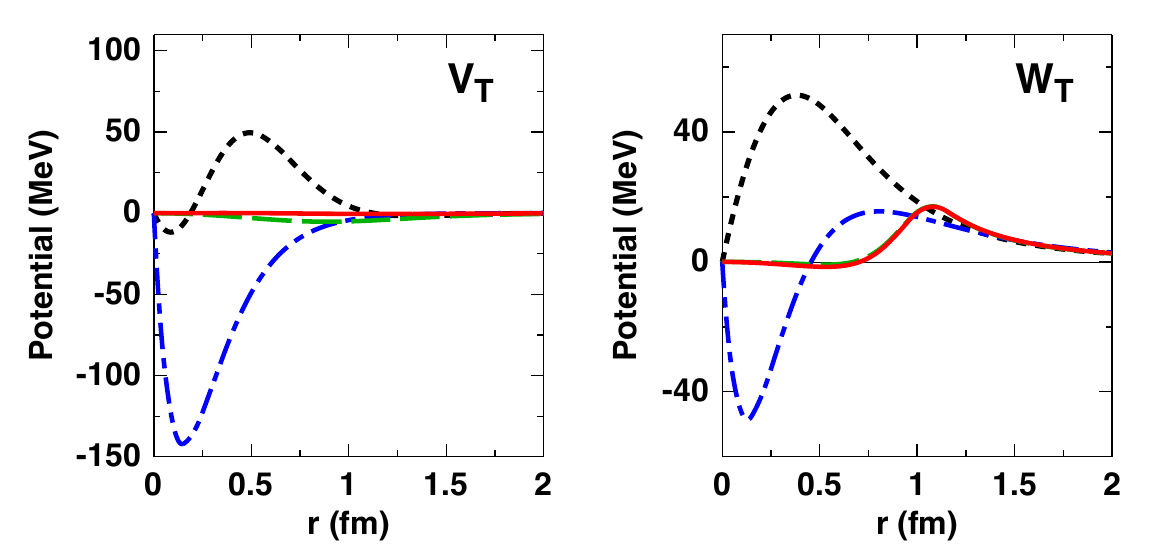}}
\caption{
The tensor potentials. Notation as in Fig.~\ref{fig_0a}.
}
\label{fig_0f1}
\end{figure}

Tensor potentials as shown in Fig.~\ref{fig_0f1}.
It is clearly seen that the chiral tensor potentials are much weaker than, particularly, the AV18.
Note that, in the case of OBEP (blue dash-dotted lines), the negative short-range potential of
$V_T$ is essentially due to the $\omega$ meson and a similar curve in $W_T$ is
due to the $\rho$ meson. Both these heavy vector mesons have no place in
chiral EFT, which is why the chiral EFT predictions are essentially flat in the short-range region
(unless there were large tensor contact contributions, which our chiral potentials do not carry).
The $W_T$ tensor force in the intermediate- and long-range region is
generated from 1PE for all models, which is why there is agreement between all models
above $r>1$ fm.

\begin{figure}
\scalebox{1.01}{\includegraphics{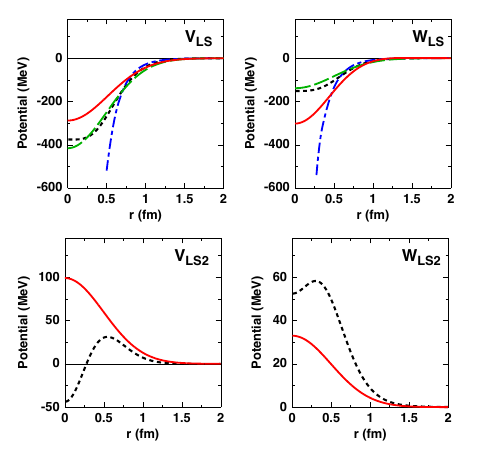}}
\scalebox{0.4}{\includegraphics{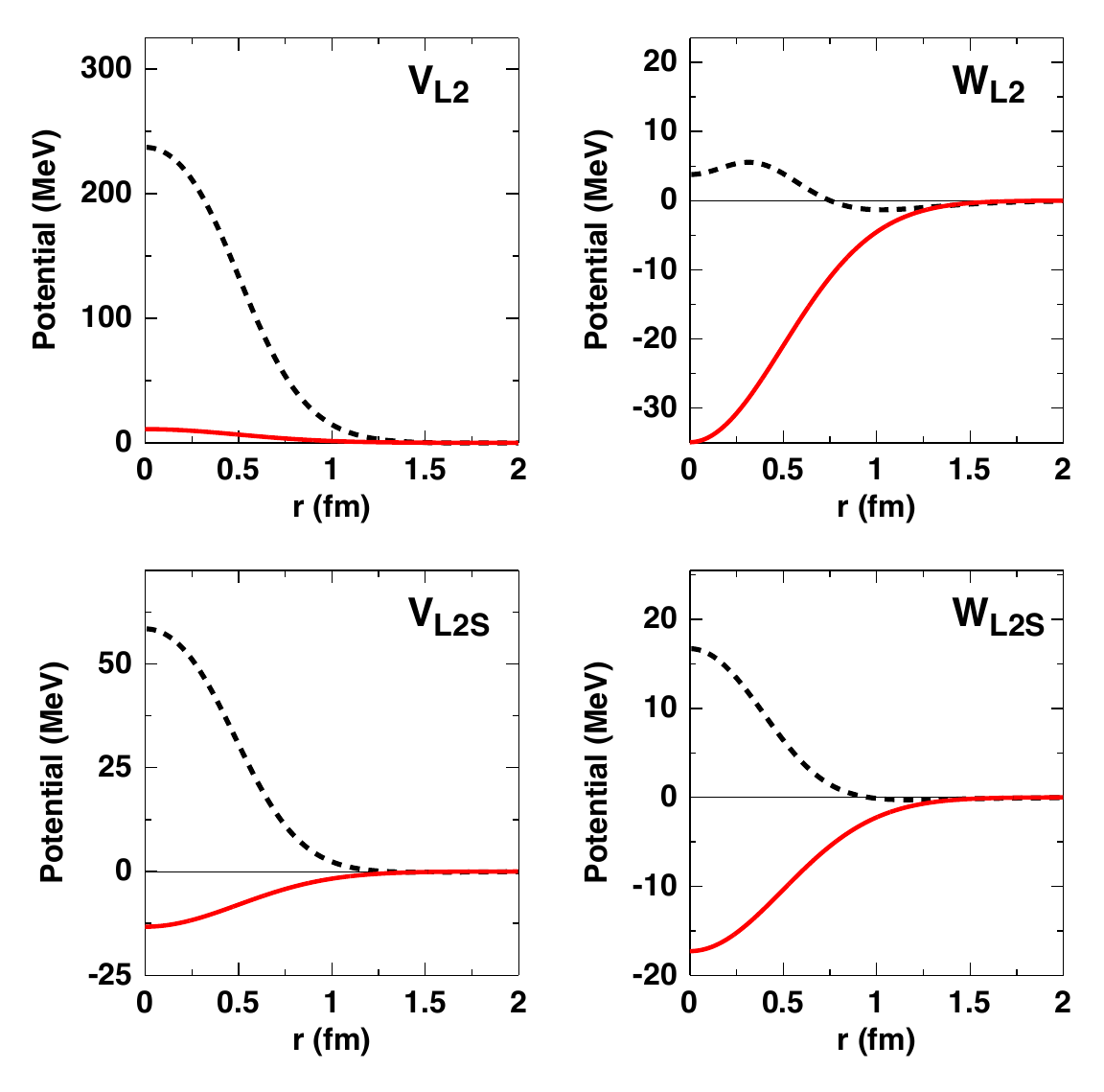}}
\caption{The eight $\vec L$-dependent potentials.
Notation as in Fig.~\ref{fig_0a}.
Moreover, 
$LS2$ stands for $(\vec L \cdot \vec S)^2$,
$L2$ for ${\vec L}^2$, and
$L2S$ for ${\vec L}^2  \vec\sigma_1 \cdot \vec \sigma_2 $.
Note that the OBEP and the chiral potential at NNLO do not include $(\vec L \cdot \vec S)^2$ and ${\vec L}^2$ components.
}
\label{fig_chr1a}
\end{figure}

The eight potential components that depend on the orbital angular momentum operator 
$\vec L$ are displayed in Fig.~\ref{fig_chr1a}. For $V_{LS}$ and $W_{LS}$ there is 
qualitative agreement between all models. Triplet $P$ waves cannot be described
quantitatively without a proper strong spin-orbit force, which is presumably 
the reason for this agreement.
Note that the chiral potential at NNLO and OBEP do not have 
 $(\vec L \cdot \vec S)^2$ and ${\vec L}^2$ components.
 For the  $(\vec L \cdot \vec S)^2$ potentials, there is rough agreement
 between N$^3$LO and AV18 for $r>0.5$ fm.
 On the other hand, the four ${\vec L}^2$ potentials appear erratic.
 Obviously, these components of the nuclear force are not well pinned down.
 They are also small, which may be why they are not so relevant and hard to pin down.

Some important partial-wave potentials are shown in Fig.~\ref{fig_0c}.
In the $^1S_0$ state, all models exhibit a strong short-range repulsion, the size
of which, however, differs dramatically. Nevertheless, there is agreement between the models 
in the (more relevant) range above 0.5 fm as demonstrated in the second $^1S_0$ frame
of the figure. 
The differences in size of the tensor forces of different models is
best demonstrated by way of the $^3S_1$-$^3D_1$ transition potential,
which we show in the third panel of Fig.~\ref{fig_0c}.
The  AV18 potential
has the strongest tensor force, OBEP is second, and NNLO and N$^3$LO have the weakest.
As discussed, for $r>1$ fm, 1PE is the dominant tensor force in all models, which is why all models agree
in that region.

Finally, we also wish to provide some idea for the cutoff dependence of the chiral potentials.
For that purpose we show, in Fig.~\ref{fig_chr7}, the
$^1S_0$ and $^3S_1$-$^3D_1$  potentials at N$^3$LO for 
the cutoff combinations $(R_\pi,R_{\rm ct})=(1.0,0.70)$ fm, $(1.1,0.72)$ fm, and $(1.2,0.75)$ fm
(solid, dashed, and dotted curves, respectively). 
The short-range parts of the $^1S_0$ potentials exemplify the effect of the short-range
cutoff on the central forces,
Eqs.~(\ref{eq_cut0}), (\ref{eq_cut2}), and (\ref{eq_cut4}),
  (ruled by $R_{\rm ct}$), while the $^3S_1$-$^3D_1$ potentials demonstrate
the impact of the long-range regulator function, Eq.~(\ref{eq_reg1pe}), (governed by $R_\pi$).

%

\begin{figure}
\scalebox{0.5}{\includegraphics{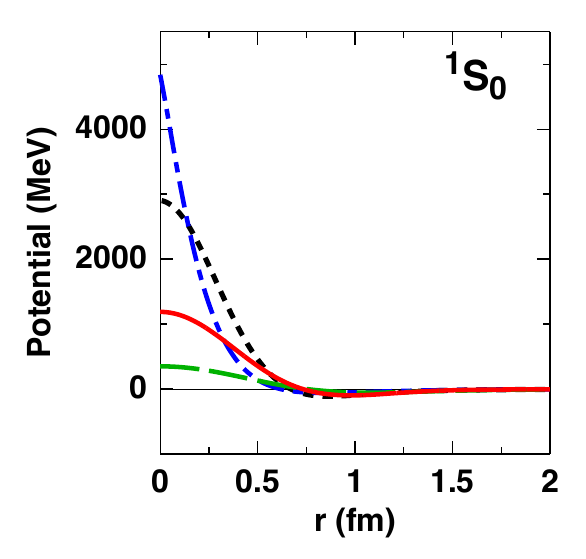}}
\scalebox{0.5}{\includegraphics{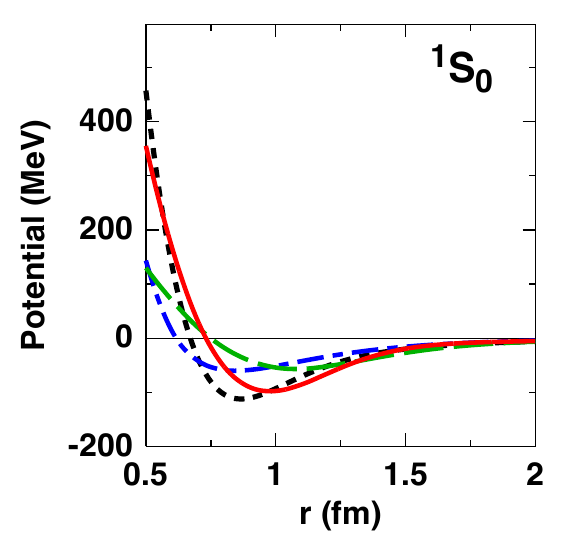}}
\scalebox{0.5}{\includegraphics{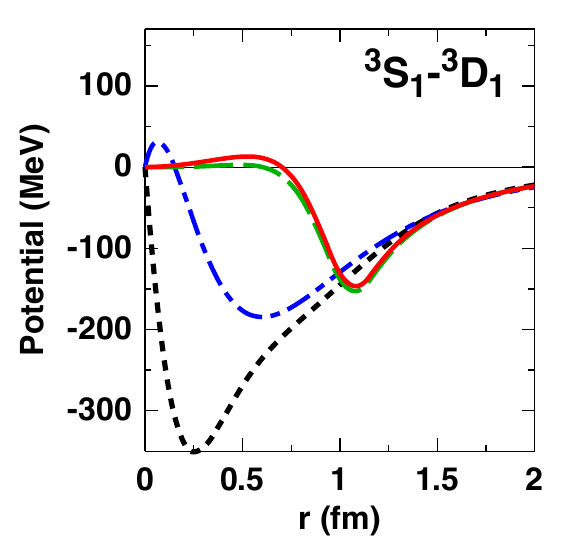}}
\caption{The $^1S_0$ potential in the range 0 to 2 fm and 0.5 to 2 fm, as well as
the $^3S_1$-$^3D_1$ potential in the range 0 to 2 fm.
Notation as in Fig.~\ref{fig_0a}.
}
\label{fig_0c}
\end{figure}

\begin{figure}\centering
\scalebox{0.55}{\includegraphics{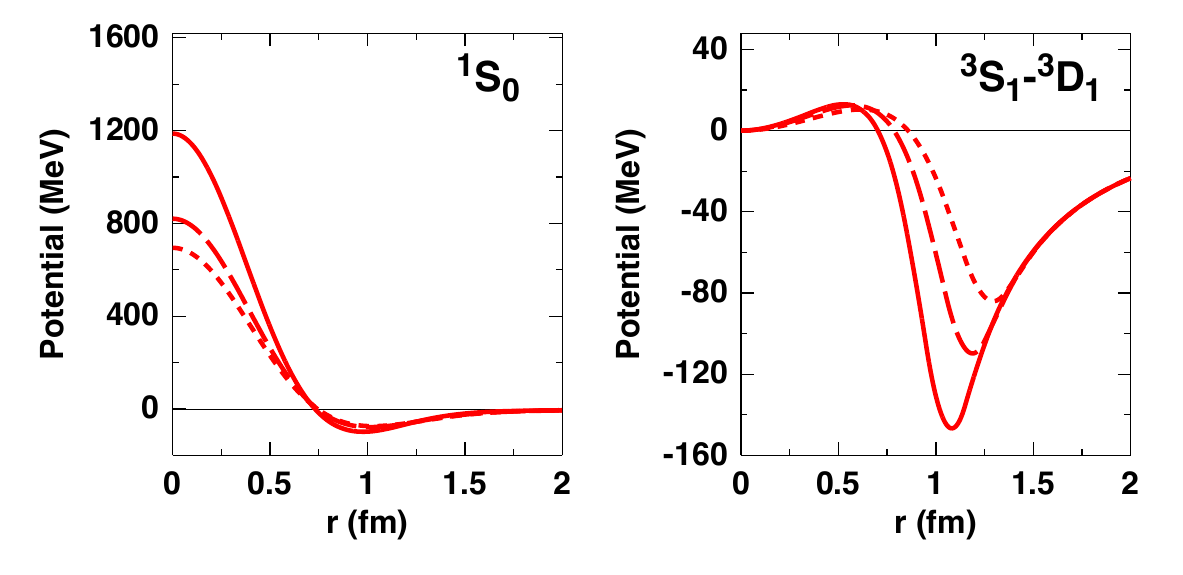}}
\caption{Cutoff dependence of the $^1S_0$ and $^3S_1$-$^3D_1$ chiral potentials
at N$^3$LO.
The cutoff combinations $(R_\pi,R_{\rm ct})=(1.0,0.70)$ fm, $(1.1,0.72)$ fm, and $(1.2,0.75)$ fm
are shown by the solid, dashed, and dotted curves.
}
\label{fig_chr7}
\end{figure}


\begin{thebibliography}{99}
\bibitem{note2}
A user-friendly FORTRAN code for all $NN$ potentials presented in this paper can be obtained 
from one of the authors (R.M.) upon request.
\end{thebibliography}
\end{document}